\documentclass[12pt]{article}

\ifx\pdfoutput\undefined
\usepackage[dvips,bookmarks]{hyperref}
\else
\usepackage{hyperref}
\fi
\hypersetup{colorlinks=false,bookmarksopen,bookmarksnumbered,citecolor=blue,
   pdfstartview=FitH}

\usepackage[pdftex]{graphicx}	
\usepackage{latexsym}

\usepackage{amssymb,amsfonts,amsmath}
\usepackage{graphicx} 
\usepackage{indentfirst}

 \usepackage{bbm}

\parskip=\medskipamount

\newcommand {\cB}{{\cal B}}
\newcommand {\cC}{{\cal C}}
\newcommand {\cD}{{\cal D}}
\newcommand {\cE}{{\cal E}}
\newcommand {\cF}{{\cal F}}
\newcommand {\cG}{{\cal G}}
\newcommand {\cH}{{\cal H}}

\newcommand {\cJ}{{\cal J}}
\newcommand {\cK}{{\cal K}}
\newcommand {\cL}{{\cal L}}
\newcommand {\cM}{{\cal M}}
\newcommand {\cN}{{\cal N}}
\newcommand {\cO}{{\cal O}}

\newcommand {\cR}{{\cal R}}
\newcommand {\cS}{{\cal S}}
\newcommand {\cT}{{\cal T}}

\newcommand {\cW}{{\cal W}}

\newcommand {\cY}{{\cal Y}}
\newcommand {\cZ}{{\cal Z}}

\newcommand{\bD}{{\bf D}}

\newcommand{\bH}{{\bf H}}
\newcommand{\bI}{{\bf I}}
\newcommand{\bJ}{{\bf J}}
\newcommand{\bK}{{\bf K}}
\newcommand{\bL}{{\bf L}}

\def\a{\alpha}

\def\b{\beta}
\def\c{\chi}
\def\d{\delta}
\def\e{\epsilon}
\def\f{\phi}
\def\g{\gamma}
\def\G{\Gamma}

\def\k{\kappa}
\def\l{\lambda}
\def\m{\mu}

\def\o{\omega}

\def\q{\theta}
\def\r{\rho}
\def\s{\sigma}
\def\t{\tau}
\def\u{\upsilon}
\def\x{\xi}

\def\F{\Phi}

\def\L{\Lambda}
\def\O{\Omega}

\def\Q{\Theta}
\def\S{\Sigma}
\def\U{\Upsilon}
\def\X{\Xi}

\def\rd{{\rm d}}
\def\ri{{\rm i}}
\def\re{{\rm e}}

\newcommand{\ve}{\varepsilon}                            
\newcommand{\cDB}{{\bar\cD}}                            

\newcommand{\pa}{\partial}                           
\newcommand{\hf}{\frac12}

\newcommand{\vf}{\varphi}

\newcommand{\be}{\begin{equation}}
\newcommand{\ee}{\end{equation}}
\newcommand{\bea}{\begin{eqnarray}}
\newcommand{\eea}{\end{eqnarray}}
\newcommand{\non}{\nonumber}
\newcommand{\1}{{\underline{1}}}

\newcommand{\bm}[1]{\mbox{\boldmath$#1$}}

\def\double #1{#1{\hbox{\kern-2pt $#1$}}}

\newcommand{\teb}{{\bar{\theta}}}
\newcommand{\qb}{{\bar{\theta}}}

\newif\ifdtup

\def\de{{\nabla}}                                         

\def\deb{{\bar{\de}}}

\newcommand{\bsubeq}{\begin{subequations}}
\newcommand{\esubeq}{\end{subequations}}

\newcommand{\vfb}{{\bar{\varphi}}}

\newcommand{\lb}{{\bar{\l}}}

\newcommand{\mbG}{{\mathbb{G}}}

\renewcommand{\vfb}{{\bar{\vf}}}
\renewcommand{\bI}{{\bar{I}}}
\renewcommand{\bJ}{{\bar{J}}}
\renewcommand{\bK}{{\bar{K}}}
\renewcommand{\bL}{{\bar{L}}}

\newcommand{\Xb}{{\bar{X}}}

\newcommand{\fb}{{{\bar{\f}}}}
\newcommand{\Fb}{{{\bar{\F}}}}

\numberwithin{equation}{section}

\begin{document}

\begin{titlepage}
\begin{flushright}
UUITP-20/13\\
LMU-ASC 87/13\\
December, 2013\\
\end{flushright}
\vspace{5mm}

\begin{center}
{\Large \bf 
Three-dimensional
$\bm{\cN = 2}$  supergravity theories: From superspace to components}
\\ 
\end{center}

\begin{center}

{\bf
Sergei M. Kuzenko${}^{a}$, Ulf Lindstr\"om${}^{b}$, Martin Ro\v cek${}^{c}$, Ivo Sachs${}^d$\\
and
Gabriele Tartaglino-Mazzucchelli${}^{a}$
} \\
\vspace{5mm}

\footnotesize{
${}^{a}${\it School of Physics M013, The University of Western Australia\\
35 Stirling Highway, Crawley W.A. 6009, Australia}}  
\vspace{2mm}

\footnotesize{
${}^{b}${\it Theoretical Physics, Department of Physics and Astronomy,
Uppsala University \\ 
Box 516, SE-751 20 Uppsala, Sweden}}
\vspace{2mm}

${}^c${\it C.N.Yang Institute for Theoretical Physics, Stony Brook University \\
Stony Brook, NY 11794-3840,USA}
\vspace{2mm}

${}^d${\it Arnold Sommerfeld Center for Theoretical Physics, Ludwig-Maximilians University\\
Theresienstr. 37, D-80333 M\"unchen, Germany}

\end{center}

\begin{abstract}
\baselineskip=14pt

For general  off-shell $\cN=2$ supergravity-matter 
systems in three spacetime dimensions, 
a formalism is developed to reduce the corresponding actions from superspace 
to components. The component actions are explicitly 
computed
in the cases of Type I and Type II minimal supergravity formulations.
We describe the models for topologically massive supergravity which 
correspond to  all the known off-shell formulations
for three-dimensional  $\cN=2$ supergravity. 
We also present a universal setting to construct supersymmetric backgrounds 
associated with these off-shell supergravities.

\end{abstract}

\vfill

\vfill
\end{titlepage}

\newpage
\renewcommand{\thefootnote}{\arabic{footnote}}
\setcounter{footnote}{0}

\tableofcontents

\section{Introduction}

The simplest way to construct  $\cN=2$ locally supersymmetric systems in three  spacetime dimensions (3D) is perhaps through dimensional reduction from 4D $\cN=1$ theories 
(see \cite{WB,GGRS,BK} for reviews).
However, not all 3D theories with four supercharges can be obtained in this way. For instance,
$\cN=2$ conformal supergravity  \cite{RvN} and (2,0) anti-de Sitter (AdS) 
supergravity\footnote{In three dimensions, 
$\cN$-extended AdS supergravity exists in $[\cN/2] +1$ different versions \cite{AT}, 
with $[\cN/2]$ the integer part of $\cN/2$.
These were called the  $(p,q)$ AdS supergravity theories  
where the  non-negative integers $p \geq q$ are such that 
$\cN=p+q$. These theories  
are naturally  associated with the 3D AdS supergroups
$\rm OSp (p|2; {\mathbb R} ) \times  OSp (q|2; {\mathbb R} )$.}  
\cite{AT} can not be so constructed. 
A more systematic approach  to generate 3D $\cN=2$ supergravity-matter systems 
is clearly desirable. 

Matter couplings in three-dimensional $\cN=2$ supergravity were thoroughly studied 
in the 1990s using on-shell component approaches \cite{dWNT,IT,DKSS}
(see also \cite{DKSST}). More recently, 
off-shell formulations for general $\cN=2$ supergravity-matter systems 
have systematically been developed 
\cite{KLT-M11,KT-M11}  purely within the superspace framework, 
extending earlier off-shell constructions \cite{ZP88,NG,HIPT}.
One of the main goals of this paper is to work out techniques
to reduce any manifestly $\cN=2$ locally supersymmetric theory presented 
in \cite{KLT-M11,KT-M11} to components. 
Upon elimination of the auxiliary fields, one naturally reproduces the partial component 
results obtained earlier in \cite{dWNT,IT,DKSS}. 

The prepotential formulation for 3D $\cN=2$ conformal supergravity 
was constructed in \cite{Kuzenko12}.  In principle, 
this prepotential solution could be obtained by off-shell dimensional reduction from 
4D $\cN=1$ conformal supergravity following the procedure sketched in section 
7.9 of {\it Superspace} \cite{GGRS}. In practice, however, it is more advantageous to follow a manifestly covariant approach and derive the solution from scratch. 
In this sense the 3D story  is similar to that of $\cN=(2,2)$ supergravity in two dimensions
\cite{GrisaruW1,GrisaruW2}.

Similarly to $\cN=1$ supergravity in four dimensions 
(see \cite{Howe,deWR,GGRS,BK} for more details),
different off-shell formulations for 3D $\cN=2$ Poincar\'e and AdS
supergravity theories 
in superspace 
can be obtained by coupling conformal supergravity to 
different conformal compensators \cite{KLT-M11,KT-M11}.   
There are three inequivalent types of conformal compensator: (i) a chiral scalar; 
(ii) a real linear scalar; and (iii) a (deformed) complex linear scalar.  

Choosing the chiral  compensator 
leads to the Type I minimal supergravity \cite{KT-M11} which is the 3D analogue of 
the old minimal formulation for 4D $\cN=1$ supergravity  \cite{old}.
As in four dimensions, this formulation can be used to realize both Poincar\'e and AdS 
supergravity theories; the latter actually describes the so-called (1,1) AdS supergravity, 
following the terminology of \cite{AT}.

Choosing the real linear  compensator leads to 
the Type II minimal supergravity \cite{KT-M11} which is a natural extension of 
the new minimal formulation for 4D $\cN=1$ supergravity  \cite{new}.
Unlike the four-dimensional case, the Type II formulation is suitable to realize 
both  Poincar\'e and AdS supergravity theories (the new minimal formulation 
cannot be used to describe 4D $\cN=1$ AdS supergravity). 
The point is that in three dimensions
the real linear superfield is the field strength of an Abelian vector multiplet, 
and the corresponding Chern-Simons terms may be interpreted as 
a cosmological term \cite{HIPT}. 
Adding such a Chern-Simons term to the supergravity action results in the action 
for  (2,0) AdS supergravity. 

Finally, choosing the complex linear compensator leads to 
the non-minimal supergravity presented in \cite{KT-M11}. 
It is analogous to the non-minimal formulation for 4D $\cN=1$ 
supergravity \cite{non-min,SG}, the oldest off-shell locally supersymmetric theory. 
Both in three and four dimensions, this formulation exists in several versions
labeled by a real parameter $n\neq -1/3, 0$ in the 4D case \cite{SG} 
or, more conveniently,  by  $w = (1-n)/(3n+1)$ in the 3D case \cite{KT-M11}.
The reason for such a freedom is  that the super-Weyl transformation 
of the complex linear compensator is not fixed uniquely \cite{KLT-M11}.
With the standard constraint 
\bea
(\bar \cD^2 - 4R) \S =0
\label{1.1}
\eea 
obeyed by the complex linear compensator $\S$, the 4D $\cN=1$ non-minimal  formulation 
is only suitable, for any value of $n$, 
to describe Poincar\'e supergravity  \cite{GGRS}. 
The situation in the 3D case is completely similar \cite{KT-M11}.
However, it was shown in \cite{ButterK11} that $n=-1$ non-minimal supergravity 
can be used to describe 4D $\cN=1$ AdS supergravity 
provided the constraint \eqref{1.1} is replaced
with a  deformed one,\footnote{The constraint \eqref{1.2}  is 
super-Weyl invariant if and only if $n=-1$.}
\bea
(\bar \cD^2 - 4R) \G = -4\m\neq 0~, \qquad  \m = {\rm const} ~.
\label{1.2}
\eea
Applying the same ideas in the 3D case gives us the non-minimal formulation 
for (1,1) AdS supergravity \cite{KT-M11}.

All supergravity-matter actions introduced in \cite{KLT-M11,KT-M11} are realized as 
integrals over the full  superspace or over its chiral subspace.
The most economical way to reduce such an action to components consists in 
recasting it as an integral of a closed super three-form  
over spacetime (that is, the bosonic body of the full superspace), 
in the spirit of  the superform approach\footnote{It is also known
as the rheonomic approach  \cite{Castellani} 
or the ectoplasm formalism \cite{Ectoplasm,GGKS}.}
to the construction of supersymmetric invariants 
\cite{Castellani,Hasler,Ectoplasm,GGKS}.
The required superform construction is given in section 3. 

In this paper, we work out the component supergravity-matter actions 
in the cases of Type I and Type II minimal supergravity 
formulations.\footnote{Various aspects of the component reduction in 4D $\cN=1$ supergravity 
theories were studied in the late 1970s \cite{Zumino78,WZ2,KLR,SG}. More complete 
presentations were given in the textbooks \cite{WB,GGRS,BK}.}
The case of non-minimal supergravity can be treated in a similar way. 
As an application, we describe off-shell models for topologically massive 
$\cN=2$ supergravity\footnote{Topologically massive $\cN=1$
supergravity was introduced in \cite{DK,Deser}. Its $\cN=2$ extended version was discussed 
in \cite{RvN}.}
which 
correspond to  all the known off-shell formulations
for three-dimensional  $\cN=2$ supergravity. However, the component actions
for topologically massive supergravity are given only for the Type I and Type II minimal formulations.

Recently, supersymmetric backgrounds in the Type II supergravity 
have been studied within the component approach, 
both in the Euclidean  \cite{Festuccia3D}
and Lorentzian \cite{HTZ} signatures, building on the earlier results in four and five dimensions, 
see \cite{FS,Jia:2011hw,Samtleben:2012gy,Klare:2012gn,DFS,CDFKS1,CDFKS2,KMTZ,Liu:2012bi,Dumitrescu:2012at,Kehagias:2012fh,deMH1,deMH2,Kom13} and references therein.  
Since the authors of  \cite{Festuccia3D,HTZ} did not have access to the complete
off-shell component actions for Type II supergravity and its matter couplings, 
their analysis was based either on the considerations of linearized supergravity 
 \cite{Festuccia3D} or on the dimensional reduction $4D \to 3D$ of the new minimal 
supergravity \cite{HTZ}. Here we present a universal setting to construct supersymmetric backgrounds associated with all the known off-shell formulations
for 3D $\cN=2$ supergravity, that is the Type I and Type II minimal and the non-minimal 
supergravity theories.\footnote{After our work was completed, there appeared 
a new paper in the hep-th archive \cite{DKSS2} which also studied supersymmetric backgrounds 
in Type I supergravity.} 
Our approach will be an extension of the 4D $\cN=1$ formalism 
to determine (conformal) isometries of curved superspaces which 
was originally developed almost twenty years ago in \cite{BK} and further 
elaborated in \cite{Kuzenko13}.\footnote{This approach has been used to construct rigid supersymmetric field theories
in 5D $\cN=1$ \cite{KT-M07}, 4D $\cN=2$ \cite{KT-M08,BKLT-M}
and 3D $(p,q)$ anti-de Sitter  \cite{KT-M11,KLT-M12,BKT-M}  superspaces.}

This paper is organized as follows. In section 2 we review the superspace 
formulation for the Weyl multiplet of $\cN=2$ conformal supergravity, following 
\cite{HIPT,KLT-M11,KT-M11}.\footnote{There exists a more general 
off-shell formulation for $\cN=2$ conformal supergravity \cite{BKNT-M1}.
It will be briefly reviewed in Appendix D.} 
In section 3 we present the locally supersymmetric and super-Weyl invariant action 
principle which is based on a closed super three-form. 
The formalism for component reduction, including  the important Weyl multiplet gauge, 
is worked out in section 4. The component actions for Type I and Type II supergravity-matter systems are derived in sections  5 and 6 respectively. In section 7 we study the off-shell formulations for topologically massive $\cN=2$ supergravity. 
Section 8 is devoted to the construction of supersymmetric backgrounds in all 
the known off-shell formulations for $\cN=2$ supergravity. 

The main body of the paper is accompanied by four  appendices. 
In appendix A we give a summary of the notation and conventions used as well as include
some technical relations. In appendix B we give an alternative form for the component 
action of the most general off-shell nonlinear $\s$-model in Type I supergravity. 
Appendix C contains the component Lagrangian for the model of 
an Abelian vector multiplet in conformal supergravity. 
Appendix D is devoted to the superspace action for $\cN=2$ conformal supergravity; 
at the component level, this action reduces to that constructed 
many years ago by Ro\v{c}ek and van Nieuwenhuizen \cite{RvN}.

\section{The Weyl multiplet in U(1) superspace}
\setcounter{equation}{0}

In this section we recall the superspace description of $\cN=2$ conformal 
supergravity. The results given here are essential for the rest of the paper. 

\subsection{U(1) superspace geometry}
\label{geometry}

We consider a curved superspace  in three spacetime dimensions, 
$\cM^{3|4}$,  parametrized by
local bosonic ($x^m$) and fermionic ($\q^\m, \bar \q_\m$)
coordinates  $z^{{M}}=(x^{m},\q^{\mu},{\bar \q}_{{\mu}})$,
where $m=0,1,2$ and  $\mu=1,2$.
The Grassmann variables $\q^{\mu} $ and $\teb_{{\mu}}$
are related to each other by complex conjugation:
$\overline{\q^{\mu}}=\teb^{{\mu}}$.
The superspace structure group is chosen to be ${\rm SL}(2,{\mathbb{R}})\times {\rm U(1)}_R$,
and the covariant derivatives
$\cD_{{A}} =(\cD_{{a}}, \cD_{{\a}},\cDB^\a)$
have the form
\bea
\cD_{{A}}&=&E_{{A}}
+\O_{{A}}
+\ri \,\F_{{A}}\cJ~.
\label{CovDev}
\eea
Here $E_A= (E_a, E_\a, \bar E^\a )= E_{{A}}{}^{{M}}(z) \pa/\pa z^{{M}}$
is the inverse superspace vielbein, 
\bea
\O_A=\hf\O_{A}{}^{bc}\cM_{bc}
=\hf\O_{A}{}^{\b\g}\cM_{\b\g}
=-\O_A{}^c\cM_c~,
\label{Lorentzconnection}
\eea
is the Lorentz connection,
and $\F_{A}$ is the ${\rm U(1)}_R$-connection.
The Lorentz generators with two  vector indices ($\cM_{ab}=-\cM_{ba}$),
one vector index ($\cM_a$) and two  spinor indices
($\cM_{\a\b}=\cM_{\b\a}$) are related to each other as follows:
$$
\cM_{a}=\hf\ve_{abc}\cM^{bc}~,~~~
\cM_{ab}=-\ve_{abc}\cM^c~,~~~
\cM_{\a\b}=(\g^a)_{\a\b}\cM_{a}~,~~~
\cM_{a}=-\hf(\g_a)^{\a\b}\cM_{\a\b}~.
$$
The Levi-Civita tensor $\ve_{abc}$
and the gamma-matrices  $(\g_a)_{\a\b}$
are defined in Appendix A.
The generators of SL(2,$\mathbb{R})\times {\rm U(1)}_R$
act on the covariant derivatives as follows:
\bea
&{[}\cJ,\cD_{\a}{]}
=\cD_{\a}~,
\qquad
{[}\cJ,\cDB^{\a}{]}
=-\cDB^\a~,
\qquad
{[}\cJ,\cD_a{]}=0~,
\non \\
&{[}\cM_{\a\b},\cD_{\g}{]}
=\ve_{\g(\a}\cD_{\b)}~,\qquad
{[}\cM_{\a\b},\cDB_{\g}{]}=\ve_{\g(\a}\cDB_{\b)}~,
~~~
{[}\cM_{ab},\cD_c{]}=2\eta_{c[a}\cD_{b]}~.
\label{generators}
\eea

The supergravity gauge group 
includes
local $\cK$-transformations
of the form
\be
\d_\cK \cD_{{A}} = [\cK  , \cD_{{A}}]~,
\qquad \cK = \x^{{C}} \cD_{{C}} +\hf K^{ c d } \cM_{c d}
+\ri \, \t  \cJ  ~,
\label{tau}
\ee
with the gauge parameters
obeying natural reality conditions, but otherwise  arbitrary.
Given a tensor superfield $U(z)$, with its indices suppressed,
it transforms as follows:
\bea
\d_\cK U = \cK\, U~.
\label{tensor-K}
\eea

The  covariant derivatives obey (anti-)commutation relations of the form
\bea
{[}\cD_{{A}},\cD_{{B}}\}&=&
T_{ {A}{B} }{}^{{C}}\cD_{{C}}
+\hf R_{{A} {B}}{}^{{cd}}\cM_{{cd}}
+\ri \,R_{ {A} {B}}\cJ
~,
\label{algebra}
\eea
where $T_{{A} {B} }{}^{{C}}$ is the torsion,
and $R_{ {A} {B}}{}^{{cd}}$  and  $R_{{A} {B}}$
constitute the curvature tensors.

Unlike the 4D case, the spinor covariant derivatives $\cD_\a$ and $\bar \cD_\a$ 
transform in the same representation of the Lorentz group, and this may lead 
to misunderstandings. If there is a risk for confusion, we will underline the spinor 
indices associated with the covariant derivatives $\bar \cD$. For instance,
when the index $C$ of the torsion $T_{AB}{}^C$ takes spinor values, we will write
the corresponding components as $T_{AB}{}^\g$ and $T_{AB}{}_{\underline \g}$. 

In order to describe $\cN=2$ conformal supergravity, the torsion 
has to obey the covariant constraints given in \cite{HIPT}.
The resulting algebra of covariant derivatives is \cite{KLT-M11,KT-M11}{\allowdisplaybreaks
\bsubeq \label{algebra-final}
\bea
\{\cD_\a,\cD_\b\}
&=&
-4\bar{R}\cM_{\a\b}
~,~~~~~~
\{\cDB_\a,\cDB_\b\}
=
4{R}\cM_{\a\b}~,
\label{N=2-alg-1}
\\
\{\cD_\a,\cDB_\b\}
&=&
-2\ri (\g^c)_{\a\b} \cD_c
-2\cC_{\a\b}\cJ
-4\ri\ve_{\a\b}\cS\cJ
+4\ri\cS\cM_{\a\b}
-2\ve_{\a\b}\cC^{\g\d}\cM_{\g\d}
~,
\label{2.7b}  \\
{[}\cD_{a},\cD_\b{]}
&=&
\ri\ve_{abc}(\g^b)_\b{}^{\g}\cC^c\cD_{\g}
+(\g_a)_\b{}^{\g}\cS\cD_{\g}
-\ri(\g_a)_{\b\g}\bar{R}\cDB^{\g}
\non\\
&&
-(\g_a)_\b{}^{\g}{\bm C}_{\g\d\r}\cM^{\d\r}
-\frac{1}{3}\Big(
2\cD_{\b}\cS
+\ri\cDB_{\b}\bar{R}
\Big)\cM_a
-\frac{2}{3}\ve_{abc}(\g^b)_{\b}{}^{\a}\Big(
2\cD_{\a}\cS
+\ri\cDB_{\a}\bar{R}
\Big)\cM^c
\non\\
&&
-\hf\Big(
(\g_a)^{\a\g}{\bm C}_{\a\b\g}
+\frac{1}{3}(\g_a)_\b{}^{\g}\big(
8\cD_{\g}\cS
+\ri\cDB_{\g}\bar{R}
\big)
\Big)\cJ ~,
\label{2.7c}\\
{[}\cD_{a},\cDB_\b{]}
&=&
-\ri\ve_{abc}(\g^b)_\b{}^{\g}\cC^c\cDB_{\g}
+(\g_a)_\b{}^{\g}\cS\cDB_{\g}
-\ri(\g_a)_\b{}^{\g}{R}\cD_{\g}
\non\\
&&
-(\g_a)_\b{}^{\g}\bar{{\bm C}}_{\g\d\r}\cM^{\d\r}
-\frac{1}{3}\Big(
2\cDB_{\b}\cS
-\ri\cD_{\b}{R}
\Big)\cM_a
-\frac{2}{3}\ve_{abc}(\g^b)_{\b}{}^{\a}\Big(
2\cDB_{\a}\cS
-\ri\cD_{\a}{R}
\Big)\cM^c
\non\\
&&
+\hf\Big(
(\g_a)^{\a\g}\bar{{\bm C}}_{\a\b\g}
+\frac{1}{3}(\g_a)_\b{}^{\g}\big(
8\cDB_{\g}\cS
-\ri\cD_{\g}{R}\big)
\Big)\cJ~,
\label{2.7d}  
\\
{[}\cD_a,\cD_b]{}
&=&
\hf\ve_{abc}(\g^c)^{\a\b}\ve^{\g\d}\Big(
-\ri\bar{{\bm C}}_{\a\b\d}
+\frac{4\ri}{3}\ve_{\d(\a}\cDB_{\b)}\cS
+\frac{2}{3}\ve_{\d(\a}\cD_{\b)} R
\Big)\cD_\g
\non\\
&&
+\hf\ve_{abc}(\g^c)^{\a\b}\ve^{\g\d}
\Big(
-\ri {\bm C}_{\a\b\d}
+\frac{4\ri}{3}\ve_{\d(\a}\cD_{\b)}\cS 
-\frac{2}{3}\ve_{\d(\a}\cDB_{\b)}\bar{R}
\Big)\cDB_\g
\non\\
&&
-  \ve_{abc}\Big(\,\frac{1}{4}  (\g^c)^{\a\b}(\g_d)^{\t\d}
\big(\ri\cD_{(\t}\bar{{\bm C}}_{\d\a\b)}
+\ri\cDB_{(\t} {\bm C}_{\d\a\b)}\big)
+\frac{1}{6}
(\cD^2 R+\cDB^2 \bar{R})
\non\\
&&~~~~~~~~~
+\frac{2}{3} \ri 
\cD^\a\cDB_{\a}\cS
-4\cC^{c}\cC_d
-4\cS^2
-4\bar{R}R
\Big)\cM^d
\non\\
&&
+\ri  \ve_{abc} \Big(
\frac{1}{2}  (\g^c)^{\a\b}[\cD_\a,\cDB_{\b}]\cS
-\ve^{cef}\cD_{e}\cC_{f}
-4\cS\cC^c
\Big)\cJ
~,
\label{2.7e}
\eea
\end{subequations}}with $ {\bm C}_{\a\b\g}$ defined by 
$ {\bm C}_{\a\b\g} = -\ri \cD_{(\a}\cC_{\b\g )} $.
The algebra involves three  dimension-1 torsion superfields:
a real scalar $\cS$, a
complex scalar $R$ and its conjugate $\bar{R}$,  and a real vector $\cC_a$;
the ${\rm U(1)}_R$ charge of $R$ is $-2$. 
The torsion superfields obey differential constraints implied by the Bianchi identities. 
The constraints are
\begin{subequations}
\bea
\cDB_\a R&=&0~,
\\
(\cDB^2-4R)\cS
&=&0~,  \label{2.12b}\\
\cD_{\a}\cC_{\b\g}
&=&
\ri {\bm C}_{\a\b\g}
-\frac{1 }{ 3}\ve_{\a(\b}\Big(
 \cDB_{\g)}\bar{R}
+4\ri \cD_{\g)}{\cS}
\Big)~.  \label{2.12c}
\eea
\end{subequations}
Eq. \eqref{2.12b}
 means that $\cS$ is a covariantly linear superfield.
When doing explicit calculations, it is useful to deal with equivalent forms of the relations 
\eqref{2.7c} and \eqref{2.7d} in which the vector index of $\cD_a$ 
is replaced by a pair of spinor indices. Such identities are given in Appendix
\ref{AppendixA}.

As an immediate application of the (anti-)commutation relations
\eqref{algebra-final}, we compute a covariantly chiral d'Alembertian. 
Let $\c$ be  a covariantly chiral scalar, ${\bar \cD}_\a \c=0$,
of U$(1)_R$ charge $-w$, that is $\cJ\c=-w\c$.\footnote{The rationale for choosing 
the U$(1)_R$ charge of $\c$  to be negative is eq. \eqref{primary_chiral}.}
The covariantly chiral d'Alembertian $\Box_{\rm c} $ is defined by 
\bea
\Box_{\rm c} \c:= \frac{1}{16}( \bar \cD^2 - 4R ) \cD^2 \c~.
\eea
By construction, the scalar $\Box_{\rm c} \c$ is covariantly chiral
and has  U$(1)_R$ charge $-w$. 
 It is an instructive exercise to evaluate  the explicit form of $ \Box_{\rm c} \c $
 using the chirality of $\c$ and 
 the relations \eqref{algebra-final}. 
The result is 
\bea
\Box_{\rm c} \c &=& \Big\{ \cD^a \cD_a +\hf R \cD^2 - 2\ri (1-w) \cC^a \cD_a
+\hf (\cD^\a R) \cD_\a + 2\ri (1-w) (\bar \cD^\a \cS) \cD_\a \non \\
&& +w(2-w)( \cC^a \cC_a +4\cS^2)  -w \ri \cD^\a\bar \cD_\a \cS 
+\frac{w}{8} (\bar \cD^2 \bar R - \cD^2 R) \Big\} \c~.
\label{cd'A}
\eea
This relation turns out to be useful for the component reduction of locally supersymmetric
sigma models to be discussed later on. 

\subsection{Super-Weyl invariance}\label{section2.2}

The algebra of covariant derivatives \eqref{algebra-final}
does not change under
a super-Weyl transformation\footnote{The super-Weyl transformation \eqref{2.3}
is uniquely fixed if one (i) postulates that the components of the inverse vielbein $E_A$ 
transform as $E'_\a = \re^{\hf \s} E_\a$ and $E'_a = \re^{\s} E_a + \text{spinor terms}$; and (ii) requires that 
the transformed covariant derivatives preserve the constraints \cite{HIPT}
leading to the algebra \eqref{algebra-final}.}
of the covariant derivatives \cite{KLT-M11,KT-M11}
\bsubeq  \label{2.3}
\bea
\cD'{}_\a&=&\re^{\hf\s}\Big(\cD_\a+(\cD^{\g}\s)\cM_{\g\a}-(\cD_{\a }\s)\cJ\Big)~,
\\
\cDB'{}_{\a}&=&\re^{\hf\s}\Big(\cDB_{\a}+(\cDB^{\g}\s){\cM}_{\g\a}
+(\cDB_{\a}\s)\cJ\Big)~,
\\
\cD'{}_{a}
&=&\re^{\s}\Big(
\cD_{a}
-\frac{\ri}{2}(\g_a)^{\g\d}(\cD_{(\g}\s)\cDB_{\d)}
-\frac{\ri}{2}(\g_a)^{\g\d}(\cDB_{(\g}\s)\cD_{\d)} \non \\
&&~~~~~
+\ve_{abc}(\cD^b\s)\cM^c
+\frac{\ri}{2}(\cD_{\g}\s)(\cDB^{\g}\s)\cM_{a}
\non\\
&&~~~~~
-\frac{\ri}{8}(\g_a)^{\g\d}({[}\cD_{\g},\cDB_{\d}{]}\s)\cJ
-\frac{3\ri}{4}(\g_a)^{\g\d}(\cD_{\g}\s)(\cDB_{\d}\s)\cJ
\Big)
\eea
accompanied by the following transformation of the torsion tensors:
\bea
\cS'&=&\re^{\s}\Big(
\cS
-\frac{\ri}{4}\cD_\g\cDB^{\g}\s
\Big)~,
 \label{2.11d} \\
\cC'_{a}&=&\re^{\s}\Big(
\cC_{a}
+\frac{1}{8}(\g_a)^{\g\d}  [\cD_{\g},\cDB_{\d}]\s
+\frac{1}{4}(\g_a)^{\g\d}(\cD_{\g}\s)\cDB_{\d}\s
\Big)~,
\\
R' &=&
\re^{\s}\Big(
R
+\frac{1}{4} \cDB^2\s
-\frac{1}{4}  ( \cDB_\g\s)\cDB^{\g}\s
\Big)~. \label{2.11f}
\eea
\esubeq
The gauge group of conformal supergravity 
is defined to be generated by the $\cK$-transformation \eqref{tau} 
and the super-Weyl transformations. 
The super-Weyl invariance is the reason why the U(1) superspace 
geometry describes the Weyl multiplet. 

Using the above super-Weyl transformation laws, it is an instructive exercise to demonstrate
that the real symmetric spinor superfield \cite{Kuzenko12}
\bea
\cW_{\a\b} := \frac{\ri}{2} \big[\cD^\g ,\bar \cD_\g \big]
\cC_{\a\b} - \big[ \cD_{(\a} , \bar \cD_{\b)} \big] \cS 
- 4 \cS \cC_{\a\b}
\label{Cotton}
\eea
transforms homogeneously, 
\bea
\cW_{\a\b}' =  \re^{2\s}\, \cW_{\a\b}~.
\label{Cotton-Weyl}
\eea
This superfield is the $\cN=2$ supersymmetric generalization of the Cotton tensor. 
Using the Bianchi identities, one can 
obtain an equivalent expression for this super Cotton tensor:
\bea
\cW_a 
= 
-\hf (\g_a)^{\a\b} \cW_{\a\b} 
=
 -\hf (\g_a)^{\a\b} [\cD_{(\a},\cDB_{\b)}]\cS
+2\ve_{abc}\cD^{b}\cC^{c}
+4\cS\cC_{a}
~.
\label{Cotton-2}
\eea
An application of this relation will be given in Appendix \ref{AppendixC}. 
A curved superspace is conformally flat if and only if $\cW_{\a\b} =0$, 
see \cite{BKNT-M1} for the proof.

For our subsequent consideration, it is important to recall one of the results obtained in \cite{KLT-M11}.
Let $\c$ be  a covariantly chiral scalar, ${\bar \cD}_\a \c=0$, which is primary 
under the super-Weyl group, $\d_\s\c=w\s\c$. Then its super-Weyl weight $w$ 
and its U(1)$_R$ charge are equal and opposite
 \cite{KLT-M11},
\bea
{\bar \cD}_\a \c=0 ~, \qquad
\cJ\c=-w\c~,\qquad 
\c'=\re^{w \s}\c~.
\label{primary_chiral}
\eea
Unlike $\c$ itself, its chiral d'Alembertian $\Box_{\rm c} \c$, eq \eqref{cd'A}, is not 
a primary superfield under the super-Weyl group. 


In what follows, we often consider the infinitesimal super-Weyl transformation
and denote  the corresponding variation   by $\d_\s$.

\section{Supersymmetric and super-Weyl invariant action}

There are two (closely related) locally supersymmetric and super-Weyl invariant actions
in $\cN=2$ supergravity  \cite{KLT-M11}.
Given a real scalar Lagrangian $\cL =\bar \cL$ with the 
 super-Weyl transformation law
\bea
\d_\s\cL=\s\cL~,
\label{4.21}
\eea
the action
\bea
S&=&\int\rd^3x\rd^2\q\rd^2\qb \,E \,\cL
~,~~~
\qquad E^{-1}= {\rm Ber}(E_A{}^M)
~,
\label{N=2Ac}
\eea
is invariant under the supergravity gauge group. 
It is also  super-Weyl invariant due to the transformation law
\bea
\d_\s E=-\s E~.
\label{SW-Ber}
\eea
Given a covariantly chiral scalar Lagrangian $\cL_{\rm c}$ of super-Weyl weight two, 
\bea
\cDB_\a\cL_{\rm c}=0~,\qquad \cJ \cL_{\rm c} = -2 \cL_{\rm c}~, \qquad
\d_\s\cL_{\rm c}=2\s\cL_{\rm c}~,
\eea
the following {\rm chiral} action
\bea
S_{\rm c}&=&\int\rd^3x\rd^2\q\rd^2\qb\, E \,\frac{\cL_{\rm c}}{R}
=\int\rd^3x\rd^2\q\, \cE \,\cL_{\rm c} 
\label{4.26}
\eea
is locally supersymmetric and super-Weyl invariant. The first representation in  (\ref{4.26}), 
which is only valid when $R\neq 0$,
is analogous to that derived by Zumino \cite{Zumino78} in 4D $\cN=1$ supergravity.
The second representation in  (\ref{4.26}) involves integration over the chiral subspace
of the full superspace, with $\cE$ the chiral density possessing the properties
\bea
\cJ  \cE = 2 \cE~, \qquad 
\d_\s\cE=-2\s\cE~.
\eea
The explicit expression for $\cE$ in terms of the supergravity prepotentials is given in 
\cite{Kuzenko12}.  Complex conjugating (\ref{4.26}) gives the action $\bar S_{\rm c}$
generated by the antichiral Lagrangian $\bar \cL_{\rm c}$. 

The two actions, \eqref{N=2Ac} and \eqref{4.26}, 
are related to each other as follows
\bea
\int\rd^3x\rd^2\q\rd^2\qb\, E \,\cL
= \int\rd^3x\rd^2\q\, \cE \,\cL_{\rm c} ~, \qquad \cL_{\rm c} :=
-\frac{1}{4}(\cDB^2-4R)\cL~.
\label{N=2Ac-2}
\eea
This relation shows that the chiral action, or its conjugate antichiral action,  
is more fundamental than  \eqref{N=2Ac}.

The chiral action can be reduced to component fields by making use of the prepotential 
formulation for $\cN=2$ conformal supergravity \cite{Kuzenko12} and following the component 
reduction procedure 
developed in \cite{BK} for $\cN=1$ supergravity in four dimensions. 
Being conceptually straightforward, however, this procedure is technically rather tedious 
and time consuming. 
A simpler way to reduce $S_{\rm c}$ to components 
consists in making use of  the superform approach to 
the construction of supersymmetric invariants \cite{Castellani,Hasler,Ectoplasm,GGKS}.
In conjunction with the requirement of super-Weyl invariance, the latter approach turns 
out to be extremely powerful. 
As a matter of taste, here we prefer to deal with  $\bar S_{\rm c}$, 
because it turns out that  the corresponding closed three-form 
involves no one-forms $ \bar E_\a $.

The super-Weyl transformation laws of the components of the 
superspace vielbein 
\bea
E^A:= \rd z^M E_M{}^A = (E^a, E^\a , \bar E_\a )~,
\eea
are:
\begin{subequations}
\bea
\d_\s E^a &=&- \s E^a ~, \\
\d_\s E^\a &=& -\hf \s E^\a +\frac{\ri}{2} E^b (\g_b)^{\a\g} \bar \cD_\g \s~, \non \\
\d_\s \bar E_\a &=& -\hf \s \bar E_\a +\frac{\ri}{2} E^b (\g_b)_{\a\g}  \cD^\g \s~.
\eea
\end{subequations}
We are looking for a dimensionless three-form,  
$\X (\bar \cL_{\rm c}) = \frac{1}{6} E^C \wedge E^B \wedge E^A\, \X_{ABC} $, 
such that (i) its components $ \X_{ABC} $ are linear functions of  $\bar \cL_{\rm c}$ and 
 covariant derivatives thereof; and (ii) $\X (\bar \cL_{\rm c})$ is super-Weyl invariant, 
 $\d_\s \X (\bar \cL_{\rm c}) =0$. Modulo an overall numerical factor, 
 such a form is uniquely determined to be 
\bea
\X (\bar \cL_{\rm c}) 
&=&\hf E^\g \wedge E^\b \wedge E^a\, \X_{a\b\g} 
+\hf E^\g \wedge E^b \wedge E^a\, \X_{ab\g} 
+\frac{1}{6} E^c \wedge E^b \wedge E^a \, \X_{abc} ~,
\eea
where
\begin{subequations}
\bea
 \X_{a\b\g} &=& 4 (\g_a)_{\b\g} \bar \cL_{\rm c}~, \\
 \X_{ab\g} &=& -\ri \ve_{abd} (\g^d)_{\g\d} \bar \cD^\d \bar \cL_{\rm c} ~,\\
\X_{a b c} &=&\frac{1}{4} \ve_{abc}(\bar \cD^2 -16 R) \bar  \cL_{\rm c}~.
\eea
\end{subequations}
It is easy to check that this three-form is closed, 
\bea
\rd \, \X (\bar \cL_{\rm c}) =0~,
\eea
and therefore $\X (\bar \cL_{\rm c}) $ generates a locally supersymmetric action.

The  locally supersymmetric and super-Weyl invariant action associated with 
$\X (\bar \cL_{\rm c}) $ is 
\bea
\bar S_{\rm c}=-\int\rd^3x\,e\,
\Big{[}
\frac{1}{4}\cDB^2- 4R
-\frac{\ri}{2}(\g^a)_{\g\r}\psi_a{}^\g\cDB^\r
+\hf\ve^{abc}(\g_a)_{\b\g}\psi_b{}^\b \psi_c{}^\g
\Big{]}\bar{\cL}_{\rm c}\Big|~,
\label{comp-ac-1}
\eea
with $e := \det (e_m{}^a)$. 
Here we have used definitions introduced in the next section.


\section{Component reduction} 
\label{components}

In this section we develop a simple universal setup to carry out the component reduction
of the general $\cN=2$ supergravity-matter systems presented in  \cite{KLT-M11,KT-M11}.  
Our consideration below is very similar to that given in standard textbooks on 
four-dimensional $\cN=1$ supergravity \cite{GGRS,BK}.

Given a superfield $U(z)$ we define its bar-projection $U|$ to be 
the $\q, \bar \q$-independent term in the expansion of $U(x,\q, \bar \q)$ in powers of $\q$'s
and $\bar \q$'s,
\bea
U|:=U(x,\q, \bar \q )|_{\q=\bar \q=0}
~.
\eea
Thus $U |$ is a field on the spacetime $\cM^3$ which is the bosonic body of the curved superspace
$\cM^{3|4}$.

In a similar way we  define the bar-projection of the covariant derivatives: 
\bea
\cD_A|:=E_A{}^M |\pa_M+\hf\O_A{}^{bc} | \cM_{bc}+\ri\F_A |\cJ~.
\eea
More generally, given a differential operator $\hat \cO:= \cD_{A_1} \dots \cD_{A_n}$, 
we define its bar-projection, ${\hat \cO}|$, by the rule 
$ ( \hat \cO |  U )|:= ( \cD_{A_1} \dots \cD_{A_n} U)|$, for any tensor superfield $U$.

Of special importance is the bar-projection of a vector covariant 
derivative,\footnote{The definition of the gravitino agrees with that used in the 4D case in  \cite{WB}.} 
\bea
\cD_a|=\bD_a 
-\hf\psi_a{}^\g \cD_\g|
-\hf\bar{\psi}_a{}_\g \cDB^\g|~,
\label{stcd}
\eea
where 
\bea
\bD_a=e_a+\hf\o_a{}^{bc} \cM_{bc}
+\ri b_a \cJ,~~~~~~
e_a:=e_a{}^m \pa_m
\label{stcd-2}
\eea
is a spacetime covariant derivative with Lorentz
and  U(1)${}_R$ connections.  
For some calculations, it will be useful to work with 
a spacetime covariant derivative
without U(1)${}_R$ connection, ${\mathfrak D}_a$, defined by
\bea
 {\mathfrak D}_a = \bD_a-\ri b_a  \cJ ~.
\label{dev000}
\eea

\subsection{The Wess-Zumino and normal gauges} 

The freedom to perform  general coordinate and local 
Lorentz
transformations  
can be used to choose a  Wess-Zumino (WZ) gauge of the form: 
\bea
&&\cD_\a|=\d_\a{}^\m {\pa\over\pa\q^\m}~,\qquad
\cDB^\a|=\d^\a{}_\m {\pa\over\pa\qb_\m}
~.
\label{WZ}
\eea
In this gauge, it is easy to see that 
\begin{subequations} \label{fields}
\bea
&E_a{}^m |=e_a{}^m~,\qquad
E_a{}^\m |=-\hf\psi_a{}^\g \d_\g{}^\m ~,\qquad
\bar E_a{}_{ \m} |=-\hf\bar{\psi}_a{}_\g \d^\g{}_\m~,\\
&
\O_a{}^{bc} |=\o_{a}{}^{bc}~,\qquad
\F_a|=b_{a}~.
\eea
\end{subequations}
The gauge condition \eqref{WZ} will be used in what follows.

In the WZ gauge, we still have a tail of component fields 
which originates at higher orders in the $\q, \bar \q$-expansion of $E_A{}^M$, 
$\O_A{}^{bc}$ and $\F_A$ and which are pure gauge (that is, they 
may be completely gauged away).  A way to get rid of such a 
tail of redundant fields is to impose a normal gauge around the bosonic
body $\cM^3$ of the curved superspace $\cM^{3|4}$, see \cite{KT-Maction} 
for more details. This gauge is defined by the conditions:
\begin{subequations}\label{normal_gauge}
\bea
\Q^M E_M{}^A (x,\Q)&=& \Q^M \d_M{}^A~, \\
\Q^M \O_M{}^{cd} (x, \Q) &=& 0~, \\
\Q^M \F_M (x,\Q) &=& 0~,
\eea
\end{subequations}
where we have introduced 
\bea
\Q^M \equiv (\Q^m, \Q^\m , \bar \Q_\m) := (0, \q^\m, \bar \q_\m)~.
\eea
In \eqref{normal_gauge} the connections with world indices are defined 
in the standard way: $\O_M{}^{cd} = E_M{}^A  \O_A{}^{cd} $ and $\F_M = E_M{}^A  \F_A{}$. It can be  proved 
\cite{KT-Maction} 
 that the normal gauge conditions \eqref{normal_gauge} 
allow one to reconstruct 
the vielbein  $E_{M}{}^A(x, \Q)$ and 
the connections $ \O_M{}^{cd} (x, \Q)$ and $ \F_M (x,\Q) $
as Taylor series in $\Q$, in which all the coefficients
(except the leading $\Q$-independent terms given by the relations 
\eqref{WZ} and \eqref{fields}) 
are tensor functions of the torsion, the curvature and their covariant derivatives 
evaluated at  $ \Q=0$. 

In principle, there is no need to introduce the normal gauge which eliminates 
the tail of superfluous fields. Such fields (once properly defined) are pure gauge and do not show up in the gauge 
invariant action. This is similar to the concept of 
double-bar projection, see e.g. \cite{Binetruy:2000zx}.

\subsection{The component field strengths}

The spacetime covariant derivatives $\bD_a$ defined by \eqref{stcd}
obey commutation relations of the form 
\bea
[\bD_a,\bD_b]&=&\cT_{ab}{}^c \bD_c
+\hf\cR_{ab}{}^{cd} \cM_{cd}
+\ri \cF_{ab} \cJ
~,
\eea
where $\cT_{ab}{}^c $ is the torsion, $\cR_{ab}{}^{cd}$ the Lorentz curvature,
and $\cF_{ab}$ the U(1)${}_R$ field strength.
These tensors can be related to the superspace geometric objects
by bar-projecting the (anti-)commutation relations \eqref{algebra}. 
A short calculation gives the torsion 
\bea
\cT_{ab}{}^c = -\frac{\ri}{2} (\bar \psi_a \g^c \psi_b - \bar \psi_b \g^c \psi_a)~.
\label{comtor}
\eea
The  Lorentz connection is 
\bea
\o_{abc}=\o_{abc}(e) +
\hf\big{[} 
\cT_{abc}
-\cT_{bca}
+\cT_{cab}
\big{]}~,
\eea
where $\o_{abc}(e)$ denotes the torsion-free connection, 
\bea
\o_{abc}(e) = 
-\hf\big{[} 
\cC_{abc}
-\cC_{bca}
+\cC_{cab}
\big{]}~,
\qquad
\cC_{ab}{}^c:=(e_{a} e_{b}{}^m  - e_{b} e_{a}{}^m )\,e_m{}^c
~.
\eea

For the gravitino field strength defined by
\bea
{\bm \psi}_{ab}{}^\g := \bD_a \psi_b{}^\g - \bD_b \psi_a{}^\g - \cT_{ab}{}^c \psi_c{}^\g
\label{gfs}
\eea
we read off 
{\allowdisplaybreaks
\bea
{\bm \psi}_{ab}{}^\g &=&
\Big(
\ri\ve_{abc}(\g^c)^{\a\b}\bar{{\bm C}}_{\a\b}{}^{\g}
-\frac{4\ri}{3}\ve_{abc}(\g^c)^{\g\d}\cDB_{\d}\cS
-\frac{2}{3}\ve_{abc}(\g^c)^{\g\d}  \cD_{\d}R
\non\\
&&
+2\ri\ve_{cd[a}(\g^c)^{\g\d}\psi_{b]}{}_\d\cC^d
+2(\g_{[a})^{\g\d}\psi_{b]}{}_\d\cS
+2\ri(\g_{[a})^{\g\d}\bar{\psi}_{b]}{}_\d R
\Big)\Big|~.
\eea
This tells us how the gravitino field-strength is
embedded in the superspace curvature and torsion.
A longer calculation is to derive an explicit expression for 
the Lorentz curvature
\bea
\cR_{ab}{}^{cd}
&=&
2e_{[a}\o_{b]}{}^{cd}
+2\o_{[a}{}^{cf}\o_{b]}{}_f{}^d
-\cC_{ab}{}^f\o_f{}^{cd}~.
\eea
The result is
\bea
\cR_{ab}{}^{cd}=
\Big{\{}\!\!
&&-\frac{\ri}{4}\ve_{abe}(\g^e)^{\a\b}\ve^{cdf}(\g_f)^{\t\d}
\big(\cD_{(\t}\bar{{\bm C}}_{\d\a\b)}
+\cDB_{(\t} {\bm C}_{\d\a\b)}\big)
\non\\
&&
+\d_{[a}^c\d_{b]}^d\Big{[}
\frac{1}{3}(\cD^2 R+\cDB^2 \bar{R})
+\frac{4\ri}{3}\cD^\a\cDB_{\a}\cS
-8\bar{R}R
-8\cS^2
\Big{]}
+4\ve_{abe}\ve^{cdf}\cC^{e}\cC_f
\non\\
&&
+\psi_{[a}{}^\b\Big{[}
(\g_{b]})_\b{}^{\g}{\bm C}_{\g\d\r}\ve^{cde}(\g_e)^{\d\r}
+\frac{1}{3}\ve_{b]}{}^{cd}\Big(
2\cD_{\b}\cS
+\ri\cDB_{\b}\bar{R}
\Big) \non \\
&& \qquad \qquad \qquad \qquad  \qquad 
-\frac{4}{3}\d_{b]}^{[c}(\g^{d]})_{\b}{}^{\g}\Big(
2\cD_{\g}\cS
+\ri\cDB_{\g}\bar{R}
\Big)
\Big{]}
\non\\
&&
+\bar{\psi}_{[a}{}_\b\Big{[}
(\g_{b]})^{\b\g}\bar{{\bm C}}_{\g\d\r}\ve^{cde}(\g_e)^{\d\r}
+\frac{1}{3}\ve_{b]}{}^{cd}\Big(
2\cDB^{\b}\cS
-\ri\cD^{\b}{R}
\Big) \non\\
&&\qquad \qquad \qquad \qquad \qquad 
-\frac{4}{3}\d_{b]}^{[c}(\g^{d]})^{\b\g}\Big(
2\cDB_{\g}\cS
-\ri\cD_{\g}{R}
\Big)
\Big{]}
\non\\
&&
+\ve^{cde}(\g_e)_{\g\d}\psi_{[a}{}^\g\psi_{b]}{}^\d\bar{R}
-\ve^{cde}(\g_e)^{\g\d}\bar{\psi}_{[a}{}_\g\bar{\psi}_{b]}{}_\d R
\non\\
&&
+2\ri\ve^{cde}(\g_e)^{\g\d}\psi_{[a}{}_\g\bar{\psi}_{b]}{}_\d \cS
+2\psi_{[a}{}^\g\bar{\psi}_{b]}{}_\g \ve^{cde}\cC_{e}
\Big{\}}\Big|~.
\eea
}
Finally, for the U(1)${}_R$ field strength 
\bea
\cF_{ab } = {\mathfrak D}_a b_b - {\mathfrak D}_b b_a - \cT_{ab}{}^c b_c
\eea
we obtain 
\bea
\cF_{ab}&=&\ve_{abc}
\Big\{\,
\frac{1}{2} (\g^c)^{\a\b}  [\cD_\a,\cDB_{\b}]\cS 
- \ve^{cef}{\mathfrak D}_{e}\cC_{f} 
-4\cS  \cC^c 
\non\\
&&~~~~~~
+\ri\ve^{cef}(\g_e)^{\g\r}\psi_f{}_\g \cD_\r\cS 
+\ri\ve^{cef}(\g_{e})^{\b\g}\bar{\psi}_{f}{}_\b \cDB_{\g}\cS  \Big\} \Big|
\non\\
&&~
+\ri(\g_c)^{\g\d}\psi_{[a}{}_\g\bar{\psi}_{b]}{}_\d \cC^c \big|
-2\psi_{[a}{}^\g\bar{\psi}_{b]}{}_\g \cS \big|~. 
\eea

It turns out that the expressions for ${\bm \psi}_{ab}{}^\g$, 
$\cR_{ab}{}^{cd}$ and $\cF_{ab}$
drastically simplify if we also partially gauge fix 
the super-Weyl invariance to choose the so-called Weyl multiplet gauge 
that will be introduced in subsection \ref{subsection4.3}.

\subsection{Residual gauge transformations}

In the WZ gauge, 
there remains 
a subset of gauge transformations 
which preserve the conditions \eqref{WZ}.  To work out the structure of this residual 
gauge freedom, we start from the transformation laws of the inverse vielbein $E_A{}^M$ and of the connections  
$\O_A{}^{cd}$ and $\F_A$ under the gauge group of conformal supergravity. 

Under the $\cK$-transformation (\ref{tau}), the gauge fields vary as follows:
\bsubeq \label{K-tr}
\bea
\d_\cK E_A{}^M&=&
\x^CT_C{}_A{}^BE_B{}^M
-(\cD_A \x^B)E_B{}^M
+\hf K^{cd}(\cM_{cd})_A{}^BE_B{}^M
+\ri \t(\cJ)_A{}^BE_B{}^{M}
~,~~~~~~~~~
\\
\d_\cK \O_A{}^{cd}&=&
\x^C T_C{}_A{}^B\O_B{}^{cd}
+\x^BR_B{}_A{}^{cd}
-(\cD_A \x^B)\O_B{}^{cd} 
+ K^{cd}(\cM_{cd})_A{}^B\O_B{}^{cd}
\non\\
&&
-(\cD_A K^{cd})
+\ri \t(\cJ)_A{}^B\O_B{}^{cd}
~,
\\
\d_\cK\F_A&=&
\x^CT_C{}_A{}^B\F_B
+\x^BR_B{}_A
-(\cD_A \x^B)\F_B 
+\hf K^{cd}(\cM_{cd})_A{}^B\F_B
\non\\
&&
+\ri \t(\cJ)_A{}^B\F_B
-\cD_A \t~.
\eea
\esubeq
Here we have introduced the Lorentz and U$(1)_R$ generators 
$(\cM^{cd})_A{}^B$ and $(\cJ)_A{}^B$, respectively,   defined by
$$
{[}\cM^{cd},\cD_A{]}=(\cM^{cd})_A{}^B\cD_B~, 
\qquad {[}\cJ,\cD_A{]}=(\cJ)_A{}^B\cD_B~.
$$
The super-Weyl transformation \eqref{2.3} acts on the gauge fields
as follows:
\bsubeq \label{sigma-tr}
\bea
\d_\s E_{a}{}^{M}
&=&
\s E_a{}^M
-\frac{\ri}{2}(\g_a)^{\g\d}(\cD_{(\g}\s)\bar{E}_{\d)}{}^M
-\frac{\ri}{2}(\g_a)^{\g\d}(\cDB_{(\g}\s)E_{\d)}{}^M
~,
\\
\d_\s E_\a{}^{M}&=&\hf\s E_\a{}^M~,
\\
\d_\s\O_{a}{}^{bc}
&=&
\s\O_{a}{}^{bc}
-\frac{\ri}{2}(\g_a)^{\g\d}(\cD_{(\g}\s)\bar{\O}_{\d)}{}^{bc}
-\frac{\ri}{2}(\g_a)^{\g\d}(\cDB_{(\g}\s){\O}_{\d)}{}^{bc}
+2(\cD^{[b}\s)\d_a^{c]}
~,~~~
\\
\d_\s\O_\a{}^{bc}&=&\hf\s\O_\a{}^{bc}+(\cD^{\g}\s)(\g_a)_{\g\a}\ve^{abc}~,
\\
\d_\s\F_{a}
&=&
\s\F_{a}
-\frac{\ri}{2}(\g_a)^{\g\d}(\cD_{(\g}\s)\bar{\F}_{\d)}
-\frac{\ri}{2}(\g_a)^{\g\d}(\cDB_{(\g}\s)\F_{\d)}
-\frac{1}{8}(\g_a)^{\g\d}  {[}\cD_{\g},\cDB_{\d}{]}\s~,~~~~~~
\\
\d_\s \F_\a&=&\hf\s \F_\a
+\ri\cD_{\a }\s~.
\eea
\esubeq
Requiring the WZ gauge to be preserved, $(\d_\cK + \d_\s) \cD_\a|=0$, gives
\bsubeq \label{WZ-imply}
\bea
\cD_\a \x^b |&=& \x^CT_C{}_\a{}^b \big|~,
\label{WZ-imply-1}
\\
\cD_\a \x^\b |&=&\big(
\x^CT_C{}_\a{}^\b
+\hf K_\a{}^\b
+\ri \t\d_\a^\b
+\hf\s\d_\a^\b
\big)\big|
~,
\label{WZ-imply-2}
\\
\cD_\a \bar{\x}_{\underline{\b}}|&=&
\x^CT_C{}_\a{}_{\underline{\b}} \big|
~,
\label{WZ-imply-3}
\\
\cD_\a K^{cd} |&=&
\big( \x^BR_B{}_\a{}^{cd}
+(\g_a)_{\a\g}\ve^{acd} \cD^{\g}\s \big) \big|
~,
\label{WZ-imply-4}
\\
\cD_\a \t |
&=&
\big( \x^BR_B{}_\a
+\ri \cD_{\a }\s \big) \big|
~.
\label{WZ-imply-5}
\eea
\esubeq
We see that the residual gauge transformations are constrained. 
More specifically, only the parameters
\bsubeq \label{lowest}
\bea
v^a:=\x^a|~,~~~\e^{\a}&:=&\x^\a|~,~~~
w_{ab}:=K_{ab}|~,~~~
{\bm\t}:=\t| ~,\label{lowest-a} 
\eea
are completely unrestricted in the WZ gauge. 
Here the bosonic parameters correspond to general coordinate ($v^a$), 
local Lorentz $(w_{ab})$ and local $R$-symmetry ($\bm \t$) 
 transformations; the fermionic parameter $\e^\a$ 
generates a local $Q$-supersymmetry transformation. 
However, the parameters $\cD_\a \x^A|$, $\cD_\a K^{cd} |$ and $\cD_\a \t |$ are 
fully determined in terms of those listed in \eqref{lowest-a}
and the following ones:
\bea
{\bm \s}&:=&\s|~,~~~~~
\eta_\a:=\cD_\a\s| \label{lowest-b}~.
\eea
\esubeq
Here the parameter 
 $\bm \s$ and $\eta_\a$ generate the Weyl and local $S$-supersymmetry transformations
 respectively. 
 It should be pointed out that there is no parameter generating a local special conformal transformation. 
As compared with the 3D $\cN=2$ superconformal tensor calculus in superspace \cite{BKNT-M1},
our formulation corresponds to a gauge in which the dilatation gauge field is switched off by making use of 
the local special conformal transformations.

The relations \eqref{WZ-imply} comprise all the conditions  
on the residual gauge transformations, which are implied by the WZ gauge. 
If in addition we also choose the normal gauge  \eqref{normal_gauge}, 
then all higher-order  terms in the $\Q$-expansion of the gauge parameters
will be determined in terms of  those listed in \eqref{lowest}.

In what follows, we will be interested in local $Q$-supersymmetry
 transformations of the 
gauge fields $e_m{}^a$, $\psi_m{}^\g=e_m{}^a\psi_a{}^\g$ 
and $b_m=e_m{}^ab_a$. Yet we introduce a more general transformation 
 \bea
 \d:=\d_Q+\d_S +\d_W +  \d_{R} 
 \label{delta3}
 \eea 
which includes the local $Q$-supersymmetry ($\e_\a$) 
and $S$-supersymmetry  ($\eta_\a$) transformations, 
as well as  the Weyl (${\bm\s}$) and
local $R$-symmetry (${\bm \t}$)    transformations.
There is a simple reason for considering this combination 
of  four transformations. As will be shown in the next two sections, 
in any off-shell formulation for Poincar\'e or AdS supergravity, 
the $Q$-supersymmetry transformation has to be accompanied by 
a special $S$-supersymmetry transformation with parameter $\eta_\a(\e)$
and, in some case, by a special U$(1)_R$ transformation with 
parameter ${\bm \t} (\e)$.  Typically, it will hold that ${\bm \s}(\e) =0$.  
 However, since $\d_\eta $ is part of the super-Weyl transformation, 
 it makes sense to include $\d_W$ into \eqref{delta3}.

Making use of the relations \eqref{K-tr}, \eqref{sigma-tr}, \eqref{WZ-imply} and \eqref{lowest}, we read off the 
transformation  laws of the gauge fields under 
  \eqref{delta3}:
\bsubeq
{\allowdisplaybreaks
\bea
\d e_m{}^a
&=&
\ri\big(\e\g_a\bar{\psi}_m +\bar{\e}\g_a\psi_m \big)
-{\bm\s}e_m{}^a
~,
\label{susy-transformations-1}
\\
\d\psi_m{}^\a&=&
2\bD_m\e^\a
+2e_m{}^a\big(
\e^\b T_a{}_\b{}^\a|
+\bar{\e}_{\underline{\b}} T_a{}^{\underline{\b}}{}^\a|\big)
+\ri(\g_m)^{\a\b}\bar{\eta}_\b
-\ri {\bm \t}\psi_m{}^\a
+\frac{1}{2}{\bm \s}\psi_m{}^\a 
~,~~~~~~
\label{susy-transformations-3}
\\
\d b_m&=&
\Big\{
- \hf e_m{}^a\e^\b\Big{[}
\ri(\g_a)^{\g\d}{\bm C}_{\b\g\d}|
+\frac{1}{3}(\g_a)_\b{}^{\g}\big(
8\ri\cD_{\g}\cS|
-\cDB_{\g}\bar{R}|
\big)
\Big{]}
\non\\
&&~~\,
+\e^\b\bar{\psi}_m{}_\d
\big(
\ri\cC_{\b}{}^\d|
+2\d_\b^\d\cS|
\big)
+\frac{\ri}{2}\psi_m{}^\d \eta_\d
+{\rm c.c.}\Big\}
\non\\
&&
-\bD_m{\bm\t} 
-\frac{1}{8}(\g_m)^{\g\d}  { [}\cD_{\g},\cDB_{\d}{]}\s|
~.
\label{susy-transformations-4}
\eea}\esubeq

The superspace torsion and curvature  
transform as tensors under the $\cK$-gauge group, 
eqs. \eqref{tau} and \eqref{tensor-K}. 
Their super-Weyl transformations
follow from
the transformation rules of the dimension-1 torsion superfields 
given in the previous section, eqs. \eqref{2.11d}--\eqref{2.11f}. 
This allows one to compute the variations of
the component  field strengths     
under the supersymmetry transformation \eqref{delta3}.

\subsection{The Weyl multiplet gauge}\label{subsection4.3}
 
 The super-Weyl invariance given by eqs. \eqref{2.3} preserves
the WZ gauge,
so we can eliminate a number of component fields.
We choose the gauge conditions 
\bea
\cS|=0~, \quad \cC_{\a\b}|=0~, \quad R|=\bar R|=0~, \quad \cD^2 R| +\bar \cD^2 \bar R| =0~,
\label{Wmg}
\eea
which constitute the Weyl multiplet gauge.
In Table 1, we identify those components  of the super-Weyl parameter $\s$ 
which have to be used in order to impose the Weyl multiplet gauge.
\begin{table}[htpb]
\begin{center}
\begin{tabular}{|c|c|}
\hline
Gauge Choice & $\s$-component \\[2mm]
\hline
$\cS|=0$ & $[\cD^\a,\cDB_\a]\s| $\\
[2mm]
\hline
$\cC_{\a\b}|=0$ & $[\cD_{(\a},\cDB_{\b)}]\s| $\\
[2mm]
\hline
$R|=\bar R|=0$ & $\cDB^2\s|~,~\cD^2 \s|$\\
[2mm] 
\hline
$\cD_\a R|=\cDB_\a\bar R|=0$ & $\cD_\a\cDB^2\s|~,~\cDB_\a\cD^2 \s|$\\[2mm]
\hline
$\cD^2 R|+\cDB^2\bar R|=0$ & $\{\cD^2, \cDB^2\}\s|$\\[2mm]
\hline
\end{tabular}
\caption{WZ-gauge choices and the parameters used to achieve them.}
\end{center}
\label{swz}
\end{table}

In the gauge \eqref{Wmg}, the super-Weyl gauge freedom is not  fixed completely. 
We stay with unbroken Weyl and local $S$-supersymmetry transformations
corresponding to the parameters ${\bm\s}$ and $\eta_\a$, $\bar \eta_\a$ respectively. 
The only independent component fields are the vielbein $e_m{}^a$, 
the two gravitini $\psi_m{}^\a $ and $\bar \psi_m{}^\a$, 
and the U(1)${}_R$ gauge field 
$b_m$. These fields and the associated local symmetries 
correspond to those describing the $\cN=2$ Weyl multiplet \cite{RvN}.

In the Weyl multiplet gauge, the explicit expressions for the gravitino field strength and the curvature tensors 
simplify drastically. 
The gravitino field strength becomes
\bea
{\bm\psi}_{ab}{}^\g &=&
\ri\ve_{abc}(\g^c)^{\a\b}\bar{{\bm C}}_{\a\b}{}^{\g}|
-\frac{4\ri}{3}\ve_{abc}(\g^c)^{\g\d}\cDB_{\d}\cS|~.
\eea
The Lorentz curvature takes the form:
\bea
\cR_{ab}{}^{cd}=
\Big{\{}\!\!
&&-\frac{\ri}{4}\ve_{abe}(\g^e)^{\a\b}\ve^{cdf}(\g_f)^{\t\d}
\big(\cD_{(\t}\bar{{\bm C}}_{\d\a\b)}
+\cDB_{(\t} {\bm C}_{\d\a\b)}\big)
+ \frac{4\ri}{3}  \d_{[a}^c\d_{b]}^d
\cD^\a\cDB_{\a}\cS
\non\\
&&
+\psi_{[a}{}^\b\Big{[}
(\g_{b]})_\b{}^{\g}{\bm C}_{\g\d\r}\ve^{cde}(\g_e)^{\d\r}
+\frac{2}{3}\ve_{b]}{}^{cd}\cD_{\b}\cS 
-\frac{8}{3}\d_{b]}^{[c}(\g^{d]})_{\b}{}^{\g} \cD_{\g}\cS
\Big{]}
\non\\
&&
+\bar{\psi}_{[a}{}_\b\Big{[}
(\g_{b]})^{\b\g}\bar{{\bm C}}_{\g\d\r}\ve^{cde}(\g_e)^{\d\r}
+\frac{2}{3}\ve_{b]}{}^{cd} \cDB^{\b}\cS 
-\frac{8}{3}\d_{b]}^{[c}(\g^{d]})^{\b\g} \cDB_{\g}\cS
\Big{]}
\Big{\}}\Big|~.~~~
\eea
From here we read off the scalar curvature
\bea
\cR (e, \psi) &=&
4\ri\cD^\a\cDB_{\a}\cS|
+\Big{\{}
\psi_{a}{}^\b\Big(
(\g^a)^{\g\d}C_{\b\g\d}|
+\frac{8}{3}(\g^a)_{\b}{}^{\g} \cD_{\g}\cS|
\Big)
+{\rm c.c.}
\Big{\}}~.~~~
\eea
An equivalent form for this  result is
\bea
\ri \cD^\a\cDB_\a\cS|=
\frac{1}{4}\Big(\cR(e,\psi)
+\ri\bar{\psi}_a\g_b{\bm\psi}^{ab}
+\ri\psi_a\g_b\bar{{\bm\psi}}^{ab}\Big)
~.
\eea
The U(1)${}_R$ field strength becomes
\bea
\cF_{ab}&=&\ve_{abc}
\Big\{
\frac{1}{2} (\g^c)^{\a\b} [\cD_\a,\cDB_{\b}]\cS
+\ri\ve^{cde}(\g_d)^{\b\g} \Big{[} \psi_e{}_\b \cD_\g\cS 
+\bar{\psi}_e{}_\b \cDB_{\g}\cS  \Big{]} \Big\} \Big|~.
\eea
An equivalent form for this result is
\bea[\cD_{(\a},\cDB_{\b)}]\cS|
&=&
(\g^a)_{\a\b}\Big\{
\cF_a
+\frac{1}{4}\psi^b{\bm{\bar{\psi}}}_{ab}
-\frac{1}{4}\bar{\psi}^{b}{\bm\psi}_{ab}
+\frac{1}{4}\ve^{abc}\big(\psi_b\g^d{\bm{\bar{\psi}}}_{cd}-\bar{\psi}_{b}\g^d{\bm\psi}_{cd}\big)
\Big\}
~,~~~~~~
\eea
 where $ \cF_a:=\hf\ve_{abc}\cF^{bc}$.

We need to determine those residual gauge transformations 
which leave invariant the Weyl multiplet gauge.
Imposing the conditions $\d\cC_{\a\b}|=\d\cS|=\d R|=0$, 
with the transformation $\d$ defined by \eqref{delta3}, 
we obtain
\bsubeq
\bea
{[}\cD_{(\a},\cDB_{\b)}{]}\s|
&=&
-\ve_{cab}(\g^c)_{\a\b}\big(\e{\bm{\bar{\psi}}}^{ab}-\bar{\e}{\bm\psi}^{ab}\big)
~,
\label{sWcond-111}
\\
\ri\cD^\g\cDB_\g\s|
&=&
\frac{\ri}{2}\ve_{cab}\big(
\e\g^c{\bm{\bar{\psi}}}^{ab}
+\bar{\e}\g^c{\bm\psi}^{ab}\big)
~,
\label{sWcond-112}
\\
\cD^2\s|
&=&
\cDB^2\s|=0
~.
\label{sWcond-113}
\eea
\esubeq
Using these results
 in (\ref{susy-transformations-1})--(\ref{susy-transformations-4}),
together with the fact that the bar-projections of all the
dimension-1 curvature superfields vanish,
we derive the transformations of the gauge fields in the Weyl multiplet gauge:
\bsubeq \label{susy-transformations-Wg}
{\allowdisplaybreaks
\bea
\d e_m{}^a
&=&
\ri\big(
\e\g^a\bar{\psi}_m
+\bar{\e}\g^a\psi_m
\big)
-{\bm\s}e_m{}^a
~,
\label{susy-transformations-1-Wg}
\\
\d\psi_m{}^\a&=&
2\bD_m\e^\a
+\ri( \tilde{\g}_m \bar{\eta})^\a
-\ri{\bm \t}\psi_m{}^\a 
-\hf{\bm \s}\psi_m{}^\a 
~,
\label{susy-transformations-3-Wg}
\\
\d b_m&=&
-\hf e_m{}^a\Big{[}\,
\e\g^b{\bm {\bar{\psi}}}_{ab}
+\frac{1}{2}\ve_{abc}\,\e{\bm {\bar{\psi}}}^{bc}
-\ri\psi_a\eta
+{\rm c.c.}
\Big{]}
- \bD_m{\bm\t} 
~,
\label{susy-transformations-4-Wg}
\eea}\esubeq
with the $\g$-matrices with world indices defined by $\g_m := e_m{}^a \g_a$
and similarly for $\tilde{\g}_m$.

The above description of the Weyl multiplet agrees with that given in \cite{RvN}.

\subsection{Alternative gauge fixings} 

There exist different schemes for component reduction that correspond to 
alternative choices of fixing the supergravity gauge freedom. Here we mention
two possible options that are most useful in the context of Type I or Type II supergravity 
formulations. 

The super-Weyl and local U$(1)_R$ gauge freedom can be used to impose the 
gauge condition \cite{KLT-M11} 
\bea
\cS =0~, \qquad \F_\a =0~, \qquad \F_a = \cC_a~.
\label{TypeIgf}
\eea
Since the resulting U$(1)_R$ connection is a tensor superfield, 
we may  equally well work with covariant derivatives $\nabla_A$  without 
 U$(1)_R$ connection, which are defined by 
 \bea
 \nabla_\a: = \cD_\a ~, \qquad \nabla_a :=  \cD_a - \ri \cC_a \cJ~.
 \eea
The gauge condition \eqref{TypeIgf} does not fix completely the super-Weyl and local U$(1)_R$ gauge freedom. 
The residual transformation is generated by a covariantly chiral scalar 
parameter $\l$, $\bar \nabla_\a \l=0$, and has the form \cite{KT-M11}
\bsubeq \label{sW+U(1)}
\bea
\nabla'{}_\a&=&
\re^{\frac{1}{2}(3\bar{\l}-\l)}
\Big(\nabla_\a+(\nabla^{\g}\l)\cM_{\g\a}\Big)~,
\\
\nabla'{}_{a}&=&\re^{\l+\bar{\l}}\Big(
\nabla_{a}
-\frac{\ri}{2}(\g_a)^{\a\b}(\nabla_{\a}\l)\bar \nabla_{\b}
-\frac{\ri}{2}(\g_a)^{\a\b}(\bar \nabla_{\a}\bar{\l})\nabla_{\b}
\non\\
&&~~~~~~~
+\ve_{abc}\big(\nabla^b(\l+\bar{\l})\big)\cM^c
-\frac{\ri}{2}(\nabla^{\g}\l)(\bar \nabla_{\g}\bar{\l})\cM_{a}
\Big)
~.
\eea
\esubeq
The dimension-1 torsion superfields transform as
\bsubeq \label{sW+U(1)-torsion}
\bea
\cC'_{a}&=&\re^{\l+\bar{\l}}\Big(
\cC_{a}
-\frac{\ri}{2}\nabla_{a}(\l-\bar{\l})
+\frac{1}{4}(\g_a)^{\a\b}(\nabla_{\a}\l)\bar \nabla_{\b}\bar{\l}
\Big)
~,
\label{sW-C-Type-I}
\\
R'
&=&
-\frac{1}{4}\re^{3\l}\Big(
\bar \nabla^2
-4R
\Big)
\re^{-\bar{\l}}
~.
\label{sW-R-Type-I}
\eea
\esubeq
This formulation is very similar to the old minimal 4D $\cN=1$ supergravity, 
see e.g. \cite{BK} for a review. It is best suited when dealing with 
Type I minimal supergravity-matter systems. 

The super-Weyl freedom can be used to impose the 
gauge condition \cite{KLT-M11} 
\bea
R =0~,
\label{TypeIIgf}
\eea
with the local U$(1)_R$ group being unbroken.
This superspace geometry is most suitable for the Type II minimal supergravity. 
The gauge condition \eqref{TypeIIgf} does not completely fix  the
super-Weyl group.
The residual super-Weyl transformation is generated by a real superfield $\s$
constrained by $\cD^2\re^{-\s}=\cDB^2\re^{-\s}=0$.

Each of the two restricted superspace geometries considered, 
\eqref{TypeIgf} and \eqref{TypeIIgf}, is suitable for describing the Weyl multiplet 
of conformal supergravity. In each case, we can define a Wess-Zumino gauge 
and a Weyl multiplet gauge. 

Some alternative gauge conditions will be used in section \ref{sym-curv-sup}.


\section{Type I minimal supergravity}

This off-shell supergravity theory and its matter couplings
are analogous to 
the old minimal formulation for
4D $\cN=1$ supergravity, see \cite{WB,GGRS,BK} for  reviews.
Its specific feature  is that its
conformal compensators are  a covariantly chiral superfield  $\F$ of super-Weyl weight $w=1/2$,
\bea
{\bar \cD}_\a \F=0 ~, \qquad
\cJ\F=-\hf\F~,\qquad
\d_\s\F=\hf \s\F~,
\label{chiral51}
\eea
and its conjugate $\bar \F$.

\subsection{Pure supergravity}
As a warm-up exercise, we first analyze the action for pure Type I supergravity
with a cosmological term. 
It is obtained from
\eqref{s-m} by switching off the matter sector, that is by setting $K=0$ and $W=\m ={\rm const}$,
\begin{align}
S_{\rm SG} = -4 \int {\rm d}^3x \rd^2\q\rd^2\qb
\,E\,
 \bar \F  \,\F
       + \m \int {\rm d}^3x {\rm d}^2 \q \,\cE\,
       \F^4
       + \bar \m \int {\rm d}^3x  {\rm d}^2 {\bar \q}\,\bar \cE\, \bar\F^4 ~.
       \label{SUGRA-I}
\end{align}
The second and third term in the action generate a 
supersymmetric cosmological term, with 
the parameter $|\m|^2$ being proportional to the cosmological constant.
The dynamics of this theory was analyzed in superspace in \cite{KT-M11}.
Here we reduce the action \eqref{SUGRA-I} to components.

In the Weyl multiplet gauge, the super-Weyl gauge freedom is not fixed completely.
We can use the residual Weyl and local U$(1)_R$ symmetries to impose
the gauge condition
\begin{subequations}\label{gauge-I}
\bea
\F| &=&1~.
\label{I-1}
\eea
In addition, the local  $S$-supersymmetry invariance can be used to make the gauge choice
\bea
\cD_\a \F|&=&0~.
\label{I-2}
\eea
\end{subequations}
The only surviving component field of $\F$ may be defined as
\bea
M:= \cD^2 \F|~.
\eea

To perform the component reduction of the kinetic term in  \eqref{SUGRA-I},
the first step is to  associate with it, by applying the relation \eqref{N=2Ac-2},
the equivalent antichiral Lagrangian
$\bar \cL_{\rm c} = (\cD^2 -4\bar R) ( \bar \F \F)$.
After that we can use   \eqref{comp-ac-1} to reduce the action to components.
The antichiral Lagrangian corresponding to the $\bar \m$-term in \eqref{SUGRA-I}
is $\bar \cL_{\rm c} = \bar \m \bar \F^4$. Finally,  the component version of the $\m$-term
in  \eqref{SUGRA-I} is the complex conjugate of the $\bar \m$-term.

Direct calculations lead to the supergravity Lagrangian
\bea
L_{\rm SG} 
&=& 
\hf \cR (e,\psi)
+\frac{\ri}{4} \ve^{abc} \Big( { \bar \psi}_{ab} \psi_c + \bar \psi_a {\psi}_{bc}\Big)
-\frac{1}{4} \bar { M} { M} +{ b}^a { b}_a \non \\
&&
- \bar \m \Big(\bar { M}
-\hf\ve^{abc}\psi_a\g_b \psi_c \Big)
- \m \Big({{ M} }
+\hf\ve^{abc}{\bar{\psi}}_a\g_b{\bar{\psi}}_c \Big) ~,
\label{SUGRA-I-com}
\eea
where the  gravitino field strength is defined as
\bea
\psi_{ab} := {\frak D}_a \psi_b -  {\frak D}_b \psi_a - \cT_{ab}{}^c \psi_c~,
\label{gfs5.8}
\eea
which differs from \eqref{gfs}.
We recall that the covariant derivative ${\frak D}_a$,
eq. \eqref{dev000},  has no U$(1)_R$
connection. It is natural to use   ${\frak D}_a$ since
the local U$(1)_R$ symmetry has been fixed.
The Type I supergravity multiplet consists of the following fields:
the dreibein $e_m{}^a$,  the gravitini $\psi_m{}^\a $ and $\bar \psi_{m \a}$,
and the auxiliary fields $M$, $\bar M$ and $b_m$.

Upon elimination of  the auxiliary fields, the Lagrangian becomes
\bea
L_{\text{SG}} &=& \hf \cR (e,\psi)
+ \frac{\ri}{4} \ve^{abc} \Big( { \bar \psi}_{ab} \psi_c + \bar \psi_a {\psi}_{bc}\Big)
\non \\
&&+ 4\bar \m \m
+ \frac{\bar \m}{2}  \ve^{abc}\psi_a\g_b \psi_c
- \frac{\m}{2} \ve^{abc}{\bar{\psi}}_a\g_b{\bar{\psi}}_c  .
\eea
This Lagrangian
describes (1,1) anti-de Sitter supergravity for $\m \neq 0$ \cite{AT}.

\subsection{Supersymmetry transformations}\label{section5.2}

The gauge conditions \eqref{I-1} and \eqref{I-2} completely fix 
the Weyl,  local U(1)$_R$ and $S$-supersymmetry invariances. 
However, performing just a single $Q$-supersymmetry transformation, 
with $\e_\a$ and $\bar \e_\a$ the only non-zero parameters in 
\eqref{susy-transformations-Wg}, 
does not preserve these gauge conditions. 
To restore the gauge defined by  \eqref{I-1} and \eqref{I-2}, 
the $Q$-supersymmetry transformation  
has to be accompanied by a compensating
$S$-supersymmetry transformation.
Indeed, applying 
the transformation \eqref{delta3} to  $\F |$ gives
\bea
\d \F | = \e^\b \cD_\b \F | 
+\hf\big({\bm\s}-\ri{\bm \t}\big) \F|
=  
\hf\big({\bm\s}
-\ri  {\bm \t}\big)
~,
\eea
where we have used eqs. \eqref{I-1} and \eqref{I-2}.
Setting $\d \F|=0$ gives
\bea
{\bm \s}={\bm \t} =0~.
\label{sigmatau}
\eea 
On the other hand, 
the transformation of  $\cD_\a\F|$ is
\bea
\d\cD_\a\F|
=
\e^\b\cD_\b\cD_\a\F|
+\bar{\e}_\b\cDB^\b\cD_\a\F|
-\eta_\a\Big(\cJ\F|
-\hf\F|\Big)
=
-\hf\e_\a M
+(\g^a\bar{\e})_\a b_a
+\eta_\a
~,~~~~~~
\eea
where here we have used \eqref{I-1} and \eqref{I-2}.
Setting $\d\cD_\a\F| =0$ gives
\bea
\eta_\a(\e)
&=&
\hf\e_\a M
-(\g^a\bar{\e})_\a b_a
~.
\eea
Using
these results 
in \eqref{susy-transformations-Wg}, 
we obtain the supersymmetry transformation laws of the gauge  fields:
\bsubeq
\bea
\d_\e e_m{}^a
&=&
\ri\big(
\e\g^a\bar{\psi}_m
+\bar{\e}\g^a\psi_m
\big)
~,
\\
\d_\e\psi_m{}^\a&=&
2\frak{D}_m\e^\a
-\ri b_m\e^\a
+\ri e_m{}^a\ve_{abc}\,b^b(\tilde{\g}^c\e)^{\a}
+\frac{\ri}{2}\bar{M}(\tilde{\g}_m\bar{\e})^{\a}
~,
\label{grav-var-I_0}
\\
\d_\e b_m&=&
-\hf e_m{}^a\Big\{\,
\e \g^b{\bar{\psi}}_{ab}
+\frac{1}{2}\ve_{abc}\e{\bar{\psi}}^{bc}
+\ri\ve_{abc}b^{b} \e {\bar{\psi}}^{c}
\non \\
&&  \qquad \qquad +\ri\big(
b_{a}\e\g^b{\bar{\psi}}_{b}
-2b_b\e\g^b{\bar{\psi}}_a
\big)
-\frac{\ri}{2}M\e\psi_a
\Big\}
+{\rm c.c.}
\eea\esubeq

The  supergravity multiplet also includes the auxiliary scalar $M=\cD^2\F|$.
Due to \eqref{sigmatau} and since 
$\cD^2\s|=0$, eq. \eqref{sWcond-113},
the supersymmetry transformation of $M$ is 
\bea
\d_\e M
&=&
\e^\b\cD_\b\cD^2\F|
+\bar{\e}_\b\cDB^\b\cD^2\F|
=
\bar{\e}_\b{[}\cDB^\b,\cD^2{]}\F|
~.
\eea
Making use of the algebra of covariant derivatives gives
\bea
\d_\e M
=
- \ve_{cab} \bar \e{\tilde{\g}}^c{\bar{\psi}}^{ab}
-\ri M \bar \e \tilde{\g}^a\psi_a
-2\ri b_a \bar \e \bar{\psi}^a{}
~.
\eea

\subsection{Matter-coupled supergravity}

We  consider a general locally supersymmetric nonlinear $\s$-model
\begin{align}
S 
= 
-4 \int {\rm d}^3x \rd^2\q\rd^2\qb
\,E\,
 \bar \F  \,{\rm e}^{-K /4}\F
       + \int {\rm d}^3x {\rm d}^2 \q \,\cE\,
       \F^4 W
       + \int {\rm d}^3x  {\rm d}^2 {\bar \q}\,\bar \cE\, \bar\F^4 \bar W~.
       \label{s-m}
\end{align}
Here the K\"ahler potential,  $K= K(\vf^I, \bar \vf^{\bar J})$, is a real function
of the covariantly chiral superfields $\vf^I$ and their conjugates
$\bar \vf^{\bar I}$,  $\bar \cD_\a \vf^I =0$.
The superpotential,  $W = W(\vf^I)$,  is a holomorphic function of $\vf^I$ alone.
The matter superfields $\vf^I$ and $\bar \vf^{\bar J}$ are chosen to be inert under the super-Weyl
and local ${\rm U(1)}_R$
transformations. This guarantees the super-Weyl invariance of the action.
In Appendix \ref{AppendixB}, we describe a different parametrization of the nonlinear $\s$-model \eqref{s-m} 
in which 
the dynamical variables $\F$ and $\vf^I$ are replaced by 
covariantly chiral superfields $\f^i= (\f^0, \f^I)$ 
of super-Weyl weight $w=1/2$ that parametrize a K\"ahler cone.  

The action (\ref{s-m}) is also invariant under a target-space K\"ahler transformation
\begin{subequations}
\bea \label{3.16}
K (\vf , \bar \vf)    &\rightarrow & K  (\vf , \bar \vf)  + \L  (\vf ) + \bar \L  ( \bar \vf)~ , \label{Kahler1} \\
W (\vf) & \rightarrow & {\rm e}^{-\L (\vf )} \,W(\vf)~,
\eea
provided the compensator changes as
\bea
\F \rightarrow {\rm e}^{\L(\vf)/4}\,\F~,
\eea
\end{subequations}
with  $\L(\vf^I)$ an arbitrary holomorphic function.

We first compute the  component form  
of  \eqref{s-m}
in the special case $W=0$,
\bea
S_{\rm kinetic} = -4 \int {\rm d}^3x \rd^2\q\rd^2\qb\,E\,
 \bar \F  \,{\rm e}^{-K(\vf, \bar \vf) /4}\F~.
\label{5.10}
\eea
Associated with $S_{\rm kinetic} $
is the antichiral Lagrangian
\bea
\bar \cL_{\rm c} = (\cD^2 -4\bar R) ( \bar \F {\rm e}^{-K /4} \F)~,
\eea
which has to be used for computing the component action using
the general formula  \eqref{comp-ac-1}.

Our consideration in this and next sections is similar to that in 4D $\cN=1$ supergravity
\cite{KU,KMcC2}.
To reduce the action to components, we impose the following
Weyl and local $S$-supersymmetry gauge conditions:
\begin{subequations} \label{F-Weyl}
\bea
( \bar \F {\rm e}^{-K /4} \F) \big| &=& 1~, \label{Weyl-gauge}\\
\cD_\a ( \bar \F {\rm e}^{-K /4} \F) \big|&=&0~. \label{S-susy-gauge}
\eea
Both gauge conditions are manifestly K\"ahler invariant.
It turns out that the condition \eqref{Weyl-gauge}
leads  to the correct Einstein-Hilbert gravitational Lagrangian at
the component level. On the other hand, the  condition \eqref{S-susy-gauge}
guarantees that no cross terms
$\cD^\a \cS | \bar \cD_\a K |$ are generated at the component level.
See appendix B for more details.
Finally we fix the local U(1)${}_R$ symmetry by imposing the gauge condition
\bea
\F |=\bar \F |= {\rm e}^{K /8}~.
\label{U(1)-gauge}
\eea
\end{subequations}
The auxiliary scalar fields contained in $\F$ and $\bar \F$ may be defined
in a manifestly K\"ahler-invariant way as
\bea
\mathbb{M}:=\cD^2(\Fb\re^{-\frac{1}{4} K} \F)|~, \qquad
\bar{\mathbb{M}}:=\cDB^2(\Fb\re^{-\frac{1}{4} K} \F)|~.
\label{MbarM}
\eea

To make the gauge condition \eqref{U(1)-gauge} K\"ahler invariant,
the K\"ahler transformation generated by a parameter $\L$ has to be accompanied
by a special U$(1)_R$-transformation with parameter $ \t = \frac{\ri}{4} ( \bar \L - \L)$
such that the component vector field  $b_a$, which belongs to the Weyl multiplet
and is defined by eq. \eqref{stcd-2},  transforms as
\bea
b_a  \rightarrow b_a + \frac{\ri}{4} {\frak D}_a ( \L - \bar \L)~.
\label{Kahler-B}
\eea

We define the component fields of $\vf^I$ as follows:
\begin{subequations} \label{vf-comfields}
\bea
X^I &:=& \vf^I |~, 
\label{comp-chiral-1}
\\
\l_\a^I&:=&\cD_\a\vf^I |~, 
\label{comp-chiral-2}\\
F^I &:=&
-\frac{1}{4} \big{[} \cD^2 \vf^I + \G^{I}_{JK} (\cD^\a \vf^{J}) \cD_\a \vf^{K}\big{]}|
~.
\label{comp-chiral-3}
\eea
\end{subequations}
Under a holomorphic reparametrization,  $X^I \to f^I (X)$,
of the target K\"ahler space,
the fields $\l^I_\a$ and $F^I$ transform as holomorphic vector fields.
Direct calculations lead to the following component Lagrangian:
\bea
L_{\rm kinetic} &=&
 \hf \cR (e,\psi) 
 + \frac{\ri}{4} \ve^{abc} \Big( \widetilde{ \bar \psi}_{ab} \psi_c + \bar \psi_a \widetilde{\psi}_{bc}\Big)
-\frac{1}{4} \bar {\mathbb M} {\mathbb M} 
+{\mathbb B}^a {\mathbb B}_a 
\non   \\
&&
+ g_{I\bar J}\Big{[} \,
 F^I \bar F^{\bar J}
-({\mathfrak D}^a X^I) {\mathfrak D}_a \bar X^{\bar J}
-\frac{\ri}{4} \l^I \g^a \stackrel{\longleftrightarrow}{\widetilde{\mathfrak D}_a} \bar \l^{\bar J}
+\frac{1}{8}\l^I \bar \l^{\bar J} \big( \ve^{abc} \bar \psi_a \g_b \psi_c - \bar \psi^a \psi_a\big)
 \non \\
&&~~~
-\frac{1}{8}  \l^I  \g^a \bar \l^{\bar J} \big(\bar \psi^b \g_a \psi_b+ \ve_{abc}  \bar \psi^b \psi^c\big) 
-\hf \psi^a \g_b\tilde{\g}_a \l^I {\mathfrak D}^b \bar X^{\bar J}
-\hf \bar \l^{\bar J} \tilde{\g}_a \g_b \bar \psi^a {\mathfrak D}^b X^I 
\Big{]}
\non \\
&&
+ \frac{1}{16} R_{I \bar K J \bar L} \l^{ I} \l^J\,\bar \l^{\bar K} \bar \l^{\bar L}
-\frac{1}{64} \big( g_{I\bar J} \l^{I} \bar \l^{\bar J} \big)^2 
~,
\label{Type-I-comp} 
\eea
where the auxiliary vector field ${\mathbb B}_a $ is defined by the rule
\bea
{\mathbb B}_a := b_a - \frac{1}{8} g_{I\bar J} \l^I \g_a \bar \l^{\bar J}
-\frac{\ri}{4}( K_I {\mathfrak D}_a X^I  - K_{\bar I} {\mathfrak D}_a \bar X^{\bar I})
~,
\label{auxVect}
\eea
and is invariant under the K\"ahler transformations, in accordance with \eqref{Kahler-B}.
The gravitino field strength in \eqref{Type-I-comp} 
differs from that introduced earlier in
\eqref{gfs5.8}:
\bea
\widetilde{\psi}_{ab} &=& \widetilde{\mathfrak D}_a \psi_b
-\widetilde{\mathfrak D}_b \psi_a -\cT_{ab}{}^c \psi_c ~,
\label{gravFS-1}
\eea
where the K\"ahler-covariant derivative 
$ \widetilde{\mathfrak D}_a$ 
is defined (similarly to the 4D case, see e.g. \cite{WB})
as follows
\begin{subequations}
\bea
\widetilde{\mathfrak D}_a \psi_b &:=& {\mathfrak D}_a \psi_b
+ \frac{1}{4}( K_J {\mathfrak D}_a X^J  - K_{\bar J} {\mathfrak D}_a \bar X^{\bar J}) \psi_b~, 
\\
\widetilde{\mathfrak D}_a \l^I &:=& {\mathfrak D}_a \l^I
-  \frac{1}{4}( K_J {\mathfrak D}_a X^J  - K_{\bar J} {\mathfrak D}_a \bar X^{\bar J}) \l^I
+  \l^J \G^I_{JK}  {\mathfrak D}_a X^K~.
\label{covDevL}
\eea
\end{subequations}
In \eqref{Type-I-comp}, as usual $g_{I \bar J} $
denotes the K\"ahler metric, $g_{I \bar J} =K_{I \bar J}$,
and $R_{I\bK J\bL}$
the Riemann curvature,
\bea
R_{I\bK J\bL}= K_{IJ \bK \bL} - g_{M\bar N} \G^M_{IJ} \G^{\bar N}_{\bK \bL}~,
\eea
with $ \G_{JK}^I  = g^{I \bL } K_{JK \bL}$ the Christoffel symbol.

We now turn to the third term in \eqref{s-m}. The corresponding antichiral
Lagrangian is $\bar \cL_{\rm c} = \bar \F^4 \bar W (\bar \vf) $.
To reduce this functional to components, we again
make use of the general rule  \eqref{comp-ac-1} in conjunction with
the relations  \eqref{F-Weyl} and  \eqref{vf-comfields}
which define
the component fields of $\F$ and $\vf^I$.
The second term in \eqref{s-m} is just the complex conjugate of the third term.

The component Lagrangian corresponding to the second and third terms
in \eqref{s-m} is
\bea
L_{\text{potential}}
&=&
-\re^{K/2}\Big{[}
\Big(\bar {\mathbb M}
-\hf\ve^{abc}\psi_a\g_b \psi_c \Big) \bar W
+\Big({{\mathbb M} }
+\hf\ve^{abc}{\bar{\psi}}_a\g_b{\bar{\psi}}_c \Big) W
\non\\
&&\qquad~~~
-\Big(\bar{F}^\bI+\frac{\ri}{2}\psi_a\g^a\lb^{\bI}\Big)\de_\bI \bar W
-\Big(F^I+\frac{\ri}{2}{\bar{\psi}}_a\g^a\l^{I}\Big)\de_I W
\non\\
&& \qquad~~~
+\frac{1}{4}\lb^\bI\lb^{\bJ}\,\de_\bI\de_{\bJ} \bar W
+\frac{1}{4}\l^{I}\l^{J}\,\de_I\de_{J} W\Big{]}
~.
\label{Type-I-comp-potential}
\eea
Here we have introduced the K\"ahler-covariant   derivatives
\bsubeq
\bea
\de_I {W}
&:=&
{W}_I+K_I {W}
~,
\\
\de_I\de_J {W}
&:=&
W_{I J}
+2K_{(I}\de_{J)}  W
- \G^L_{I J}\de_L W
+ K_{I J}  W
+K_{I}K_{J}  W ~.
\eea
\esubeq

The component Lagrangian corresponding to the supergravity-matter system
\eqref{s-m} is $L=L_{\text{kinetic}} + L_{\text{potential}} $.
Putting together \eqref{Type-I-comp} and
\eqref{Type-I-comp-potential} gives
\bea
L 
&=& 
\hf \cR (e,\psi) 
+ \frac{\ri}{4} \ve^{abc} \Big( \widetilde{ \bar \psi}_{ab} \psi_c + \bar \psi_a \widetilde{\psi}_{bc}\Big)
-\frac{1}{4} \bar {\mathbb M} {\mathbb M} 
+{\mathbb B}^a {\mathbb B}_a \non   \\
&&
+ g_{I\bar J}\Big{[} ~
 F^I \bar F^{\bar J}
-({\mathfrak D}^a X^I) {\mathfrak D}_a \bar X^{\bar J}
-\frac{\ri}{4} \l^I \g^a \stackrel{\longleftrightarrow}{\widetilde{\mathfrak D}_a} \bar \l^{\bar J}
+\frac{1}{8} \l^I \bar \l^{\bar J} \big( \ve^{abc} \bar \psi_a \g_b \psi_c - \bar \psi^a \psi_a\big)
 \non \\
&&~~~~~~~~
-\frac{1}{8}  \l^I  \g^a \bar \l^{\bar J} \big(\bar \psi^b \g_a \psi_b+ \ve_{abc}  \bar \psi^b \psi^c\big) 
-\hf \psi^a \g_b\tilde{\g}_a \l^I {\mathfrak D}^b \bar X^{\bar J}
-\hf \bar \l^{\bar J} \tilde{\g}_a \g_b \bar \psi^a {\mathfrak D}^b X^I 
\Big{]}
\non \\
&&
+ \frac{1}{16} R_{I \bar K J \bar L} \l^{ I} \l^J\,\bar \l^{\bar K} \bar \l^{\bar L}
-\frac{1}{64} \big( g_{I\bar J} \l^{I} \bar \l^{\bar J} \big)^2 
\non\\
&&
-\re^{K/2}\Big{[}
\Big(\bar {\mathbb M}
-\hf\ve^{abc}\psi_a\g_b \psi_c \Big) \bar W
+\Big({{\mathbb M} }
+\hf\ve^{abc}{\bar{\psi}}_a\g_b{\bar{\psi}}_c \Big) W
\non\\
&&\qquad~~~
-\Big(\bar{F}^\bI+\frac{\ri}{2}\psi_a\g^a\lb^{\bI}\Big)\de_\bI \bar W
-\Big(F^I+\frac{\ri}{2}{\bar{\psi}}_a\g^a\l^{I}\Big)\de_I W
\non\\
&& \qquad~~~
+\frac{1}{4}\lb^\bI\lb^\bJ\,\de_\bI\de_{\bJ} \bar W
+\frac{1}{4}\l^{I}\l^{J}\,\de_I\de_{J} W\Big{]}~.
\eea
Upon elimination of the auxiliary fields, the Lagrangian turns into
\bea
L 
&=& 
\hf \cR (e,\psi) 
+\frac{\ri}{4} \ve^{abc} \Big( \widetilde{ \bar \psi}_{ab} \psi_c
+ \bar \psi_a \widetilde{\psi}_{bc}\Big)
+ \frac{1}{16} R_{I \bar K J \bar L} \l^{I} \l^J\, \bar \l^{\bar K} \bar \l^{\bar L}
-\frac{1}{64} \big( g_{I\bar J} \l^{ I} \bar \l^{\bar J} \big)^2
 \non   \\
&&
+ g_{I\bar J}\Big{[} 
-({\mathfrak D}^a X^I) {\mathfrak D}_a \bar X^{\bar J}
-\frac{\ri}{4} \l^I \g^a \stackrel{\longleftrightarrow}{\widetilde{\mathfrak D}_a} \bar \l^{\bar J}
+\frac{1}{8}  \l^I \bar \l^{\bar J} \big( \ve^{abc} \bar \psi_a \g_b \psi_c - \bar \psi^a \psi_a\big)
 \non \\
&&~~~~~~~~
-\frac{1}{8} \l^I  \g^a \bar \l^{\bar J} \big(\bar \psi^b \g_a \psi_b+ \ve_{abc}  \bar \psi^b \psi^c\big) 
-\hf \psi^a \g_b\tilde{\g}_a \l^I {\mathfrak D}^b \bar X^{\bar J}
-\hf \bar \l^{\bar J} \tilde{\g}_a \g_b \bar \psi^a {\mathfrak D}^b X^I 
\Big{]}
\non\\
&&
+\re^{K/2}\Big{[}~
\hf\ve^{abc} \Big( \psi_a\g_b \psi_c  \bar W
- {\bar{\psi}}_a\g_b{\bar{\psi}}_c W  \Big)
+\frac{\ri}{2}\psi_a\g^a\lb^{\bI} \de_\bI \bar W
+\frac{\ri}{2}{\bar{\psi}}_a\g^a\l^{I}\de_I W
\non\\
&& \qquad~~~
-\frac{1}{4}\lb^\bI\lb^{\bJ}\,\de_\bI\de_{\bJ} \bar W
-\frac{1}{4}\l^{I}\l^{J}\,\de_I\de_{J} W\Big{]}  \non \\
&& - \re^K (g^{I \bJ} \nabla_I W \bar \nabla_{\bJ} \bar W -4 W \bar W) ~.
\label{5.23}
\eea
The potential generated,
$P_{\rm 3D} =  \re^K (g^{I \bJ} \nabla_I W \bar \nabla_{\bJ} \bar W -4 W \bar W)$,
differs slightly from the famous four-dimensional result
$P_{\rm 4D} =  \re^K (g^{I \bJ} \nabla_I W \bar \nabla_{\bJ} \bar W -3 W \bar W)$,
see e.g. \cite{WB}.

\subsection{Supersymmetry transformations in Einstein frame}

In matter coupled supergravity, the gauge conditions  \eqref{F-Weyl}
depend on 
the matter fields. 
As a consequence, the supersymmetry transformation laws of the supergravity fields 
will differ from those given in 
subsection \ref{section5.2}.
To preserve the gauge condition $\F|=\re^{K/8}$, 
we have to choose
\bea
{\bm\s}(\e)=0~,~~~~~~
{\bm\t}(\e)
=
-\frac{\ri}{4}\big(K_I \e\l^I  - K_\bI \bar{\e}\lb^{\bI} 
\big)
~.
\label{TIgc-1}
\eea
To preserve the gauge condition $\cD_\a(\Fb\re^{-K/4}\F)|=0$, we have to apply  
the compensating $S$-supersymmetry transformation with parameter
\bea
\eta_\a(\e)
&=&
\hf\e_\a{\mathbb M}
+\bar{\e}^\b\Big{[}
-b_{\a\b}
+\frac{\ri}{4}\big(K_I {\mathfrak D}_{\a\b}X^I
-   K_\bI {\mathfrak D}_{\a\b}\Xb^\bI
\big)
\non\\
&&~~~~~~~~~~~~~~~~~
+\frac{1}{8} g_{I \bar J} \big( \ve_{\a\b}\l^{I}\lb^{\bJ} 
+2\l_{(\a}^I\lb_{\b)}^{\bJ}\big)
\Big{]}
~.
\label{TIgc-2}
\eea

Making use of the parameters ${\bm\t}(\e)$ and $\eta_\a(\e)$
in \eqref{susy-transformations-Wg},
one may derive the supersymmetry transformations of the 
supergravity fields $e_m{}^a$ and $\psi_m{}^\g$ 
and $b_m$. These expressions are not illuminating, 
and here we do not  give them.  
We only comment upon the derivation of the supersymmetry transformation of  
${\mathbb M}$.
Its transformation follows from the fact that ${\mathbb M}$ is defined to be the lowest component of
the scalar superfield $\cD^2(\Fb\re^{-\frac{1}{4} K} \F)$. 
Making use of \eqref{TIgc-1} and \eqref{TIgc-2}, after some algebra we get
\bea
\d_\e {\mathbb M}
&=& \,
-\ve_{cab} \bar{\e} \tilde{\g}^c{\bar{\psi}}^{ab}
-\ri \bar{\e}  \tilde{\g}^a\psi_a {\mathbb M}
-2\ri b_a \bar{\e}\bar{\psi}^a
+g_{I\bJ} F^I \bar{\e}\bar{\l}^{\bJ}
-\ri g_{I\bJ} \bar{\e}\tilde{\g}^a \l^I{\mathfrak D}_a\Xb^\bJ
\non\\
&&
-\hf\bar{\e}\tilde{\g}^a\g^b\bar{\psi}_a
\big(
  K_\bI  {\mathfrak D}_b\Xb^\bI
-   K_I {\mathfrak D}_bX^I
\big)
+2\ri{\bm \t}(\e){\mathbb M}
~.
\eea

In conclusion, we give 
the transformation rules of the component fields of  $\vf^I$: 
\bsubeq
\bea
\d_\e X^I&=&\e \l^I~,
\\
\d_\e \l_\a^I&=&
2\e_\a\Big(
F^I
+\frac{1}{4}\G^I_{JK} \l^{J} \l^K
\Big)
+2\ri(\g^a\bar{\e})_\a\Big(
{\mathfrak D}_aX^I
-\hf\psi_a\l^I
\Big)
+\ri{\bm\t}(\e)\l_\a^I
~,~~~~~~
\label{vfvf2}
\\
\d_\e F^I&=&
-\e\l^{J}\G^I_{JK} F^K
+\hf\l^I\eta(\e)
+2\ri{\bm \t}(\e)F^I
+\ri\bar{\e}\g^a\big({\mathfrak D}_a
+\ri b_a\big)\l^I 
-\frac{1}{4} g^{I\bar{L}}R_{J\bar{L} K\bar{P}}\,\bar{\e}\bar{\l}^{\bar{P}}\,\l^{J}\l^K
\non\\
&&
+\ri\bar{\e}\g^a\l^{J}\G^I_{JK} {\mathfrak D}_aX^K
-\ri\bar{\e}\g^a\psi_a F^I
-\bar{\e}\g^a\tilde{\g}^b\bar{\psi}_a\Big({\mathfrak D}_bX^I-\hf\psi_b\l^I\Big)
~.
\eea
\esubeq
These can be derived by using the definition of the components 
of $\vf^I$ \eqref{vf-comfields}.


\section{Type II minimal supergravity}

This supergravity theory is a 3D analogue of the new minimal formulation 
for 4D $\cN=1$ supergravity \cite{new} (see \cite{GGRS,BK} for reviews).
Its conformal compensator is a real covariantly linear scalar 
$\mathbb G$, 
\bea
(\cD^2 - 4 \bar R ) \mathbb G =  
(\cDB^2-4R)\mathbb G= 0~,
\label{N=2realLinear}
\eea
 chosen to be nowhere vanishing, $\mathbb G \neq 0$. 
The super-Weyl transformation of $\mathbb G$ is uniquely fixed 
by the constraint  (\ref{N=2realLinear}) to be
\bea
\d_\s {\mathbb G}=\s \mathbb G~.
\label{N=2sWrealLinear}
\eea

\subsection{Real linear scalar}

A general solution of the off-shell constraint 
(\ref{N=2realLinear}) is
\bea
\mathbb G= \ri {\bar \cD}^\a \cD_\a G =  \ri { \cD}^\a \bar \cD_\a G ~,
\label{G-prep}
\eea
where 
the real {\it unconstrained} scalar $G$ is defined modulo  
gauge transformations of the form:
\bea 
\d G = \L + \bar \L~, \qquad  \cJ \L =0~, \quad {\bar \cD}_\a \L =0~.
\label{6.4}
\eea
This gauge freedom allows us to interpret  $G$ as the gauge prepotential for an Abelian massless vector 
multiplet, 
and $\mathbb G$ as the gauge invariant field strength.\footnote{In four 
dimensions, the real linear superfield is naturally interpreted 
as the gauge invariant field strength  
of a massless tensor multiplet \cite{Siegel-tensor}.}
The  prepotential can be chosen to be  inert under the 
super-Weyl transformations,\footnote{The transformation law  
\eqref{N=2sWrealLinear0} is consistent 
with the requirement that the gauge parameter $\L$ in \eqref{6.4} 
be super-Weyl inert.}
\bea
 \d_\s G =0~.
\label{N=2sWrealLinear0}
\eea
Then the field strength $\mathbb G$, defined by eq. \eqref{G-prep}, transforms according 
to \eqref{N=2sWrealLinear}.

Making use of the constraint \eqref{N=2realLinear},
we deduce the important identity
\bea
\cD^{\a \b} {\mathbb G}_{\a\b} = 
8 \Big\{  (\cD^\a \cS ) \bar \cD_\a -  (\bar \cD^\a \cS)  \cD_\a \Big\} {\mathbb G}
+4\ri  \Big\{ (\cD^\a R)  \cD_\a +  (\bar \cD^\a \bar R)  \bar \cD_\a \Big\} {\mathbb G}~,
\label{6.7}
\eea
where we have denoted 
\bea
{\mathbb G}_{\a \b} = (\g^a)_{\a\b} {\mathbb G}_a 
:=  \big[ \cD_{ (\a } , \bar \cD_{\b)} \big] {\mathbb G}
+4\cC_{\a\b} {\mathbb G} ~.
\eea
In the Weyl multiplet gauge \eqref{Wmg}, it follows from \eqref{6.7} that 
\bea
{\mathbb G}^a | 
= 
\cH^a 
-\ve^{abc} \bar  \psi_b \psi_c {\mathbb G}| 
-\ri \ve^{abc} \big( \psi_b \g_c \cD{\mathbb G}| + \bar \psi_b \g_c \bar \cD {\mathbb G}| \big)~,
\label{6.8}
\eea
where $\cH^a$ denotes the Hodge-dual of the field strength of a U(1) gauge field $ a_a$, 
\bea
\cH^a = \hf \ve^{abc}\cH_{bc}~, \qquad 
\cH_{ab} = {\mathfrak D}_a a_b -  {\mathfrak D}_b a_a - \cT_{ab}{}^c a_c~.
\eea
The other independent component fields of $\mathbb G$ may be chosen as follows: 
\bea
\mathbb G |~, \qquad 
\cD_\a {\mathbb G}|~,  \qquad \bar \cD_\a {\mathbb G}|~, \qquad
\ri \cD^\a \bar \cD_\a {\mathbb G}|~.
\eea

\subsection{Poincar\'e supergravity}

The off-shell action for Type II supergravity without a cosmological term  \cite{KLT-M11,KT-M11} is 
\bea
S_{\text{Poincar\'e}}
=4\int {\rm d}^3x \rd^2\q\rd^2\qb
\,E\,
\big(  \mathbb G \ln \mathbb G - 4G \cS \big)~.
\label{TypeII_action}
\eea
The action can be written in a different but equivalent  form:
\bea
S_{\text{Poincar\'e}}
=4 \ri \int {\rm d}^3x \rd^2\q\rd^2\qb
\,E\, G \, \cD^\a \bar \cD_\a \ln \frac{\mathbb G}{\bar \F \F}~,
\eea
where $\F$ is a nowhere vanishing covariantly chiral superfield 
of the type eq. \eqref{chiral51}. One may see that the variables $\F$ and 
$\bar \F$ are purely gauge degrees of freedom. 

The theory \eqref{TypeII_action} was shown in \cite{KT-M11} to be 
 classically equivalent to Type I supergravity without a cosmological term,  
the latter being defined by eq.  \eqref{SUGRA-I} with $\m=0$.
The above action can equivalently be described by the antichiral Lagrangian 
\bea
\bar \cL_{\rm c} = -(\cD^2 - 4 \bar R) 
\big(  \mathbb G \ln \mathbb G - 4G \cS \big)~,
\eea
which has to be used to carry out the component reduction 
of \eqref{TypeII_action} by applying the general rule \eqref{comp-ac-1}. 

Component reduction is often greatly simplified if suitable gauge conditions 
are imposed.
Making use of the Weyl and local $S$-supersymmetry transformations allow us to choose
the gauge conditions
\begin{subequations}  \label{6.11}
\bea
{\mathbb G}\big|&=& 1~,\\
\cD_\a {\mathbb G}\big| &=&0~.
\eea
\end{subequations} 
The compensator also contains a real scalar component field that can be defined as
\bea
Z:= \ri \cD^\a \bar \cD_\a {\mathbb G}|~.
\label{6.15}
\eea
It is also useful to choose a WZ gauge for the 
U(1) gauge symmetry \eqref{6.4}. A standard choice is 
\begin{subequations}  \label{6.12}
\bea
{G}\big|&=& 0~,\\
\cD_\a  G \big| &=&0~, \\
\cD^2 G \big|&=&0~.
\eea
\end{subequations} 
It then follows from \eqref{6.11} and \eqref{6.12} that 
\begin{subequations} 
\bea
\cD^2 \bar \cD_\a {G}\big|
&=& 
0~,
\\
\bar \cD^\a \cD^2  G \big| 
&=& 
\frac{\ri}{2} ( \bar \psi^b \g_a \tilde{\g}_b)^\a a^a
+(\bar \psi^a  \tilde{\g}_a)^\a~, 
\\
 -\frac{1}{4} \bar \cD^2 \cD^2 G| 
 &=& 
 \frac{\ri}{2} \bD_a a^a 
+\frac{1}{4} \bar \psi^b \g_a \psi_b a^a 
+ \frac{\ri}{2} \bar \psi^a \psi_a +\hf Z~.
\label{barD2D2}
\eea
\end{subequations}
The only
independent component fields  of $G$ are 
\begin{subequations}  
\bea
 \big[ \cD_{ (\a } , \bar \cD_{\b)} \big] G| &=& \hf a_{\a\b}~, \label{a-def} \\
( {\bar \cD}^\a \cD_\a )^2 G|&=& -Z~.
\eea
\end{subequations}  
By construction, the  scalar $Z $ is invariant under the gauge 
transformations \eqref{6.4}.

The component supergravity Lagrangian is
\bea
L_{\text{Poincar\'e}} &=& \hf \cR (e,\psi) 
+ \frac{\ri}{4}\ve^{abc} \Big( { \bar {\bm \psi}}_{ab} \psi_c + \bar \psi_a {\bm \psi}_{bc}\Big)
+a_a \cF^a - \frac{1}{4} \tilde{\cH}_a \tilde{\cH}^a -\frac{1}{4}Z^2~,
\label{6.14}
\eea
where we have introduced the combination
\bea
\tilde{\cH}^a:=\cH^a -  \ve^{abc} \bar \psi_b \psi_c~.
\eea
The gravitino field strength is defined as in \eqref{gfs},
with $\bD_a = \bD_a (e,\psi, b)$ the covariant derivative containing 
the U(1)${}_R$ connection $b_a$. 
In this formulation, the supergravity multiplet consists of the following fields: 
the dreibein $e_m{}^a$,  the gravitini $\psi_m{}^\a $ and $\bar \psi_{m \a}$, 
the two gauge fields $a_m$ and $b_m$, 
and the auxiliary  scalar $Z$.

It is not difficult to demonstrate that the vector fields $a_a$ and $b_a$ 
have no propagating degrees of freedom 
for the dynamical system \eqref{6.14}.
To see this, let us work out the equation of motion 
for the U(1)${}_R$ gauge field $b_a$. 
In the supergravity Lagrangian \eqref{6.14},
this field  appears both in the Rarita-Schwinger and  Chern-Simons terms. 
We note that 
\bea
\int \rd^3 x \,e\, a_a \cF^a = \int \rd^3 x \,e\,b_a \cH^a
~,
\label{aFbH}
\eea
modulo a total derivative. 
Another relevant observation is that  the Rarita-Schwinger Lagrangian depends on $b_a$ 
only via the linear term  $-\ve^{abc}b_a \bar \psi_b \psi_c$.  
As a result, the equation of motion for $b_a$ is 
\bea
\tilde{\cH}^a
=0~.
\label{6.16}
\eea
This equation tells us that $a_a$ has no independent degrees of freedom 
on the mass shell.
Now, varying \eqref{6.14} with respect to $a_a$ and making use of \eqref{6.16} gives
\bea
\cF^a=0~,
\eea
and therefore the U(1)${}_R$ connection $b_a$ is flat and may  completely be gauged away. 

The off-shell Lagrangian \eqref{6.14} does not coincide with that proposed in \cite{HIPT} 
to describe $(2,0)$ Poincar\'e supergravity 
(in our terminology, Type II supergravity without a cosmological term), 
see eq. (4.1) in \cite{HIPT}.
In particular, the Lagrangian given in \cite{HIPT} contains no $\cH_a \cH^a$  term. 
The two Lagrangians are actually equivalent 
modulo a  total derivative and a redefinition of the  $b_a$
field.\footnote{GT-M is grateful to Daniel Butter for pointing out the same situation in
the new minimal formulation for 4D $\cN=1$ supergravity (see, e.g., \cite{Muller86,ButterConf1} for the relevant  discussions).}
Indeed, making use of 
\eqref{aFbH} and defining 
\bea
b_a ~\to ~ b'_a = b_a
-\frac{1}{4}\tilde{\cH}_a
~,
\eea
the Lagrangian \eqref{6.14} takes the form
\bea
L_{\text{Poincar\'e}}
=
 \hf \cR (e,\psi) 
+ \frac{\ri}{4}\ve^{abc} \Big( { \bar {\bm \psi}}'_{ab} \psi_c + \bar \psi_a {\bm\psi}'_{bc}\Big)
+ b'_a\cH^a
 -\frac{1}{4}Z^2
~,
\label{6.14bis}
\eea
where the gravitino field strength ${\bm\psi}'_{ab}$ is defined as \eqref{gfs} but with the 
U(1)$_R$ connection $b_a$ replaced by $b'_a$.
The Lagrangian \eqref{6.14bis} is equivalent to the one given  in \cite{HIPT}.

\subsection{(2,0) anti-de Sitter supergravity}

The main difference between Type II supergravity and the new minimal formulation for 
$\cN=1$ supergravity in four dimensions  is that 
the action \eqref{TypeII_action} can be deformed by adding a gauge-invariant cosmological term 
\bea
S_{\rm cosm} =-4 \x\int {\rm d}^3x \rd^2\q\rd^2\qb
\,E\, G \, \mathbb G ~.
\label{(2,0)cosmological}
\eea
To evaluate its component form, we have to make use of the supersymmetric action principle \eqref{comp-ac-1} with 
\bea
\bar \cL_{\rm c} =\x (\cD^2 - 4 \bar R) (G \, \mathbb G )
=\x {\mathbb G} \cD^2 G + 2 \x(\cD^\a {\mathbb G}) \cD_\a G~.
\eea
A short calculation that makes use of \eqref{barD2D2} leads to 
\bea
L_{\rm cosm} 
= 
\x\Big( 
Z
+\frac{1}{4} a_a {\cH}^a  
-\frac{\ri}{2} \ve^{abc}\bar \psi_a \g_b \psi_c 
\Big)~.
\label{6.24}
\eea

The  superfield action for (2,0) AdS supergravity is 
\bea
S_{\text{AdS}}
=4\int {\rm d}^3x \rd^2\q\rd^2\qb
\,E\,
\big(  \mathbb G \ln \mathbb G - 4G \cS - \x G \mathbb G\big)~.
\label{(2,0)-action}
\eea
The component Lagrangian for  off-shell (2,0) AdS supergravity is
\bea
L_{\rm AdS} = \hf \cR (e,\psi) 
&+& 
\frac{\ri}{4} \ve^{abc} \Big( { \bar {\bm \psi}}_{ab} \psi_c + \bar \psi_a {\bm \psi}_{bc}\Big)
+a_a \cF^a 
- \frac{1}{4} \tilde{\cH}_a \tilde{\cH}^a -\frac{1}{4}Z^2 \non \\
&+& \x\Big( Z+  \frac{1}{4} a_a {\cH}^a  -\frac{\ri}{2} \ve^{abc}\bar \psi_a \g_b \psi_c \Big)~.
\label{TypeII-off-shell}
\eea
In this theory, the equation of motion for the U$(1)_R$ gauge field $b_a$ 
is still given by \eqref{6.16}.
As concerns the equation of motion for $a_a$, it becomes 
\bea
\cF^a +\hf \x \cH^a=0~.
\eea
We see that the local U$(1)_R$ gauge freedom can be completely fixed by imposing the condition 
$a_a = -\frac{2}{ \x} b_a$. 
 
 Dynamics described by the off-shell  theory 
 \eqref{TypeII-off-shell} is equivalent to that generated by 
 \bea
\widetilde{L}_{\text{AdS}} &=& \hf \cR (e,\psi) 
+\frac{\ri}{4}\ve^{abc} \Big(
{\bar{\bm \psi}}_{ab} \psi_c 
+ \bar \psi_a {\bm \psi}_{bc} 
-2 \x \bar \psi_a \g_b \psi_c 
\Big)
+\x^2 -\frac{1}{\x}  b_a {\cF}^a  ~.~~~~
\label{TypeII-on-shell}
\eea
One can recognize 
\eqref{TypeII-on-shell}
to be  the standard on-shell Lagrangian for (2,0) AdS supergravity
\cite{AT} (see also  \cite{IT}).
The third term in the parentheses in \eqref{TypeII-on-shell} may be absorbed into the
gravitino field strength by introducing a modified covariant derivative
\bea
\hat{\bf D}_a \psi_b{}^\b = {\bf D}_a \psi_b{}^\b - \hf \x (\g_a){}^\b{}_\g \psi_b{}^\g~.
\eea

\subsection{Supersymmetry transformations}

The gauge conditions \eqref{6.11} 
completely fix the Weyl and local $S$-supersymmetry freedom.
To preserve the condition ${\mathbb G}|=1$, no residual Weyl invariance remains, 
 ${\bm \s}=0$. However, each $Q$-supersymmetry transformation 
 has to be accompanied by a compensating $S$-supersymmetry transformation 
in order to preserve the condition $\cD_\a {\mathbb G}\big| =0$. 
Indeed, the field $\cD_\a{\mathbb{G}}|$ transforms as 
\bea
(\d_Q +\d_S)  \cD_\a{\mathbb{G}}|
=
\e^\b\cD_\b\cD_\a{\mathbb{G}}|
+\bar{\e}_\b\cDB^\b\cD_\a{\mathbb{G}}|
+\eta_\a{\mathbb{G}}|
=
\hf \tilde{\cH}_a(\g^a\bar{\e})_{\a}
-\frac{\ri}{2}Z\bar{\e}_\a 
+\eta_\a
~,~~~~~~
\eea
where here we have used the identities
 \eqref{6.7}--\eqref{6.8}, \eqref{6.11} and \eqref{6.15}.
We have to require  $(\d_Q +\d_S)  \cD_\a{\mathbb{G}}| =0$, and therefore
\bea
\eta_\a(\e)
&=&
-\hf \tilde{\cH}_a(\g^a\bar{\e})_{\a}
+\frac{\ri}{2}Z\bar{\e}_\a ~.
\label{eta31}
\eea

Choosing ${\bm \s}={\bm \t}=0$ and $\eta_\a = \eta_\a(\e)$  
in \eqref{susy-transformations-Wg}, 
 we obtain the supersymmetry transformations of 
 the  gauge fields $e_m{}^a$, $\psi_m{}^\g$ and $b_m$:
\bsubeq
{\allowdisplaybreaks
\bea
\d_\e e_m{}^a
&=&
\ri\big(
\e\g^a\bar{\psi}_m
+\bar{\e}\g^a\psi_m
\big)
~,
\\
\d_\e\psi_m{}^\a&=&
2\bD_m\e^\a
+\frac{\ri}{2}\tilde{\cH}_m\e^\a
-\frac{\ri}{2} e_m{}^a\ve_{abc}\tilde{\cH}^c(\tilde{\g}^b\e)^{\a}
+\frac{1}{2}Z(\tilde{\g}_m\e)^{\a}
~,
\label{grav-var-II_0}
\\
\d_\e b_m&=&
-\frac{1}{4} e_m{}^a\Big\{
\ve_{abc} \e{\bm{\bar{\psi}}}^{bc}
+ 2\e\g^b  {\bm{\bar{\psi}}}_{ab}{}
- \ri\e\g^b{\bar{\psi}}_a\tilde{\cH}_b
-\e{\bar{\psi}}_a Z
\Big\}
+{\rm c.c.}
~~~~~~
\eea}\esubeq

The supergravity multiplet also includes the fields $a_m$ and $Z$.
The supersymmetry transformation of $a_m$ follows from its definition 
$a_m = e_m{}^a a_a $, with $a_a$ originating as a component field 
of $G$, eq. \eqref{a-def}. 
Note that, in order to preserve  the WZ gauge \eqref{6.12},
in computing the supersymmetry transformations of $a_m$ 
it is necessary to include a
compensating $\e$-dependent U(1) gauge transformation \eqref{6.4} with parameter $\L(\e)$ 
such that\bsubeq
\bea
\L(\e)|&=&0
~,
\\
\cD_\a\L(\e)|
&=&
-\frac{1}{4}(\bar{\e}\g^b)_{\a} a_b
+\frac{\ri}{2}\bar{\e}_\a
~,
\\
\cD^2\L(\e)|
&=&
\frac{\ri}{2} \bar{\e}\g^a\tilde{\g}^b \bar{\psi}_a a_b
-\bar{\e}\g^a\bar{\psi}_a
~.
\eea
\esubeq
We then obtain 
\bea
\d_\e a_m
&=&
(\d_\e e_m{}^a)a_a
-e_m{}^a(\g_a)^{\a\b}\big{(}
\e^\g\cD_\g{[}\cD_{\a},\cDB_{\b}{]}G|
+2\ri\cD_{\a\b}\L(\e)|
+{\rm c.c.}
\big)
\non\\
&=&
\ri\e\g^a\bar{\psi}_ma_a
+(\g_m)^{\a\b}\e^\g\cD_\g\{\cD_{\a},\cDB_{\b}\}G|
+4\ri \psi_m{}^\a\cD_\a\L(\e)|
+{\rm c.c.}
\eea
Evaluating this variation gives 
\bea
\d_\e a_m
&=&
-2\big(\e{\bar{\psi}}_{m} +\bar{\e}\psi_{m}\big)
~.
\eea
The scalar field $Z$ originates as a component field of $\mathbb G$, 
eq. \eqref{6.15}, and therefore its supersymmetry transformation is 
\bea
\d_\e Z
&=&
\frac{\ri}{2}\e_\a\cD^2\cDB^\a{\mathbb G}|
+\frac{\ri}{2}\bar{\e}_\a\cDB^2\cD^\a{\mathbb G}|
+\ri(\cD^\a\cDB_\a\s){\mathbb G}|~.
\eea
Making use of \eqref{sWcond-112}, we then derive
\bea
\d_\e Z
&=&
-\frac{\ri}{2}\e\g^a\bar{\psi}_a Z
-\hf\ve^{abc}\e\g_a\bar{\psi}_b\tilde{\cH}_c
+\hf\e{\bar{\psi}}_a\tilde{\cH}^a
+\frac{\ri}{2}\ve^{abc}\e\g_a{\bm{\bar{\psi}}}_{bc}
+{\rm c.c.}
~~~~~~~~~~
\eea

For completeness, let us also work out the supersymmetry transformation 
of the field strength $\tilde{\cH}_a$. Making use of the definition of 
$\tilde{\cH}_a$ gives
\bea
\d_\e \tilde{\cH}^a
&=&
-\frac{1}{2}(\g^a)^{\a\b}\Big{\{}
\e^\g\cD_\g{[}\cD_{\a},\cDB_{\b}{]}\mbG|
+\bar{\e}_\g\cDB^\g{[}\cD_{\a},\cDB_{\b}{]}\mbG|
+({[}\cD_{\a},\cDB_{\b}{]}\s)\mbG|
\Big{\}}
~.
\eea
With the aid of 
 \eqref{sWcond-111} we obtain
\bea
\d_\e
\tilde{\cH}^a
&=&
-\frac{\ri}{2}\ve^{abc}\e{\bar{\psi}}_b\tilde{\cH}_c
+\ri\e\g^{[a}{\bar{\psi}}_b\tilde{\cH}^{b]}
+\hf\ve^{abc}\e\g_b\bar{\psi}_c Z
-\ve^{abc}\e{\bm{\bar{\psi}}}_{bc}
+{\rm c.c.}
~~~~~~~
\eea

\subsection{Matter-coupled supergravity}

The action for a locally supersymmetric $\sigma$-model 
coupled to Type II supergravity is 
\begin{align}
S_{\rm matter} =\int {\rm d}^3x \rd^2\q\rd^2\qb
\,E\,\mbG K(\vf,\bar\vf)~.
\label{s-m-II}
\end{align}
Here the K\"ahler potential $ K(\vf,\bar\vf)$ and the matter 
superfields are the same as in section 5. 
In particular,  the covariantly chiral superfields $\vf^I$
are super-Weyl and U$(1)_R$ neutral, 
$\d_\s \vf^I = \cJ \vf^I=0$.
The action is invariant under the K\"ahler transformations \eqref{Kahler1} due to the identity
\bea
 \int {\rm d}^3x \rd^2\q\rd^2\qb \,E\,\mbG \, \L(\vf)=0~.
 \eea
 
In order to carry out the component reduction of $S_{\rm matter}$,
we associate with \eqref{s-m-II} 
the  antichiral Lagrangian
\bea
\bar{\cL}_c
&=&
-\frac{1}{4}(\cD^2-4\bar{R})(\mbG K )= -\frac{1}{4} {\mathbb G} \cD^2 K 
-\hf (\cD^\a {\mathbb G}) \cD_\a K
~.
\eea
The component fields of $\vf^I$ are defined as in 
\eqref{vf-comfields}.
Unlike the Type I supergravity case, now  we do not have 
to modify the gauge conditions on the compensator 
in the presence of matter.
Direct calculations lead to the following component Lagrangian:
\bea
L_{\rm matter} 
&=&
g_{I\bJ}\Big{[}\,
F^I\bar{F}^\bJ
-({\mathfrak D}_aX^I) {\mathfrak D}^a\Xb^\bJ
-\frac{\ri}{4}  \l^I \g^a \stackrel{\longleftrightarrow}{\widetilde{\bD}_a} \bar \l^{\bar J} 
-\hf{\bar{\psi}}^a\lb^{\bJ} {\mathfrak D}_aX^I 
+\hf\psi_a\l^I {\mathfrak D}^a\Xb^\bJ
\non\\
&&~~~~~~
-\hf\ve^{abc} \big(\psi_a\g_b\l^I {\mathfrak D}_c \Xb^\bJ-\bar{\psi}_a\g_b \lb^\bJ {\mathfrak D}_c X^I \big)
+\frac{1}{8}\psi_a\g^b{\bar{\psi}}^a \, \l^I\g_b\lb^\bJ 
-\frac{1}{8}\psi_a{\bar{\psi}}^a \,\l^{I}\lb^\bJ 
\non\\
&&
~~~~~~
+\frac{1}{8}\ve^{abc} ( \psi_a{\bar{\psi}}_b\, \l^I\g_c\lb^\bJ +\psi_a\g_b{\bar{\psi}}_c\, \l^{I}\lb^\bJ )
+\frac{1}{8}   \l^I\g_a\lb^\bJ \tilde{\cH}^a
-\frac{\ri}{8}Z \l^{I}\lb^\bJ
\Big{]}
\non \\
&&+ \frac{\ri}{4} {\cH}^a\Big{[}  K_\bI {\mathfrak D}_a\Xb^\bI
- K_I  {\mathfrak D}_aX^I \Big{]}
+\frac{1}{16}R_{I\bK J\bL} \l^{I}\l^J\,\lb^\bK \lb^{\bL}
~. \label{6.33}
\eea
Here we have introduced the K\"ahler-covariant derivative 
\bea
\widetilde{\bD}_a \l^I &:=& {\bD}_a \l^I +  \l^J \G^I_{JK}  {\mathfrak D}_a X^K
= {\mathfrak D}_a \l^I + \ri b_a \l^I + \l^J \G^I_{JK}  {\mathfrak D}_a X^K
~.
\eea
The $\s$-model action generated by the Lagrangian \eqref{6.33} proves to be invariant under the K\"ahler 
transformations. The first term in the fourth line of \eqref{6.33} is the only one which varies under 
the K\"ahler transformations. The corresponding contribution to the action is indeed K\"ahler invariant  
 due to the identity $\int \rd^3 x\,e\, {\cH}^a {\mathfrak D}_a \L =0$. 

As may be seen from  \eqref{6.33}, the gauge fields $b_a$ and $a_a$ couple to conserved currents
of completely different types. The U$(1)_R$ gauge field couples to the U$(1)_R$ Noether current 
\bea
\cJ^a_{\rm Noether} = \ve^{abc}{\bar{\psi}}_b{}\psi_{c }+ \hf g_{I\bar J} \l^{ I} \g^a\lb^\bJ ~.
\eea
As regards the gauge field $a_a$, it couples to the topological current
\bea
\cJ^a_{\rm top} = \hf \ve^{abc} (
{\mathfrak D}_b {\mathfrak R}_c -  {\mathfrak D}_c {\mathfrak R}_b - \cT_{bc}{}^d \,{\mathfrak R}_d) ~,
\qquad 
{\mathfrak R}_a := \ri ( 
K_\bI {\mathfrak D}_a\Xb^\bI
- K_I  {\mathfrak D}_aX^I )~,
\eea
 which is identically conserved. These properties were pointed out in \cite{HIPT}.
 
Now, we consider a  complete supergravity-matter 
system described by the action \cite{KT-M11}
\bea
S= 4  \int {\rm d}^3x \rd^2\q\rd^2\qb \,E\,
\Big( \mathbb G \Big\{  \ln \mathbb G + \frac{1}{4} K(\vf, \bar \vf)  \Big\}
 - 4G \cS  \Big)~.
 \label{TypeII-matter36}
\eea
It describes Poincar\'e supergravity coupled to the locally supersymmetric $\s$-model. 
As shown in \cite{KT-M11}, this theory is dual to the Type I supergravity-matter system \eqref{5.10}. 
To compute the corresponding component Lagrangian, we
combine $L_{\rm matter}$ given by \eqref{6.33}  with the Type II supergravity Lagrangian
without cosmological term, eq. \eqref{6.14}. The result is 
\bea
L 
&=&
 \hf \cR (e,\psi)
  + \frac{\ri}{4} \ve^{abc} \Big( { \bar {\bm \psi}}_{ab} \psi_c + \bar \psi_a {\bm \psi}_{bc}\Big) 
+a_a \cF^a - \frac{1}{4} \tilde{\cH}_a \tilde{\cH}^a -\frac{1}{4}{\mathbb Z}^2 \non \\
&&
+g_{I\bJ}\Big{[}~
F^I\bar{F}^\bJ 
-({\mathfrak D}_aX^I) {\mathfrak D}^a\Xb^\bJ
-\frac{\ri}{4}  \l^I \g^a \stackrel{\longleftrightarrow}{\widetilde{\bD}_a} \bar \l^{\bar J} 
-\hf{\bar{\psi}}^a\lb^{\bJ} {\mathfrak D}_aX^I 
+\hf  \psi_a\l^I {\mathfrak D}^a\Xb^\bJ
\non \\
&&~~~~~~~
-\hf\ve^{abc} \big(\psi_a\g_b\l^I {\mathfrak D}_c \Xb^\bJ-\bar{\psi}_a\g_b \lb^\bJ {\mathfrak D}_c X^I \big)
+\frac{1}{8}\psi_a\g^b{\bar{\psi}}^a \, \l^I\g_b\lb^\bJ 
-\frac{1}{8}\psi_a{\bar{\psi}}^a \,\l^{I}\lb^\bJ 
\non\\
&&~~~~~~~
+\frac{1}{8}\ve^{abc} ( \psi_a{\bar{\psi}}_b\, \l^I\g_c\lb^\bJ +\psi_a\g_b{\bar{\psi}}_c\, \l^{I}\lb^\bJ )
+\frac{1}{8}   g_{I\bJ}\l^I\g_a\lb^\bJ \tilde{\cH}^a 
\Big{]} \non \\
&&
+ \frac{\ri}{4} {\cH}^a\Big{[}  K_\bI {\mathfrak D}_a\Xb^\bI- K_I  {\mathfrak D}_aX^I \Big{]}
+\frac{1}{16}R_{I\bK J\bL}\,\l^{I}\l^J\,\lb^\bK \lb^{\bL}
-\frac{1}{64}\big(g_{I\bJ}\l^{I}\lb^\bJ\big)^2~,
\label{6.35}
\eea
where we have defined 
\bea 
\mathbb{Z}
&:=& Z+\frac{\ri}{4}g_{I\bJ}\l^{I}\lb^{\bJ}~.
\eea

Let us show that the dynamical system \eqref{6.35} is equivalent to 
the Type I supergravity-matter system \eqref{5.23} with $W=0$. 
Integrating out  $ Z$ gives 
\bea
{\mathbb Z}=0~.
\label{6.37}
\eea
The equation of motion for the gauge field $b_a$ is 
\bea
{\mathbb{H}}^a
&:=& 
\tilde{\cH}^a 
- \hf g_{I\bar J} \l^{ I} \g^a\lb^\bJ 
=
\cH^a
-\ve^{abc}{\bar{\psi}}_b{}\psi_{c }
- \hf g_{I\bar J} \l^{ I} \g^a\lb^\bJ =0
~.
\label{6.38}
\eea
Let us consider the equation of motion for the gauge field $a_a$. 
It can be represented in the form 
\bea
{\mathfrak D}_a {\mathbb B}_b -  {\mathfrak D}_b {\mathbb B}_a - \cT_{ab}{}^c \,{\mathbb B}_c =0~,
\eea
 where  ${\mathbb B}_a$ is defined in \eqref{auxVect}. 
 This equation tells us that the local U$(1)_R$ gauge freedom can be completely fixed by 
 choosing the condition 
 \bea
 b_a = \frac{1}{8} g_{I\bar J} \l^I \g_a \bar \l^{\bar J}
+\frac{\ri}{4}( K_I {\mathfrak D}_a X^I  - K_{\bar I} {\mathfrak D}_a \bar X^{\bar I})~.
\label{6.40}
\eea
Making use of the equations \eqref{6.37}, \eqref{6.38} and \eqref{6.40} reduces 
the supergravity-matter system \eqref{6.35} 
to that described by the Lagrangian  \eqref{5.23} with $W=0$. 

To preserve the gauge condition \eqref{6.40},  any K\"ahler transformation generated by a parameter $\L$ has to 
be accompanied 
by a special U$(1)_R$-transformation with parameter $ \t = \frac{\ri}{4} ( \bar \L - \L)$, see also 
eq. \eqref{Kahler-B}.

Finally, we generalize the supergravity-matter system \eqref{TypeII-matter36}
to include a cosmological term. The manifestly supersymmetric action is
\bea
S= 4  \int {\rm d}^3x \rd^2\q\rd^2\qb \,E\,
\Big( \mathbb G \Big\{  \ln \mathbb G + \frac{1}{4} K(\vf, \bar \vf)  \Big\}
 - 4G \cS- \x G \, \mathbb G  \Big)~. 
\eea
The corresponding component Lagrangian is obtained 
from the supergravity-matter Lagrangian \eqref{6.35}
by adding the cosmological term \eqref{6.24}. The result is
\bea
L 
&=& \hf \cR (e,\psi) 
+\frac{\ri}{4} \ve^{abc} \Big( { \bar {\bm \psi}}_{ab} \psi_c + \bar \psi_a {\bm \psi}_{bc}\Big) 
+a_a \cF^a 
- \frac{1}{4} \tilde{\cH}_a \tilde{\cH}^a -\frac{1}{4}({\mathbb Z} -2\x)^2 \non \\
&& 
+   \frac{1}{4} \x a_a {\cH}^a  
-\frac{\ri}{2} \x \ve^{abc}\bar \psi_a \g_b \psi_c   
+\x^2
 \non \\
&&
+g_{I\bJ}\Big{[}\,
F^I\bar{F}^\bJ 
-({\mathfrak D}_aX^I) {\mathfrak D}^a\Xb^\bJ
-\frac{\ri}{4}  \l^I \g^a \stackrel{\longleftrightarrow}{\widetilde{\bD}_a} \bar \l^{\bar J} 
-\hf{\bar{\psi}}^a\lb^{\bJ} {\mathfrak D}_aX^I 
+\hf\psi_a\l^I {\mathfrak D}^a\Xb^\bJ
\non\\
&&~~~~~~~
-\hf\ve^{abc} \big(\psi_a\g_b\l^I {\mathfrak D}_c \Xb^\bJ-\bar{\psi}_a\g_b \lb^\bJ {\mathfrak D}_c X^I \big)
+\frac{1}{8}\psi_a\g^b{\bar{\psi}}^a \, \l^I\g_b\lb^\bJ 
-\frac{1}{8}\psi_a{\bar{\psi}}^a \,\l^{I}\lb^\bJ 
\non\\
&&~~~~~~~
+\frac{1}{8}\ve^{abc} ( \psi_a{\bar{\psi}}_b\, \l^I\g_c\lb^\bJ +\psi_a\g_b{\bar{\psi}}_c\, \l^{I}\lb^\bJ )
+\frac{1}{8}   g_{I\bJ}\l^I\g_a\lb^\bJ \tilde{\cH}^a
- \frac{\ri}{4} \x \l^{I}\lb^\bJ
\Big{]} 
\non \\
&&
+ \frac{\ri}{4} {\cH}^a\Big{[}  K_\bI {\mathfrak D}_a\Xb^\bI- K_I  {\mathfrak D}_aX^I \Big{]}
+\frac{1}{16}R_{I\bK J\bL}\l^{I}\l^J\,\lb^\bK \lb^{\bL}
-\frac{1}{64}\big(g_{I\bJ}\l^{I}\lb^\bJ\big)^2 
~.
\eea

We conclude this section by giving 
the supersymmetry transformations of the component field of 
$\vf^I$:
\bsubeq
\bea
\d_\e X^I&=&\e\l^I
~,
\\
\d_\e \l_\a^I&=&
2\e_\a\Big(
 F^I
 +\frac{1}{4}\G^I_{JK} \l^{J} \l^K
 \Big)
+2\ri(\g^a\bar{\e})_{\a}\Big(
\bD_aX^I
-\hf\psi_a\l^I
\Big)
~,~~~~~~
\\
\d_\e F^I&=&
-\e\l^{ J}\G^I_{JK}  F^K
+\hf\l^I\eta(\e)
+\ri\bar{\e}\g^a\bD_a\l^I
-\frac{1}{4} g^{I\bar{L}}R_{J\bar{L} K\bar{P}}\,\bar{\e}\bar{\l}^{\bar{P}}\,\l^{J}\l^K
 \non\\
&&
+\ri\bar{\e}\g^a\l^{J}\G^I_{JK} \bD_aX^K
-\frac{\ri}{2}\bar{\e}\g^a\psi_a F^I
-\bar{\e}\g^a\tilde{\g}^b\bar{\psi}_a\Big(\bD_bX^I-\hf\psi_b\l^I\Big)
~.
\eea
\esubeq
It is a useful exercise for the reader to derive these 
transformation laws.

\subsection{$R$-invariant sigma models}

Type II minimal supergravity admits more general matter couplings \cite{KT-M11} 
than those we have so far studied. In particular, it
can be coupled to $R$-invariant $\s$-models, 
similarly to  the new minimal $\cN=1$ supergravity in four dimensions 
(see, e.g.,  \cite{Ferrara:1983dh} for more details).
Here we briefly discuss such theories.  

We consider a system of covariantly chiral scalars $\f^I$ 
of super-Weyl weights $r_I$, 
\bea
\bar \cD_\a \f^I=0~, \qquad 
\cJ \f^I= -r_I \f^I ~, \qquad
\d_\s \phi^I=r_I \s\phi^I~.
\eea
We introduce a supergravity-matter system of the form:
\bea
S&=& 4  \int {\rm d}^3x \rd^2\q\rd^2\qb \,E\,
\Big( 
\mathbb G \Big\{  \ln \mathbb G + \frac{1}{4} 
{\bm K} 
\big(\phi^I/{\mathbb G}^{r_I}, \bar {\phi}^{\bar J}/{\mathbb G}^{r_J}\big)\Big\}
 - 4G \cS  \Big) \non \\
&& \qquad +  \Big{\{} \int {\rm d}^3x {\rm d}^2 \q \,\cE\,
{\bm W}(\phi^I)
\, + \,{\rm c.c.}~\Big{\} } ~.
\label{6.57}
\eea
This action is  super-Weyl invariant if  the superpotential ${\bm W} (\f^I)$ obeys the 
homogeneity equation
\bea
 \sum_I r_I \f^I \bm W_I=2\bm W~.
\eea
The action is invariant under the local U$(1)_R$ transformations if
the K\"ahler potential $\bm K(\f^I , \bar \f^{\bar J} )$ obeys the equation
\bea
 \sum_I r_I \f^I \bm K_I=\sum_{\bar{I}} r_I \bar \f^{\bar{I}} \bm K_{\bar{I}}~.
\eea
In a flat superspace limit, the theory \eqref{6.57} reduces to a general 
$R$-invariant nonlinear $\s$-model. 

The action  \eqref{6.57}  may be reduced to components using the
formalism developed above. In general, however, 
the Weyl and $S$-supersymmetry 
gauge conditions \eqref{6.11} have to be replaced with 
matter-dependent ones
(similar to the gauge conditions \eqref{F-Weyl} in Type I supergravity)
if we want the gravitational action to be given in Einstein frame. 
We will not give such an analysis here. 


\section{Topologically massive supergravity}

Consider $\cN=2$ conformal supergravity coupled to matter supermultiplets. 
The supergravity-matter action is
\bea
S= \frac{1}{g} S_{\rm CSG} + S_{\rm matter}~,
\label{7.1}
\eea
where $S_{\rm CSG}$ denotes the conformal supergravity action 
\cite{RvN,BKNT-M2} and $S_{\rm matter}$ the matter action 
\cite{KLT-M11,KT-M11}. Both terms in \eqref{7.1} must be super-Weyl invariant. 
As regards  $S_{\rm CSG}$, the formulation given in \cite{RvN} is purely component, 
and the concept of super-Weyl transformations is not defined within this approach. 
However, the  super-Weyl invariance of $S_{\rm CSG}$ is manifest in the 
superspace formulation  given recently in \cite{BKNT-M2}, see 
Appendix \ref{AppendixC} for a review.
Requiring the
super-Weyl invariance of $ S_{\rm matter}$ 
is equivalent to the fact  
that this action will describe an $\cN=2$ superconformal field theory 
in a flat superspace limit.  

The equation of motion for conformal supergravity is 
\bea
-\frac{4}{g} \cW_{\a\b} + \cJ_{\a\b} = 0~,
\label{7.2}
\eea
where $\cW_{\a\b}$ is the $\cN=2$ super Cotton tensor, eq.  \eqref{Cotton}, 
and $\cJ_{\a\b}$ is the matter supercurrent. 
This equation 
is obtained by varying $S$ 
with respect to the real vector  prepotential 
$H^{\a\b} =  H^{\b\a}$ of conformal supergravity \cite{Kuzenko12},  
\bea
\cW_{\a\b} \propto \frac{{\bm \d}  }{ {\bm \d} H^{\a\b}}S_{\rm CSG}
~, \qquad \cJ_{\a\b} \propto \frac{{\bm \d}  }{ {\bm \d} H^{\a\b}}S_{\rm matter}~,
\eea
with $ {{\bm  \d} }/{{ \bm\d}H^{\a\b} }$ a covariantized variational derivative
with respect to $H^{\a\b}$. Eq. \eqref{7.2} and the matter equations 
of motion determine the dynamics of the supergravity-matter system.

\subsection{Properties of the supercurrent}

The fundamental properties of the super Cotton tensor are: 
(i) its super-Weyl transformation law  \eqref{Cotton-Weyl};
and (ii) the transversality condition \cite{BKNT-M1}
\bea
\cD^\b \cW_{\a\b} = \bar \cD^\b \cW_{\a\b} =0~.
\eea
The matter supercurrent must have analogous properties.  
Specifically, it  is characterized by the super-Weyl transformation law
\bea
\cJ'_{\a\b} = \re^{2\s}\, \cJ_{\a\b}~
\label{SC-SW}
\eea 
and obeys the conservation equation
\bea
\cD^\b \cJ_{\a\b} = \bar \cD^\b \cJ_{\a\b} =0 ~.
\label{SC-CE}
\eea
These must hold when the matter fields are subject to their equations of motion. 
Of course, the relations \eqref{SC-SW} and \eqref{SC-CE} may be viewed as the consistency 
conditions for the equation of motion \eqref{7.2}.  However, there is an independent way 
to justify  \eqref{SC-SW} and \eqref{SC-CE} that follows from the definition of $\cJ_{\a\b}$
as the  covariantized variational derivative
with respect to $H^{\a\b}$. Here we only sketch the proof. For a more complete
derivation, it is necessary to develop a background-quantum formalism 
for 3D $\cN=2$ supergravity similar to that given by Grisaru and Siegel 
for  $\cN=1$ supergravity in four dimensions \cite{GS1,GS2} (see \cite{BK} for a pedagogical 
review).

As demonstrated in \cite{Kuzenko12}, in complete analogy with the 4D case
\cite{SiegelWZ}, 
the gravitational superfield originates 
via $ \exp (-2\ri H )$, where
\bea
H = \bar H = H^m\pa_m +H^\m D_\m +  \bar H_\m \bar D^\m
\eea
and $D_\m$ and $\bar D^\m$ are the  spinor covariant derivatives of Minkowski superspace. 
By construction, the superfields $H^M=(H^m, H^\m , \bar H_\m )$ are super-Weyl invariant. 
The supergravity gauge group can be used to gauge away $H^\m$ and its conjugate, 
leaving us with the only unconstrained prepotential $H^m$. This prepotential 
possesses a highly nonlinear gauge transformation 
\bea
\d_L H_{\a\b} = \bar D_{(\a} L_{\b)} - D_{(\a} \bar L_{\b)} +O(H)~, 
\eea
where the gauge parameter $L_\a$ is an unconstrained complex spinor. 
Due to the nonlinear nature of this transformation, the gravitational superfield is not a 
tensor object, and special care is required in order to represent the
variation of the action induced by a variation $H^m \to H^m +\d H^m$ in a covariant 
way. This is what the background-quantum splitting in supergravity  \cite{GS1,GS2} is about.

It turns out that giving the gravitational superfield a finite displacement is equivalent to 
a deformation of the covariant derivatives that can be represented,  in a chiral representation, 
as follows: 
\begin{subequations} 
\bea
\bar \cD^\a &\to& \bar \cF \bar \cD^\a +\dots~, \\
\cD_\a &\to & \re^{-2\ri \bH} \Big( \cN_\a{}^\b \cF \cD_\b +\dots\Big)   \re^{2\ri \bH}~, 
\qquad \det \,(\cN_\a{}^\b ) = 1~, 
\eea 
\end{subequations}
where 
\bea
\bH = -\hf \bH^{\a\b} \cD_{\a\b} - \frac{\ri}{6}(\cD_\b \bH^{\a\b}) \bar \cD_\a 
- \frac{\ri}{6}( \bar\cD_\b \bH^{\a\b})  \cD_\a +\dots
\eea
The ellipses in these expressions denote all terms with Lorentz and U$(1)_R$ generators. 
The deformed covariant derivatives must obey the same constraints as the original ones
$\cD_A$.  This can be shown to imply that the complex scalar $\cF$ and the 
unimodular $2\times 2$ matrix $\cN$ are determined in terms of $\bH^a$.   
The vector superfield $\bH^a$ describes the finite deformation of the gravitational superfield.
A crucial property of the first-order operator $\bH$ is that it is super-Weyl invariant when acting 
on any  super-Weyl inert real scalar $U = \bar U$, 
\bea
\d_\s \bH \cdot U = 0~,
\eea
provided $\bH_{\a\b}$ transforms as 
\bea
\d_\s \bH_{\a\b} = -\s \bH_{\a\b}~.
\label{cH-SW}
\eea
The superfield $\bH_{\a\b}$ proves to be defined modulo gauge transformations of the form
\bea
\d_L \bH_{\a\b} = \bar \cD_{(\a} L_{\b)} - \cD_{(\a} \bar L_{\b)} +O(\bH)~,
\eea
which 
are compatible with the super-Weyl transformation \eqref{cH-SW} provided
the gauge parameter is endowed with the properties 
\bea
\cJ L_\a = L_\a~, \qquad \d_\s L_\a=- \frac{3}{2} \s L_\a~.
\eea

Giving the gravitational superfield an infinitesimal displacement, 
$\bH_a= \d \bH_a$, the matter action changes as
\bea
\d S_{\rm matter} = \int {\rm d}^3x {\rm d}^2 \q  \rd^2 \bar \q\,E\,
\d \bH^a \cJ_a \equiv  \int {\rm d}^3x {\rm d}^2 \q  \rd^2 \bar \q\,E\,
\d \bH^a  \frac{{\bm \d}  }{ {\bm \d} H^{a}}S_{\rm matter}~.
\label{vari}
\eea
This functional must be super-Weyl invariant. 
Due to eqs. \eqref{SW-Ber} and \eqref{cH-SW}, 
and since the matter equations of motion hold,
we conclude that 
the super-Weyl transformation of the supercurrent is given by eq. 
\eqref{SC-SW}. Since $S_{\rm matter}$ is invariant under the supergravity
gauge transformations, choosing $ \d \bH_{\a\b} = \bar \cD_{(\a} L_{\b)} - \cD_{(\a} \bar L_{\b)}$
in \eqref{vari} should give $\d S_{\rm matter} =0$ if the matter equations of motion hold. 
Since $L_\a$ is completely arbitrary, this is possible if and only if the conservation 
equation \eqref{SC-CE} holds. 

\subsection{Topologically massive minimal supergravity: Type I}\label{section7.2}

Let us choose  $ S_{\rm matter}$  to be the superconformal sigma model 
 \eqref{suco}. The corresponding supercurrent proves to be 
 \bea
 \cJ_{\a\b} = N_{i \bar j} \cD_{(\a}\f^i \bar \cD_{\b)} \bar \f^{\bar j}
 - \frac{1}{4} \big[\cD_{(\a} , \bar \cD_{\b)} \big] N -\cC_{\a\b} N ~.
\label{7.7}
\eea
The matter equations of motion are
\bea
-\frac{1}{4} (\bar \cD^2 -4 \bar R ) N_i + P_i =0~.
\label{7.8}
\eea
The relative coefficients in \eqref{7.7} are uniquely fixed if one 
demands the transversality condition \eqref{SC-CE}
 to hold on the mass shell, 
eq. \eqref{7.8}. Alternatively, it may be shown that 
the relative coefficients in \eqref{7.7} are uniquely fixed 
if one requires the super-Weyl transformation law \eqref{SC-SW}.
In the flat superspace limit, the supercurrent 
\eqref{7.7} reduces to the one given in \cite{KT-M11}.

We now turn to considering 
topologically massive Type I supergravity. 
It is described by the action 
\bea
S_{\rm TMSG} =\frac{1}{g} S_{\rm CSG} - S_{\rm SG}~, 
\label{TMSG-I}
\eea
where $S_{\rm SG}$ is the action for Type I supergravity 
with a cosmological term, eq. \eqref{SUGRA-I}.
In topologically massive gravity \cite{DJT}
and its supersymmetric extensions
\cite{DK,Deser}, the Einstein term 
appears with the `wrong' sign.
In the context of the $\s$-model action  \eqref{suco},
the matter sector in \eqref{TMSG-I} corresponds to the choice
 $N =4 \bar \F \F $ and $P = -\m \, \F^4$. 
The equation of motion for $\F$ is 
\bea
\frac{1}{4}(\cDB^2-4R)\bar \F + \m \F^3= 0~.
\label{7.10}
\eea
The equation of motion for 
the gravitational superfield 
\eqref{7.2} becomes
\bea
-\frac{4}{g} \Big\{ \frac{\ri}{2} \big[\cD^\g ,\bar \cD_\g \big]
\cC_{\a\b} &-& \big[ \cD_{(\a} , \bar \cD_{\b)} \big] \cS 
- 4 \cS \cC_{\a\b} \Big\} \non \\
&+&4 \cD_{(\a}\F\bar \cD_{\b)} \bar \F
 - \big[\cD_{(\a} , \bar \cD_{\b)} \big] (\bar \F \F)  -4 \cC_{\a\b} \bar \F \F=0 ~.
 \label{7.11}
\eea
As shown in \cite{KLT-M11}, 
the freedom to perform the super-Weyl and local U$(1)_R$ transformations 
can be used to impose the gauge\footnote{Upon gauge-fixing $\F$ 
to become constant, 
there still remain rigid scale and U$(1)_R$ transformations that allow us to make 
$f$ in \eqref{7.12} have any given value. The choice $f=1$ leads to a canonically normalized 
Einstein-Hilbert term at the component level.}
\bea
\F=\sqrt{f} = {\rm const}~,
\label{7.12}
\eea
which implies the conditions \eqref{TypeIgf}. 
Then, the matter equation of motion \eqref{7.10} turns into 
\bea
R = \m = {\rm const}~. 
\eea
Using the identity $\cD^\b \cC_{\a\b} = -\hf \bar \cD_\a \bar R -2\ri \cD_\a \cS$,
which follows from \eqref{2.12c}, we also obtain 
\bea
\cD^\b \cC_{\a\b} = \bar \cD^\b \cC_{\a\b} = 0~.
\label{7.15-0}
\eea 
Now, the conformal supergravity equation \eqref{7.11} drastically simplifies 
\bea
\frac{\ri}{2}{[}\cD^\g,\bar \cD_\g{]}\cC_{\a\b} 
+ g f \,\cC_{\a\b} =0~.
\label{7.15}
\eea
The equations \eqref{7.15-0} and \eqref{7.15}
have a solution $\cC_{a} =0$, which corresponds to (i) a flat superspace 
for $\m =0$, or (ii)  (1,1) anti-de Sitter superspace if $\m \neq 0$. 
In the case $\m=0$, we can linearize  the equation \eqref{7.15}
around Minkowski superspace. Its obvious implication is 
$(\Box -m^2) \cC_a = 0$, where $m= \hf f g$.

Combining the Lagrangians \eqref{SUGRA-I-com} and \eqref{C.12}, 
we obtain the component Lagrangian for topologically massive 
Type I  supergravity
\bea
L_{\rm TMSG}&=&
 \frac{1}{4g} 
 \ve^{abc} \Big{[}{\cR}_{bc}{}_{fg}  \omega_{a}{}^{fg} 
+ \frac{2}{3}{\omega}_{af}{}^g {\omega}_{bg}{}^h {\omega}_{ch}{}^f 
- 4 {{\cF}}_{ab} {b}_c 
+  \ri{\bm{\bar{\psi}}}_{bc}\g_d\tilde{\g}_a\ve^{def} {\bm\psi}_{ef}
\Big{]} \non \\
&&\quad 
- \hf \cR (e,\psi)
 -\frac{\ri}{4} \ve^{abc} \Big( { \bar \psi}_{ab} \psi_c + \bar \psi_a {\psi}_{bc}\Big)
+\frac{1}{4} \bar { M} { M} -{ b}^a { b}_a \ \non \\
&&\qquad
+ \bar \m \Big(\bar { M}
-\hf\ve^{abc}\psi_a\g_b \psi_c \Big)
+ \m \Big({{ M} }
+\hf\ve^{abc}{\bar{\psi}}_a\g_b{\bar{\psi}}_c \Big) ~.
\eea
The Lagrangian is computed in the Weyl, local U(1)$_R$ and $S$-supersymmetry gauge \eqref{gauge-I}.
However, it is possible to avoid the use of \eqref{gauge-I}. To achieve this the component form of 
$S_{\rm SG}$ has to be computed using the results of Appendix B.

\subsection{Topologically massive minimal supergravity: Type II}

Topologically massive Type II supergravity
 is described by the action 
\bea
S_{\rm TMSG} = \frac{1}{g}S_{\rm CSG} - S_{\rm AdS}~, 
\label{TMSG-II}
\eea
where $S_{\rm AdS}$ is the action for (2,0) AdS supergravity, eq.  \eqref{(2,0)-action}.
We can think of the theory with action \eqref{(2,0)-action}
as a model for the vector multiplet coupled to background supergravity. 
Then, the  equation of motion for $G$ is 
\bea
\ri \cD^\a \bar \cD_\a \ln \mathbb G - 4 \cS  -  2\xi  \mathbb G =0~.
\label{7.17}
\eea
The supercurrent corresponding to the action $S_{\rm matter}=- S_{\rm AdS}$ is 
\bea
\cJ_{\a\b} = \frac{4}{\mathbb G} \cD_{(\a}\mathbb G \bar \cD_{\b)} \mathbb G
 - \big[\cD_{(\a} , \bar \cD_{\b)} \big] \mathbb G  -4\cC_{\a\b} \mathbb G ~.
 \label{7.28}
\eea
It is an instructive exercise to show that $\cJ_{\a\b} $
 possesses the super-Weyl transformation law
 \eqref{SC-SW} and obeys the conservation equation \eqref{SC-CE}
 provided \eqref{7.17} holds. 
 In the flat superspace limit, the supercurrent \eqref{7.28} reduces to the one given in \cite{KT-M11}.

Instead of \eqref{7.11}, 
now the equation of motion for the gravitational superfield is
\bea
-\frac{4}{g} \Big\{ \frac{\ri}{2} \big[\cD^\g ,\bar \cD_\g \big]
\cC_{\a\b} &-& \big[ \cD_{(\a} , \bar \cD_{\b)} \big] \cS 
- 4 \cS \cC_{\a\b} \Big\} \non \\
&+&
\frac{4}{\mathbb G} \cD_{(\a}\mathbb G \bar \cD_{\b)} \mathbb G
 - \big[\cD_{(\a} , \bar \cD_{\b)} \big] \mathbb G  - 4\cC_{\a\b} \mathbb G =0~.
 \label{7.19}
\eea
As shown in \cite{KLT-M11}, 
the freedom to perform the super-Weyl transformations 
can be used to impose the gauge
\bea
\mathbb G =f= {\rm const}~,
\eea
which implies the constraint \eqref{TypeIIgf}.
Then the  equation of motion \eqref{7.17} tells us that 
\bea
\cS = -\frac{\xi}{2} = {\rm const}~.
\eea
These properties lead to the constraint \eqref{7.15-0}. 
As a result, the conformal supergravity equation \eqref{7.19} turns into 
\bea
 \frac{\ri}{2} \big[\cD^\g ,\bar \cD_\g \big]\cC_{\a\b} 
+ (gf+2\x)\cC_{\a\b} =0~.
 \label{7.33}
\eea
The equations \eqref{7.15-0} and \eqref{7.33} 
have a solution $\cC_{a} =0$, which corresponds either to  a flat superspace 
for $\x =0$ or  (2,0) anti-de Sitter superspace if $\x \neq 0$.

Combining the Lagrangians \eqref{TypeII-off-shell} and \eqref{C.12}, 
we obtain the component Lagrangian for topologically massive 
Type II  supergravity
\bea
L_{\rm TMSG}&=&
 \frac{1}{4g} 
 \ve^{abc} \Big{[}{\cR}_{bc}{}_{fg}  \omega_{a}{}^{fg} 
+ \frac{2}{3}{\omega}_{af}{}^g {\omega}_{bg}{}^h {\omega}_{ch}{}^f 
- 4 {{\cF}}_{ab} {b}_c 
+\ri{\bm{\bar{\psi}}}_{bc}\g_d\tilde{\g}_a\ve^{def} {\bm\psi}_{ef}
\Big{]} \non \\
&&\quad
- \hf \cR (e,\psi) 
-\frac{\ri}{4} \ve^{abc} \Big( { \bar {\bm \psi}}_{ab} \psi_c + \bar \psi_a {\bm \psi}_{bc}\Big)
-a_a \cF^a + \frac{1}{4} \tilde{\cH}_a \tilde{\cH}^a +\frac{1}{4}Z^2 \non \\
&&\qquad
- \x\Big( Z+  \frac{1}{4} a_a {\cH}^a  -\frac{\ri}{2} \ve^{abc}\bar \psi_a \g_b \psi_c \Big)~.
\eea
The Lagrangian is computed in the Weyl and local $S$-supersymmetry gauge \eqref{6.11}.
However, one can avoid the use of \eqref{6.11}. To achieve this the component form of 
$S_{\rm AdS}$ has to be computed using the results of Appendix C.

\subsection{Topologically massive non-minimal supergravity}

Topologically massive non-minimal  supergravity
 is described by the action 
\bea
S_{\rm TMSG} = 
\frac{1}{g}S_{\rm CSG} - S_{\rm AdS}~, 
\eea
where $S_{\rm AdS}$ denotes  the action for non-minimal 
(1,1) AdS supergravity \cite{KT-M11}
\bea
S_{\text{AdS}} = -2 \int {\rm d}^3x {\rm d}^2 \q \rd^2 \bar \q 
\,E\,
{  { (\bar \G \, \G)}  }^{-1/2}~.
\label{7.36}
\eea
The dynamical variable $\G$ is a deformed complex linear 
scalar $\G$ obeying the constraint \eqref{1.2}.
If we think of \eqref{7.36} as the action describing the dynamics 
of matter superfields $\G$ and $\bar \G$ in a background curved superspace, 
then this theory is dual to the Type I minimal model  \eqref{SUGRA-I},
see \cite{KT-M11} for more details.
As a result, topologically massive non-minimal  supergravity
is dual to that constructed in subsection \ref{section7.2}. To relate the two theories,
it suffices to note that when $\G$ and $\bar \G$ are subject to their 
equations of motion, we can represent 
\bea
\G = \F^{-3} \bar \F~,
\eea
where $\F$ is a chiral scalar of super-Weyl weight 1/2 
under   the equation 
of motion \eqref{7.10}.


\section{Symmetries of curved superspace}
\label{sym-curv-sup}
\setcounter{equation}{0}

In this section we derive the conditions for a curved superspace to possess 
(conformal) isometries. After that we concentrate on a discussion of curved backgrounds
admitting conformal and rigid supersymmetries.

\subsection{Conformal isometries}

Consider some background superspace $\cM^{3|4}$ such that its geometry 
is of the type 
described in
section \ref{geometry}. 
In order to formulate rigid superconformal or rigid supersymmetric field theories
on  $\cM^{3|4}$, it is necessary to determine all (conformal) isometries of this superspace.  
This can be done similarly to the case of 4D $\cN=1$ supergravity
described in detail in \cite{BK} and reviewed in \cite{Kuzenko13}.
In this subsection we study the infinitesimal conformal isometries  of $\cM^{3|4}$.

Let $\x = \x^A E_A$ be a  real supervector field on
 $\cM^{3|4}$, $\x^A  \equiv (\x^a , \x^\a , \bar \x_\a) $.
It  is called  conformal Killing 
if one can associate with $\x$ a supergravity gauge 
transformation\footnote{Strictly speaking, the parameters of gauge transformations 
are usually restricted to have compact support  in spacetime, see e.g. 
\cite{DeWitt}. The (conformal) Killing vector and spinor fields do not have 
this property. }
\eqref{tau}
and an infinitesimal  super-Weyl transformation
\eqref{2.3}
such that their combined action does not change the covariant derivatives,
\bea
(\d_\cK + \d_\s) \cD_A =0~.
\label{8.1}
\eea
Since the vector covariant derivative $\cD_a$ is given in terms of an anti-commutator of two spinor 
ones, it suffices to analyze the implications of \eqref{8.1} for the case $A =\a$. 
A short calculation gives
\bea
(\d_\cK + \d_\s) \cD_\a&=&
\Big\{ \hf  ( \s +2\ri \t) \ve_{\a\b}
 + \cD_{{\a}} \x_{\b} 
 +\ri\x_{(\a}{}^{\g} \cC_{\b)\g}
-\x_{\a\b} \cS
-\hf K_{\a\b}
\Big\} \cD^\b
  \non\\
 &&
 -\Big\{
 \cD_{{\a}} \bar{\x}_{\b}
 +\ri\x_{\a\b} \bar{R}
 \Big\}
 \cDB^{\b}
+\Big\{
\hf  \cD_{{\a}}\x_{\b\g} -2\ri \ve_{\a(\b} \bar{\x}_{\g)}
\Big\}
\cD^{\b\g}
 \non\\
 &&
-\Big{[}
\ve_{\a(\b}(\cD_{\g)}\s) 
+\hf  \cD_{{\a}} K_{\b\g}
-4\ve_{\a(\b}\x_{\g)}\bar{R}
-4\ri\ve_{\a(\b}\bar{\x}_{\g)}\cS
-2\bar{\x}_{\a}\cC_{\b\g}
\non\\
&&~~~
+\x_{\a}{}^{\d} {\bm C}_{\b\g\d}
-\frac{2}{3}\ve_{\a(\b}\x_{\g)\d} \Big(
2\cD^{\d}\cS
+\ri\cDB^{\d}\bar{R}
\Big)
+\frac{1}{6}\Big(
2\cD_{\a}\cS
+\ri\cDB_{\a}\bar{R}
\Big)\x_{\b\g}
 \Big{]}
 \cM^{\b\g}
 \non\\
 &&
-\Big{[} \cD_\a (\s + \ri \t)
-2\bar{\x}^{\b}\cC_{\a\b}
-4\ri\bar{\x}_{\a}\cS
\non\\
&&~~~
+\hf\x^{\b\g} {\bm C}_{\a\b\g}
-\frac{1}{6}\x_{\a\g} \Big(
8\cD^{\g}\cS
+\ri\cDB^{\g}\bar{R}
\Big)
 \Big{]}\cJ~. 
\label{8.2}
\eea
The right-hand side of \eqref{8.2} is a linear combination of the five 
 linearly independent
operators $\cD^\b$, $\bar \cD^\b$, $\cD^{\b\g}$, $\cM^{\b \g}$ and $\cJ$. 
Therefore, demanding $(\d_\cK + \d_\s) \cD_\a=0$ gives five different equations. 
Let us first consider the equations associated with the operators $\cD^\b$ and $\cD^{\b\g}$
in the right-hand side of \eqref{8.2},
\bsubeq \label{8.3}
\bea
\cD_{{\a}} \x_{\b}
&=&
- \hf\ve_{\a\b}\big(\s
+2\ri \t\big)
 -\ri\x_{(\a}{}^{\g} \cC_{\b)\g}
+\x_{\a\b} \cS
+\hf K_{\a\b}
~, \label{8.3a}
\\
\cD_\a\x_{\b\g}&=&4\ri\ve_{\a(\b}\bar{\x}_{\g)}
~, \label{8.3b}
\eea
\esubeq
as well as their complex conjugate equations.
These relations imply, in particular,  that the  parameters $\x^\a,\,\bar{\x}_\a,\, K_{\a\b},\,\s$ and $\t$ 
are uniquely expressed in terms of  
$\x^a$ and its covariant derivatives as follows:
\bsubeq  \label{master-def}
\bea
\x^{\a}&=& - \frac{\ri}{6}\cDB_\b\x^{\b \a}
~,~~~~~~
\bar{\x}_{\a}=
-\frac{\ri}{6}\cD^\b\x_{\b\a}
 ~,
 \label{master-def-x}
\\
\s&=&
\hf\big(
\cD_{\a} \x^{\a}
+\cDB^{\a} \bar{\x}_{\a}
\big) 
~,~~~
 \label{master-def-s}
 \\
\t  
&=&
-\frac{\ri}{4}\big(
\cD_{\a} \x^{\a}
-\cDB^{\a} \bar{\x}_{\a}
\big) 
~,
 \label{master-def-t}
\\
K_{\a\b}
&=&
 \cD_{(\a} \x_{\b)}
 -\cDB_{(\a} \bar{\x}_{\b)}
-2\x_{\a\b} \cS
~.
 \label{master-def-K}
\eea
\esubeq
This is why we may also use the notation $\cK = \cK [ \x ]$ and $\s = \s[ \x ]$.
In accordance with \eqref{8.3b}, the remaining vector parameter $\x^a$ satisfies the 
equation\footnote{The equation \eqref{CKSV-master} is analogous 
to the conformal Killing equation, ${\mathfrak D}_{(\a \b} V_{\g \d )} =0$,
on a (pseudo) Riemannian three-dimensional manifold.} 
\bea
\cD_{(\a}\x_{\b\g)} &=&0 
\label{CKSV-master}
\eea
and its complex conjugate.
Immediate corollaries  of \eqref{CKSV-master} are
\bsubeq
\bea
&
(\cD^2+4\bar{R})\x_{a}
=
(\cDB^2+4R)\x_{a}
=
0
~,
\\
&
\cD_{a}\x_{b}=
\eta_{ab}\s
-\ve_{abc}K^{c}
~.
\eea
\esubeq
The latter relation  implies the conformal Killing equation
\bea
\cD_{a}\x_{b} + \cD_{b}\x_{a} = \frac{2}{3} \eta_{ab} \cD^{c}\x_{c}~.
\eea

If the equation \eqref{CKSV-master} holds and the conditions
\eqref{master-def-x}--\eqref{master-def-K} are adopted, 
it can be shown that the conditions \eqref{8.1} are satisfied identically.
Therefore, \eqref{CKSV-master} is the fundamental equation containing all the information 
about the conformal Killing supervector fields. 
As a consequence,  we can give an alternative definition of 
the conformal Killing supervector field. It is a real supervector field
\bea
\x = \x^A E_A ~, \qquad \x^A  \equiv (\x^a , \x^\a , \bar \x_\a) =
 \Big( \x^a , - \frac{\ri}{6}\cDB_\b\x^{\b \a} , 
-\frac{\ri}{6}\cD^\b\x_{\b\a} \Big) 
\eea
which obeys the master equation \eqref{CKSV-master}.

If $\x_1$ and $\x_2$ are two conformal Killing supervector fields, their 
Lie bracket $[\x_1, \x_2]$ is a conformal Killing supervector field. 
It is obvious that, for any real $c$-numbers $r_1$ and $r_2$, the linear combination
 $r_1 \x_1 + r_2 \x_2$ is a  conformal Killing supervector field. 
Thus the set of all conformal Killing supervector
fields is a super Lie algebra. The conformal Killing supervector fields generate the symmetries 
of a superconformal field theory on $\cM^{3|4}$. 


Making use of \eqref{8.2}, the condition  $(\d_\cK + \d_\s) \cD_\a=0$ 
leads to several additional relations which can be represented in the form:
\bsubeq \label{8.9}
\bea
\cD_{{\a}}\bar{\x}_{\b}
&=&
-\ri\x_{\a\b} \bar{R}
~,
\\
\cD_\a K_{\b\g}
&=&
4\cC_{(\a\b}\bar{\x}_{\g)}
-2{\bm C}_{\d(\a\b}\x_{\g)}{}^{\d}
-\frac{1}{3}\big(\ri \cDB_{(\a}\bar{R}+2\cD_{(\a}\cS \big)\x_{\b\g)}
\non\\
&&
+\ve_{\a(\b}\Big{[}
-2\cD_{\g)}\s
+8\bar{R}\x_{\g)}
+8\ri\cS\bar{\x}_{\g)}
+\frac{8}{3}\cC_{\g)\d}\bar{\x}^{\d}
-\frac{4}{3}{\bm C}_{\g)\d\r}\x^{\d\r}
\non\\
&&~~~~~~~~~
+\frac{10}{9}
\x_{\g)\d}\Big(
\ri\cDB^\d \bar{R}
+2\cD^\d\cS
\Big)
\Big{]}
~,
\\
\cD_\a\t
&=&
\ri\cD_\a\s
+4\cS\bar{\x}_\a
-2\ri\cC_{\a\d}\bar{\x}^{\d}
+\frac{\ri}{2} {\bm C}_{\a\d\r}\x^{\d\r}
+\frac{1}{6}
\Big(
\cDB^\b \bar{R}
-8\ri\cD^\b\cS
\Big)\x_{\a\b}
~.
\eea
\esubeq
Actually these relations  have nontrivial implications.
Eqs. \eqref{8.3} and \eqref{8.9} tell us that 
the spinor covariant derivatives of the parameters
$\U:= ( \x^B, K^{\b\g},  \t)$  can be represented as
linear combinations of $\U$, $\s$,  $\cD_\a\s$ and $\bar \cD_\a \s$.
It turns out that the vector covariant derivative of $ \U$ can be represented  
as a linear combination of $\U$, $\s$ and   $\cD_A\s$.
In order to prove this assertion, the key observation is that, because of \eqref{8.1}, 
the torsion tensor $T_{AB}{}^C$, the Lorentz and U$(1)_R$ curvature tensors $R_{AB}{}^{cd}$ and $R_{AB}$, all defined by eq. \eqref{algebra}, as well as their covariant derivatives
are invariant under the transformation $\d =  \d_\cK + \d_\s $
generated by the conformal Killing supervector field. In particular, 
the dimension-1 torsion tensors $\cS$, $R$ and $\cC_a$ 
are invariant, and therefore 
\begin{subequations} \label{DDs}
\bea
-\frac{\ri}{4} \cD^\b \bar \cD_\b \s
&=& (\x^B \cD_B + \s ) \cS  ~,
\label{DDs-1} \\
-\frac{1}{4} \bar \cD^2 \s &=& ( \x^B \cD_B + \s )R   -2\ri \t    R
~,  
\label{DDs-2}\\ 
-\frac{1}{8}(\g_a)^{\b\g}  [\cD_{\b},\cDB_{\g}]\s &=&(\x^B \cD_B  + \s )  \cC_a  +K_a{}^b  \cC_b   ~.
\label{DDs-3}
\eea
\end{subequations}
To complete the proof, it only remains to make use of eq. \eqref{2.7b}.

It is an instructive exercise to derive the following identity
\bea
\cD_\a\cD_{\b\g}\s
&=&
\frac{2}{3}\ve_{\a(\b}\Big\{
2\ri\cC_{\g)\d}\cD^{\d}\s
+4\cS\cD_{\g)}\s
+3\ri \bar{R}\cDB_{\g)}\s
-\frac{\ri}{4}
\cDB_{\g)}(\cD^2\s)
-\frac{\ri}{2}\cD_{\g)}(\cD^\d\cDB_\d\s)
\Big\}
\non\\
&&
-\frac{\ri}{2}\cD_{(\a}\big([\cD_{\b},\cDB_{\g)}]\s\big)
\eea
and its complex conjugate.
In conjunction with eqs. \eqref{DDs}, they tell us that $\cD_A \cD_B \s$ can be represented as  
a linear combination of $\U$, $\s$ and $\cD_C\s$. We have already established that 
$\cD_A \U$ is a linear combination of $\U$, $\s$ and   $\cD_C\s$.
These properties mean that the super Lie algebra of the conformal Killing vector fields 
on $\cM^{3|4}$ is finite dimensional. 
The number of its even and odd generators cannot exceed those in 
the $\cN=2$ superconformal algebra  ${\mathfrak{osp}}(2|4)$.

To study supersymmetry transformations at the component level, 
it is useful to spell out one of the implications of \eqref{8.1}
with $A= a$. Specifically, we consider the equation $(\d_\cK + \d_\s) \cD_{a} =0$ 
and read off its part proportional to a linear combination of the spinor covariant derivatives $\cD_\b$. 
The result is
\bea
0&=&
\cD_{a}\x_{\a}
+\frac{\ri}{2}(\g_a)_{\a}{}^{\b} \cDB_{\b}\s
-\ri\ve_{abc}(\g^b)_\a{}^{\b}\cC^c\x_{\b}
-(\g_a)_\a{}^{\b}( \x_{\b}\cS
+\bar{\x}_{\b}R)
\non\\
&&
-\hf\ve_{abc}\x^{b}(\g^c)^{\b\g}\Big(
\ri\bar{{\bm C}}_{\a\b\g}
-\frac{4\ri}{3}\ve_{\a(\b}  \cDB_{\g)}\cS 
-\frac{2}{3}\ve_{\a(\b} \cD_{\g)} R
\Big)
~.
\label{8.11}
\eea

\subsection{Conformally related superspaces}

Consider a curved superspace
${\hat{\cM}}^{3|4}$ that is conformally related to $\cM^{3|4}$. 
This means that the covariant derivatives
 $\cD_A$ and ${\hat{\cD}}_A$, which correspond to $\cM^{3|4}$ and $\hat{\cM}^{3|4}$ respectively, 
 are related to each other in accordance with \eqref{2.3},  
\bsubeq
\bea
 {\hat{\cD}}_\a&=&\re^{\hf\o}\Big(\cD_\a+(\cD^{\g}\o)\cM_{\g\a}-(\cD_{\a }\o)\cJ\Big)
 ~,
 \\
{\hat{\cD}}_{a}
&=&\re^{\o}\Big(
\cD_{a}
-\frac{\ri}{2}(\g_a)^{\g\d}(\cD_{(\g}\o)\cDB_{\d)}
-\frac{\ri}{2}(\g_a)^{\g\d}(\cDB_{(\g}\o)\cD_{\d)}
+\frac{\ri}{2}(\cD_{\g}\o)(\cDB^{\g}\o)\cM_{a}
\non\\
&&~~~~
+\ve_{abc}(\cD^b\o)\cM^c
-\frac{\ri}{8}(\g_a)^{\g\d}({[}\cD_{\g},\cDB_{\d}{]}\o)\cJ
-\frac{3\ri}{4}(\g_a)^{\g\d}(\cD_{\g}\o)(\cDB_{\d}\o)\cJ
\Big)
~,~~~~~
\eea
\esubeq
for some super-Weyl parameter $\o$.
 The two superspaces 
$\cM^{3|4}$ and $\hat{\cM}^{3|4}$ prove to 
have the same conformal Killing supervector fields. 
Given such a 
supervector field $\x$, it can be represented in two different forms
\bea
\x= \x^A   E_A = {\hat{\x}}^A {\hat{E}}_A~,
\eea
where ${\hat{ E}}_A$ is the inverse vielbein associated 
with the covariant derivatives ${\hat{\cD}}_A$. 
The parameters $\x^A$ and $\hat{\x}^A$ are related to each others as 
follows:
\bea
{\hat\x}^a&=&\re^{-\o}\x^a
~,\qquad
{\hat\x}^\a 
=\re^{-\hf\o}\Big(\x^\a
+\frac{\ri}{2}\x^{\b\a}{\cDB}_{\b}\o
\Big)
~.
\eea
One may prove that the following identities hold
\bsubeq
\bea
\s[{\hat{\x}}] &=& \s[\x] - \x\, \o
~,
\\
\t[\hat{\x}]&=&
\t[\x]
-\ri\x^\a\cD_{\a}\o
+\ri\bar{\x}_\a\cDB^{\a}\o
+\frac{1}{8}\x^{\a\b}{[}\cD_\a,\cDB_{\b}{]}\o
-\frac{1}{4}\x^{\a\b}(\cD_{\a}\o)\cDB_{\b}\o
~,
\\
 K_{\a\b}[{\hat{\x}}]
&=&
K_{\a\b}[\x]
-2\x_{(\a}\cD_{\b)}\o
+2\bar{\x}_{(\a}\cDB_{\b)}\o \non \\
&& \qquad +\ve^{abc}(\g_c)_{\a\b}\x_a\cD_b\o
+\frac{\ri}{2}\x_{\a\b}(\cD^{\g}\o){\cDB}_{\g}\o
~.
\eea
\esubeq
These identities imply the following important relation:
\bea
\cK[\hat{\x}] := \hat{\x}^A\hat{\cD}_A
+\hf K^{cd}[\hat{\x}]\cM_{cd}
+\ri \t[\hat{\x}]\cJ
=\cK[\x]
~.
\eea


\subsection{Isometries}

In order to describe $\cN=2$ Poincar\'e or anti-de Sitter supergravity theories, 
the Weyl multiplet has to be coupled to a certain conformal compensator $\X$ and its conjugate. 
In general, the latter is a scalar superfield of super-Weyl weight $w\neq 0$ and U($1)_R$ charge $q$, 
\bea
\d_\s\X=w\s\X~,\qquad 
\cJ\X=q\X
~,
\label{GenCom}
\eea
chosen to be nowhere vanishing, $\X \neq 0$. It  is assumed that $q=0$ if and only if $\X$
is real, which is the case for Type II supergravity. 
Different off-shell supergravity theories correspond to
different superfield types of $\X$.

Once $\X$ and its conjugate have been fixed, 
the off-shell supergravity multiplet is completely described in terms of the following data:
(i) the U(1) superspace geometry described earlier; (ii) the conformal compensator 
and its conjugate. 
Given a supergravity background, its isometries should preserve both of these inputs. 
This leads us to the concept of Killing supervector fields.

A conformal Killing supervector field $\x = \x^A E_A$  on $\cM^{3|4}$  is said to be Killing 
if the following conditions hold:
\begin{subequations} \label{8.13both}
\bea
\Big[\x^B \cD_B + \hf K^{bc}[\x] \cM_{bc} + \ri \t [\x]\cJ, \cD_A \Big]  + \d_{\s [\x]} \cD_A &=&0~ ,
\label{8.13a} \\ 
\Big(\x^B \cD_B  + \ri q \t  [\x]  + w \s [\x] \Big) \X& =&0~, \label{8.13b}
\eea
\end{subequations}
with the parameters $ K^{bc}[\x]$, $\t [\x]$ and $\s[\x]$ defined as in \eqref{master-def}.
The set of all Killing supervector fields on $\cM^{3|4}$ is a super Lie algebra. 
The Killing supervector fields generate the symmetries 
of rigid  supersymmetric  field theories on $\cM^{3|4}$. 

The Killing equations \eqref{8.13both} are super-Weyl invariant. 
Specifically, if $(\cD_A , \X)$ and  $(\hat{\cD}_A , \hat{\X})$ are 
conformally related supergravity backgrounds, 
\bea
 {\hat{\cD}}_\a&=&\re^{\hf\o}\Big(\cD_\a+(\cD^{\g}\o)\cM_{\g\a}-(\cD_{\a }\o)\cJ\Big)
 ~, \qquad \hat{\X} = \re^{w \s} \X~,
 \eea
then the equations  \eqref{8.13both} imply that 
$\x= \x^B   E_B = {\hat{\x}}^B {\hat{E}}_B$ is also a Killing 
supervector field with respect to $(\hat{\cD}_A , \hat{\X})$. 
In particular, it holds that 
\bea
\Big(\hat{\x}^B \hat{\cD}_B  + \ri q \t  [\hat{\x}]  + w \s [\hat{\x}] \Big) \hat{\X}& =&0~.
\eea

The super-Weyl and local U(1)$_R$ symmetries allow us
to choose a useful gauge
\bea
\X=1
\label{Xgauge1}
\eea
which characterizes the off-shell supergravity formulation chosen.
If $q \neq 0$, there remains no residual super-Weyl and local U(1)$_R$ freedom
in this gauge. 
Otherwise, the local U(1)$_R$ symmetry remains unbroken while the super-Weyl freedom 
is completely fixed.

In the gauge \eqref{Xgauge1}, the Killing equation \eqref{8.13b} becomes
\bea
\ri q\Big(\x^B\F_B
+\t [\x] \Big)+w\s[\x] 
=0
~.
\eea
Hence, the isometry transformations are generated by those conformal Killing supervector
fields which respect the conditions
\bsubeq
\bea
\s [\x]&=&0
~,
\label{constr-sigma}
\\
\t [\x]&=&
- \x^B \F_B
~,   
\qquad q\ne0~.
\label{constr-tau}
\eea
\esubeq
These properties provide the main rationale for choosing the gauge condition  \eqref{Xgauge1}
which is:
for any off-shell supergravity formulation, the isometry transformations are characterized by 
the condition $\s [\x]=0$. 

Since  for $q\ne0$ the local U(1)$_R$ symmetry is completely fixed  in the gauge \eqref{Xgauge1},
it is reasonable to switch to new covariant derivatives without U$(1)_R$ connection which are defined by
$ \cD_A \to \de_A := \cD_A-\ri\F_A\cJ $.\footnote{This is similar to the 4D procedure of de-gauging
introduced by Howe \cite{Howe} and reviewed in \cite{GGRS}.}
 The original U$(1)_R$ connection $\F_A$ 
turns into a tensor superfield.

\subsection{Charged conformal Killing spinors}

We wish to look for those curved superspace backgrounds which admit 
at least one conformal supersymmetry. 
By definition, such a superspace possesses 
a conformal Killing supervector field $\x^A$  with the property
\bea
\x^a|=0 ~, \qquad \e^\a := \x^\a | \neq 0~.
\eea
All other bosonic parameters will also be assumed to vanish, 
$\s | =  \t | = K_{\a\b}|=0$.
Our analysis will be restricted to U(1) superspace  backgrounds 
without covariant fermionic fields,  that is
\bea
\cD_\a\cS|= 0~, \qquad \cD_\a R|= 0~, \qquad \cD_\a \cC_{\b\g}|=0~.
\label{8.233} 
\eea
These conditions mean that the gravitini can completely be gauged away
such that the projection \eqref{stcd} becomes
\bea
\cD_a|=\bD_a \quad \Longleftrightarrow \quad \psi_m{}^\a =0~.
\eea
In what follows, we always assume that the gravitini have been gauged away. 

The above definitions provide a superspace realization for what is usually 
called  a ``supersymmetric spacetime.'' 
For instance, according to \cite{HIPT}, it is a supergravity background 
``for which all fermions {\it and their supersymmetry variations} vanish 
for some non-zero supersymmetry parameter.''

We introduce scalar and vector fields associated with the superfield torsion:
\bea
{s} := \cS|~, \qquad {{r}} := R|~, 
\qquad {c}_a := \cC_a|~.
\eea
We also recall that  the $S$-supersymmetry
parameter is $\eta_\a := \cD_\a \s|$.
Bar-projecting the equation \eqref{8.11} gives
\bea
0&=&
\bD_{a}\e^{\a}
+\frac{\ri}{2}(\tilde{\g}_a\bar{\eta})^\a
+\ri\ve_{abc}\,{c}^b (\tilde{\g}^c\e)^\a
-{s} (\tilde{\g}_a\e)^\a
-\ri{{r}}(\tilde{\g}_a\bar{\e})^{\a}
~.
\label{Killing-Spinor}
\eea
This is equivalent to the following two equations:
\bsubeq \label{conf-kill-spinor-full}
\bea
0&=&
\big(\bD_{(\a\b}
-\ri{c}_{(\a\b} \big)
\e_{\g)}
~,
\label{conf-kill-spinor}
\\
\bar{\eta}_\a&=&
-\frac{2\ri}{3}
\Big(
(\g^a\bD_a\e)_\a
+2\ri(\g^a\e)_\a c_a
+3{s} \e_{\a}
+3\ri {{r}}\bar{\e}_{\a}
\Big)
~.
\eea
\esubeq

Equation \eqref{conf-kill-spinor} tells us that the supersymmetry parameter is 
a {\it charged conformal Killing spinor}, since   \eqref{conf-kill-spinor} can be rewritten in the form
\bea
\widetilde{\bD}_{(\a\b} \e_{\g)} =0~, \qquad 
\widetilde{\bD}_{\a\b} \e_{\g} := {\mathfrak D}_{\a\b} \e_\g 
-\ri (  b_{\a\b} + {c}_{\a\b} )\e_\g~.
\label{8.23}
\eea
Let us choose $\e_\a$ to be  a bosonic (commuting) spinor. 
Then it follows from \eqref{8.23} that the real vector field $V_{a} :=  (\g_a)^{\a\b}\bar \e_{\a} \e_{\b} $ has 
the following properties: (i) $V_a$ is a conformal Killing vector field, 
${\mathfrak D}_{(\a\b} V_{\g\d)} =0$; and (ii) $V_a$ is null or time-like, since 
$V^a V_a = (\bar \e^\a \e_\a)^2 \leq 0$. This vector field is null if and only if 
$\bar \e_\a \propto \e_\a$. As a result, we have reproduced two of the main results of \cite{HTZ}.

By construction, the conditions \eqref{8.233} are  supersymmetric, 
that is  
\bea
( \d_\cK + \d_\s )\cD_\a\cS=0~, \qquad
(  \d_\cK + \d_\s )\cD_\a R= 0~, \qquad
( \d_\cK + \d_\s ) \cD_\a \cC_{\b\g}=0~.
\eea
Evaluating the bar-projection of these variations gives, respectively,
\bsubeq \label{8.300}
\bea
\cD^2\cDB_{\a}\s|
&=&
\bar{\e}^{\b}\Big(
8{\mathfrak D}_{\a\b}{s}
-4\ri{[}\cD_{(\a}, \cDB_{\b)}{]}\cS|
-4\ri\ve_{\a\b}\cD^{\g}\cDB_{\g}\cS|
\Big)
+16\ri\e_\a \bar{{{r}}}{s}
\non\\
&&
-8\ri\eta_\a {s}
+6\bar{\eta}_\a\bar{{r}}
~,
\\
\cD^2\cDB_\a\s|
&=&
\e^{\b}\Big(
8\ri ( {\mathfrak D}_{\a\b} +2\ri b_{\a\b}  -2\ri {c}_{\a\b} )\bar{{r}}
-32\ri\ve_{\a\b}{s} \bar{{r}}
\Big)
+2\bar{\e}_{\a} \cDB^2 \bar{R}|
\non\\
&&
-4\ri\bD_{\a\b}\eta^\b
+4\ri \eta_\a {s}
-6\bar{\eta}_\a  \bar{{r}}
~,
\\
0&=&
\e_{(\a}\big( {\mathfrak D}_{\b\g)} +2\ri b_{\b\g)}  +4\ri {c}_{\b\g)} \big)\bar{{r}}
\non\\
&&
+\bar{\e}^{\d}\Big{\{}{\mathfrak D}_{(\a\b}{c}_{\g\d)}
-\hf\big(
\cD_{(\a} \bar{{\bm C}}_{\b\g\d)}
+\cDB_{(\a}{\bm C}_{\b\g\d)}
\big)|
\non\\
&&~~~~~~
+\ve_{\d(\a}\Big{[}
\,
[\cD_{\b},\cDB_{\g)}]\cS|
-\ri{\mathfrak D}_{\b\g)} {s}
+\frac{3}{2}\ve^{cab}(\g_c)_{\b\g)}{\mathfrak D}_a {c}_b
+6{c}_{\b\g)} {s}
\Big{]}
\Big{\}}
\non\\
&&
-\frac{1}{2} \bD_{(\a\b}\eta_{\g)}
-\frac{3\ri}{2}c_{(\a\b}\eta_{\g)}
~.
\eea
\esubeq

\subsection{Supersymmetric backgrounds}

In order to describe a {\it rigid supersymmetry transformation},
the structure equations given in the previous subsection 
have to be supplemented by the additional condition 
\bea
\s[\x] =0 \quad \Longrightarrow \quad  \eta_\a =0
~,
\label{SzeroEta}
\eea
in accordance with \eqref{constr-sigma}.
Here we do not specify any particular compensator. However, we assume that some compensator has been 
chosen and the gauge condition \eqref{Xgauge1} has been imposed.

Because of \eqref{SzeroEta}, the equation \eqref{Killing-Spinor} turns into 
\bea
\bD_{a} \e^\a &=&
-\ri\ve_{abc}  {c}^b (\tilde{\g}^c\e)^\a
+ {s}(\tilde{\g}_a\e)^{\a}
+\ri r  (\tilde{\g}_a\bar{\e})^{\a} ~.
\label{Killing-Spinor32} 
\eea
In the spinor notation, this equation reads
\bea
\bD_{\a\b} \e_\g &=& \ri \,{c}_{(\a\b} \e_{\g)}
-2 \ri \ve_{\g(\a} \r_{\b)}~, \qquad
\r_\a:= \frac{2}{3}{c}_a(\g^a\e)_{\a}
-\ri {s} \e_{\a}
+ {{r}}\bar{\e}_{\a} ~.
\label{KS32}
\eea
This relation shows that, in a neighborhood of any given point $x_0$,  
the supersymmetry parameter $\e_\g (x)$ is determined by it value at $x_0$.
As a result,  any non-zero solution of eq. \eqref{Killing-Spinor32} or, equivalently,
\eqref{KS32} is nowhere vanishing if the spacetime $\cM^3$ is a connected 
manifold.\footnote{This can be proved as follows. 
Let us assume that $\e_\g (x)$ vanishes at some point $x_0 \in \cM^3$. 
We can expand $\e_\g (x)$ in 
a covariant Taylor series centered at $x_0$ (see, e.g., \cite{KT-Maction}) in an open  neighborhood 
$U $ of $x_0$. Then, due to \eqref{KS32}, 
$\e_\g (x)$ is equal to zero on $U$.  
It also clear that $\e_\g(x)$ vanishes on the closure $\overline U$ of $U$. 
Now we can introduce the subset $\cW \in \cM^3$ consisting of all zeros of $\e_\g (x)$. 
It follows that this subset is open and closed, and therefore it coincides with $\cM^3$
since the latter is connected.}

\subsection{Supersymmetric backgrounds with four supercharges}

The existence of rigid supersymmetries imposes non-trivial restrictions on 
the background fields.  
For simplicity, here we  work out these restrictions in the case of four supercharges.
Since $\s=0$, one  may deduce from \eqref{8.300} 
the following  conditions:
\bea
\cD^2R|=
\cD^{\g}\cDB_{\g}\cS|=
{[}\cD_{(\a}, \cDB_{\b)}{]}\cS|=
\big(
\cD_{(\a} \bar{{\bm C}}_{\b\g\d)}
+\cDB_{(\a}{\bm C}_{\b\g\d)}
\big)|=0
~.
\eea
It is an instructive exercise to demonstrate that these conditions
constrain the background fields ${s}$, ${r}$ and ${c}_a$
as follows:
\bsubeq
\bea
{\mathfrak D}_a 
{s}
&=& 0~, \\
{\mathfrak D}_a  {{r}}&=& 2\ri b_a {{r}}~,
\\
{\mathfrak D}_{a} {c}_b&=&
2\ve_{abc}{c}^c {s}
~, \label{8.38c}
\\
r\, {s}
&=&0~,
\\
r \,{c}_a
&=& 0 ~.
\eea
\esubeq
It follows from \eqref{8.38c} that $c_a$ is a Killing vector field, 
\bea
{\mathfrak D}_{a} {c}_b + {\mathfrak D}_{b} {c}_a =0~.
\eea
The U$(1)_R$ field strength proves to vanish.
\bea
\cF_{ab}&=&0
~.
\eea
The Einstein tensor \eqref{Einstein-Tensor} is uniquely fixed to be
\bea
\cG_{ab}&=&
4\Big{[}
{c}_a {c}_b
+\eta_{ab}\big({s}^2
+\bar {{r}} {{r}}\big)
\Big{]}~.
\eea
We recall that in three dimensions the Riemann tensor is determined in terms of the Einstein tensor 
according to eq. \eqref{Einstein-Tensor}.
For  the Cotton tensor \eqref{def-Cotton} we obtain
\bea
\cW_{ab}=
-24s  \Big{[}
{c}_{a} {c}_b
-\frac{1}{3}\eta_{ab} {c}^{d} {c}_d
\Big{]}~.
\label{Cotton37}
\eea
The spacetime is conformally flat if  ${s} \,{c}_a=0$.

So far we have not specified any compensator. 
We now turn to considering the known off-shell supergravity formulations \cite{KT-M11}.


\subsection{Type I minimal backgrounds with four supercharges}\label{section8.6}

In Type I supergravity, the conformal compensators are  
a covariantly chiral superfield  $\F$ of super-Weyl weight $w=1/2$
and its complex conjugate $\bar \F$. We recall that the properties of $\F$ 
are given by eq. \eqref{chiral51}.
The freedom to perform the super-Weyl and local 
U(1)$_R$ transformations can be used to impose the gauge
\bea
\F=1
~.
\eea
Such a gauge fixing is accompanied by  the consistency conditions \cite{KLT-M11}
\bea
0=\cD_\a\F=-\frac{\ri}{2}\F_\a~,\qquad 
0=\{\cD_\a,\cDB_\b\}\F
=
-\F_{\a\b}
+\cC_{\a\b}
-2\ri\ve_{\a\b}\cS
~,
\eea
and therefore
\bea
 \cS=0~, \qquad
\F_\a=0~,\qquad
\F_{\a\b}= \cC_{\a\b}~.
\eea
Since the local U(1)$_R$ invariance is completely fixed in this gauge,
 it is more convenient to make use of  covariant derivatives
without U$(1)_R$ connection, 
\bea
\de_A:=\cD_A-\ri\F_A\cJ
~,
\eea
which  satisfy the anti-commutation relations
\bsubeq
\bea
&\{\de_\a,\de_\b\}
=
-4\bar{R}\cM_{\a\b}
~,~~~~~~
\{\deb_\a,\deb_\b\}
=
4{R}\cM_{\a\b}~,
\\
&\{\de_\a,\deb_\b\}
=
-2\ri\de_{\a\b}
-2\ve_{\a\b}\cC^{\g\d}\cM_{\g\d}
~.
\eea
\esubeq

The Killing spinor equation \eqref{Killing-Spinor32} 
becomes
\bea
{\frak D}_a\e^\a
&=&
\ri {c}_a\e^\a
-\ri\ve_{abc} {c}^b  (\tilde{\g}^c\e)^{\a}  
+\ri{{r}}(\tilde{\g}_a\bar{\e})^\a 
~.
\label{KillingSpinor-I_0}
\eea
The supersymmetric backgrounds with four supercharges are characterized by the properties:
\bsubeq
\bea
r \,{c}_a
&=& 0 ~,
\\
{\mathfrak D}_a  {{r}}&=& 0~,
\\
{\mathfrak D}_{a} {c}_b&=&0
~.
\eea
\esubeq
The Einstein tensor is 
\bea
\cG_{ab}&=&
4\Big{[}
{c}_a {c}_b
+\eta_{ab}
\bar r   r
\Big{]}~.
\eea
Such a spacetime is necessarily conformally flat, 
\bea
\cW_{ab}
&=&
0
~.
\eea
The solution with ${c}_a =0$ corresponds to the (1,1) AdS superspace 
\cite{KT-M11}.

The Killing spinor equation \eqref{KillingSpinor-I_0} is equivalent to the condition that the gravitino variation
 \eqref{grav-var-I_0} vanishes, 
\bea
-\hf\d_\e\psi_m{}^\a=
-\frak{D}_m\e^\a
+\frac{\ri}{2} b_m\e^\a
-\frac{\ri}{2} e_m{}^a\ve_{abc}\,b^b(\tilde{\g}^c\e)^{\a}
-\frac{\ri}{4}\bar{M}(\tilde{\g}_m\bar{\e})^{\a}
=0
~,
\eea
provided we replace
\bea
c_a\to\hf b_a~,~~~
r\to-\frac{1}{4}\bar{M}
~.
\eea


\subsection{Type II minimal  backgrounds with four supercharges}

Type II minimal supergravity is obtained by coupling the Weyl multiplet 
to a real linear compensator $\mathbb{G}$ with the super-Weyl transformation
law given by eq. \eqref{N=2sWrealLinear}. 
The super-Weyl invariance allows us to choose the gauge
\bea
\mathbb{G}=1~.
\label{8.47}
\eea
Because the compensator is real, its U$(1)_R$ charge \eqref{GenCom}
is equal to zero, and thus the local U(1)$_R$ group remains unbroken in the gauge 
chosen.  
The consistency condition for \eqref{8.47} is 
\bea
R=\bar{R}=0 
~.
\eea
Then, the anti-commutators of spinor covariant derivatives become
\bsubeq
\bea
&\{\cD_\a,\cD_\b\}
=
\{\cDB_\a,\cDB_\b\}
=0
~,
\\
&\{\cD_\a,\cDB_\b\}
=
-2\ri\cD_{\a\b}
-2\cC_{\a\b}\cJ
-4\ri\ve_{\a\b}\cS\cJ
+4\ri\cS\cM_{\a\b}
-2\ve_{\a\b}\cC^{\g\d}\cM_{\g\d}
~.
\eea
\esubeq

The Killing spinor equation for Type II minimal supergravity is 
\bea
\bD_{a} \e^\a &=&
-\ri\ve_{abc}  {c}^b (\tilde{\g}^c\e)^{\a}
+ {s}(\tilde{\g}_a\e)^{\a}
~.
\label{Killing-Spinor322} 
\eea
All supersymmetric backgrounds with four supercharges are characterized by the conditions
\bsubeq
\bea
{\mathfrak D}_a {s}&=& 0~,\\
{\mathfrak D}_{a} {c}_b&=&
2\ve_{abc}{c}^c {s}
~.
\eea
\esubeq
The Einstein tensor is
\bea
\cG_{ab}&=&
4\Big{[}
{c}_a {c}_b
+\eta_{ab} {s}^2
\Big{]}~,
\eea
and the Cotton tensor is given by eq. \eqref{Cotton37}.
The solution with ${c}_a =0$ corresponds to the (2,0) AdS superspace 
\cite{KT-M11}.
In the case $c_a \neq 0$, the traceless Ricci tensor is 
\bea
\cR_{ab} - \frac{1}{3} \eta_{ab} \cR 
=
 4 \Big[ c_a c_b -  \frac{1}{3} \eta_{ab} c^2\Big]~.
\eea
From this we conclude (see, e.g.,  Table 1 in \cite{Chow:2009km}) thatÊ
spacetime is of Êtype N (for $c_a$ null), type $\rm D_s$ (for $c_a$ spacelike) or 
$\rm D_t$ (for $c_a$ timelike) Êin the Petrov-Segre classification.
 For $\rm D_t$ and $\rm D_s$ it is shown in \cite{Chow:2009km}  that spacetime is necessarily 
 biaxially squashed ${\rm AdS}_3$.Ê

The Killing spinor equation \eqref{Killing-Spinor322} is equivalent to the condition that, in the 
gauge $\psi_m{}^\a=0$, the gravitino variation
 \eqref{grav-var-II_0} vanishes, 
\bea
-\hf\d_\e\psi_m{}^\a=
-\bD_m\e^\a
-\frac{\ri}{4}\cH_m\e^\a
+\frac{\ri}{4} e_m{}^a\ve_{abc}\cH^c(\tilde{\g}^b\e)^{\a}
-\frac{1}{4}Z(\tilde{\g}_m\e)^{\a}
=0
~,
\eea
provided we make the replacements
\bea
b_a\to b_a-\frac{1}{4}\cH_a
~,~~~
c_a\to -\hf\cH_a
~,~~~
s\to
-\frac{1}{4}Z
~.
\eea

\subsection{Non-minimal backgrounds with four supercharges}

Non-minimal supergravity in three dimensions was studied in \cite{KLT-M11,KT-M11}.
It is obtained by coupling  the Weyl multiplet to 
a complex linear compensator $\S$ and its conjugate. Here $\S$ 
obeys  the constraint
\bea 
(\cDB^2-4R)\S=0
\label{complex-linear}
\eea
and is subject to no reality condition. 
By definition,  the compensator $\S$ is chosen to be nowhere vanishing 
and transforms as a primary field of weight $w\ne0,1$  under the super-Weyl group.
 Then, the   U$(1)_R$  charge of $\S$ 
 is uniquely determined 
  \cite{KLT-M11}, 
\bea
\d_\s \S=w\s\S \quad \Longrightarrow \quad \cJ\S=(1-w) \S
~.
\label{complex-linear2}
\eea

The super-Weyl and local U$(1)_R$ symmetries can be used to impose the gauge condition
\bea
\S=1~.
\label{4.50}
\eea
In this gauge, 
some restrictions on the geometry occur \cite{KLT-M11}.
To describe them, it is useful to  split the covariant derivatives as 
\bea
\cD_\a=\nabla_\a+\ri T_\a\cJ
~,~~~
\cDB_\a=\bar{\nabla}_\a+\ri \bar{T}_\a\cJ~,
\eea
where 
the original U$(1)_R$ connection $\F_\a$ has been renamed as $T_\a$.
In the gauge (\ref{4.50}), the  constraint $({\bar \cD}^2 -4R)\S=0$ turns into 
\bea
R&=&\frac{1-w}{4} \ \Big(\ri 
\bar{\nabla}_\a\bar{T}^\a
+ w\bar{T}_\a\bar{T}^\a
\Big)~.
\label{4.52}
\eea
We see that $R$ becomes a descendant of $\bar T_\a$.
Eq. (\ref{4.52})  is not the only consistency condition implied by the gauge fixing (\ref{4.50}).
Evaluating explicitly $\{ \cD_\a, \cD_\b\} \S $ and  $\{ \cD_\a, {\bar \cD}_\b\} \S $
and then setting $\S=1$
gives
\bsubeq
\bea
&&\de_{(\a}T_{\b)}=0~,~~~~~~
\cS=\frac{1}{8}\Big(
\deb^\a T_\a
-\de^\a\bar{T}_\a
+2\ri T^\a\bar{T}_\a
\Big)
~,
\label{SDT}
\\
&&~~~~~~
\F_{\a\b}
=
\cC_{\a\b}
+\frac{\ri}{2}\de_{(\a}\bar{T}_{\b)}
+\frac{\ri}{2}\deb_{(\a} T_{\b)}
+T_{(\a}\bar{T}_{\b)}
~.
\label{FCDT}
\eea
\esubeq
If we define a new vector covariant derivative $\nabla _a$ by $\cD_a = \nabla_a +\ri \F_a \cJ$, 
then the algebra of the covariant derivatives $\nabla_A =(\nabla_a, \nabla_\a , {\bar \nabla}_\a)$
proves to be 
\bsubeq
\bea
\{\de_\a,\de_\b\}&=&
-2\ri T_{(\a}\de_{\b)}
-\ri(w-1)\Big(
{\nabla}^\g{T}_\g
+\ri w{T}^\g{T}_\g
\Big)
\cM_{\a\b}
~,
\\
\{\de_\a,\deb_\b\}&=&
-2\ri\de_{\a\b}
-\ri\bar{T}_\b\de_\a
+\ri T_\a\deb_\b
-2\ve_{\a\b}\cC^{\g\d}\cM_{\g\d}
\non\\
&&
+\frac{\ri}{2}\Big(
\deb^\g T_\g
-\de^\g\bar{T}_\g
+2\ri T^\g\bar{T}_\g
\Big)
\cM_{\a\b}
~.
\eea
\esubeq

The Killing spinor equation in this case is
\bea
{\frak D}_{a}\e^{\a}
=
\ri\F_{a}|\e^{\a}
-\ri\ve_{abc}  {c}^b (\tilde{\g}^c \e)^{\a}
+ {s}(\tilde{\g}_a \e)^{\a}
+\ri {{r}}  (\tilde{\g}_a\bar{\e})^\a
~.
\label{8.60}
\eea
It should be kept in mind that $R$, $\cS$ and $\F_a$ are now 
 composite superfields constructed in term of $T_\a$, $\bar T_\a$ and their 
covariant derivatives, in accordance with eqs.  (\ref{4.52}), \eqref{SDT} and 
  \eqref{FCDT} respectively.
Supersymmetric backgrounds with four supercharges are very constrained
in the non-minimal case.
Indeed, the requirement that $T_\a| =0$ be invariant under the isometry transformations
leads to the condition
\bea
0=\e^\g\de_\g T_\a|
-\bar{\e}^\g\deb_\g T_\a|
~,
\eea
which implies $\de_\a T_\b=\deb_\a T_\b=0$. 
Due to \eqref{4.52}--\eqref{FCDT}, we deduce that
 \bea
 {{r}} =0~, \qquad {s}=0~,\qquad \F_a|= {c}_a
 ~,
 \eea
and then ${c}_b$ is covariantly constant, 
\bea
{\frak D}_{a}{c}_b&=&
0
~.
\eea
The Einstein tensor becomes
\bea
\cG_{ab}
&=&
4{c}_a {c}_b
~.
\eea
Such a spacetime is necessarily conformally flat, $\cW_{ab}=0$.

Non-minimal supergravity is the only off-shell supergravity formulation which does not 
allow for anti-de Sitter backgrounds. However, there exists an alternative
non-minimal formulation in the case $w=-1$ \cite{KT-M11}, inspired by the 4D construction 
in \cite{ButterK11}, which admits an anti-de Sitter solution.

\subsection{Non-minimal AdS backgrounds with four supercharges}

In the case $w=-1$, 
the complex linear constraint \eqref{complex-linear}  admits a nontrivial deformation. 
We introduce a new conformal compensator $\G$ that has  the transformation properties
\bea
\d_\s \G= -\s\G~,\qquad
\cJ\G=2\G
\eea
and  obeys  the {\it improved} linear constraint\footnote{In the case $w=-1$,
there exists a more general deformation, 
$ (\bar \cD^2 - 4 R) \Gamma = -4W(\vf)$,
where $W(\vf^I)$ is a matter superpotential depending on super-Weyl inert chiral 
 superfields $\vf^I$. This super-Weyl invariant constraint reduces to \eqref{8.66} 
 for $W=\m$.} 
\begin{align}
-\frac{1}{4} (\bar \cD^2 - 4 R) \Gamma = \m ={\rm const}~,
\label{8.66}
\end{align}
with the complex parameter $\m \neq 0$ inducing a cosmological constant.
This constraint is super-Weyl invariant.

The super-Weyl and local U$(1)_R$ symmetries  allow us to 
impose the gauge condition
\bea
\G=1~.
\eea
As in the previous subsection, this gauge condition
 implies some restrictions on the geometry. 
 Indeed, the  constraint $({\bar \cD}^2 -4R)\G=\mu$ 
turns into 
\bea
R&=&\mu+\frac{\ri}{2}\Big(
\bar{\nabla}_\a\bar{T}^\a
+\ri\bar{T}_\a\bar{T}^\a
\Big)~.
\label{RDT2}
\eea
We see that $R$ becomes a descendant of $\bar T_\a$.
Next, evaluating the expressions $\{ \cD_\a, \cD_\b\} \G $ and  $\{ \cD_\a, {\bar \cD}_\b\} \G $
and then setting $\G=1$, we again obtain the relations \eqref{SDT} and \eqref{FCDT}. 
As in the previous subsection, we can introduce
 covariant derivatives without U$(1)_R$ connection, 
$\nabla_A =(\nabla_a, \nabla_\a , {\bar \nabla}_\a)$. Their algebra 
proves to be 
\bsubeq
\bea
\{\de_\a,\de_\b\}&=&
-2\ri T_{(\a}\de_{\b)}
-4\bar{\mu}\cM_{\a\b}
+2\ri\Big(
{\nabla}^\g{T}_\g
-\ri {T}^\g{T}_\g
\Big)
\cM_{\a\b}
~,
\\
\{\de_\a,\deb_\b\}&=&
-2\ri\de_{\a\b}
-\ri\bar{T}_\b\de_\a
+\ri T_\a\deb_\b
-2\ve_{\a\b}\cC^{\g\d}\cM_{\g\d}
\non\\
&&
+\frac{\ri}{2}\Big(
\deb^\g T_\g
-\de^\g\bar{T}_\g
+2\ri T^\g\bar{T}_\g
\Big)
\cM_{\a\b}
~.
\eea
\esubeq

The Killing spinor equation coincides with \eqref{8.60}.
Unlike the non-minimal formulation studied in the previous subsection, 
the scalar $R$ is now given by eq.  \eqref{RDT2}.
This modified expression for $R$ leads to different 
backgrounds with four supercharges.
Due to the presence of the parameter $\m$ in \eqref{RDT2}, 
demanding the existence of four supersymmetries gives
\bea
\cS|=0~,\qquad
R|=\mu~,\qquad
\F_{a}|
=
\cC_{a}|
=c_a
~.
\eea
Moreover, one also finds the condition $\cC_c|R|=0$. Since $R|=\m \neq 0$, we conclude that 
\bea
{c}_a =0~.
\eea
The Einstein tensor is
\bea
\cG_{ab} 
&=&
4\eta_{ab} \,\bar{\m} \m~.
\eea
This background corresponds to the (1,1) anti-de Sitter superspace \cite{KT-M11}.

\vspace{1cm}
\noindent
After this work was completed, 
there appeared a new paper 
\cite{DKSS2} which has some overlap with our results in subsections 
\ref{section7.2} and \ref{section8.6}.



\bigskip
\noindent{\bf Acknowledgements:}\\
\smallskip
SMK acknowledges the generous hospitality of the Arnold Sommerfeld Center for Theoretical Physics,
 Munich and the INFN, Padua where part of this work was done.  
MR and IS acknowledge the generous hospitality and stimulating atmosphere of 
the physics department at Harvard University, where this work was begun. 
UL, MR and IS acknowledge the stimulating atmosphere of the 
``2012 Summer Simons Workshop in Mathematics and Physics'' 
where part of this work was done. 
The work of SMK and UL was supported in part by the Australian Research Council, 
grant  DP1096372.
MR acknowledges NSF Grant phy 0969739. 
 IS was supported by the DFG Transregional Collaborative Research Centre TRR 33, the DFG cluster of 
 excellence Origin and Structure of the Universe as well as the DAAD project 54446342.
The work of GT-M was supported by the Australian Research Council's Discovery Early Career 
Award (DECRA), project No. DE120101498.


\appendix

\section{Notation, conventions and some technical details} 
\label{AppendixA}

Our 3D notation and conventions follow those used in \cite{KLT-M11}.
In particular, the vector indices are denoted by lower case Latin letters 
from the beginning of the alphabet, for instance $a,b  = 0,1,2$. 
The Minkowski metric is $\eta_{ab}={\rm diag}(-1,1,1)$, 
and the Levi-Civita tensor $\ve_{abc}$ is normalized by $\ve_{012}=-1$, and hence
$\ve^{012}=1$.
The spinor indices are denoted by small Greek letters 
from the beginning of the alphabet, for instance $\a,\b  = 1,2$. 

To deal with spinors, we introduce a basis of {\it real} symmetric $2\times 2$ 
matrices
\begin{subequations}
\bea
\g_a
&=&  (\g_a )_{\a  \b} = (\g_a)_{\b\a} ~
=({\mathbbm 1}, \s_1, \s_3) ~,
\eea
and also define
\bea
\tilde{\g}_a
&=&
(\g_a )^{\a  \b} = (\g_a)^{\b\a}
:=\ve^{\a \g} \ve^{\b \d} (\g_a)_{\g \d} ~,
\eea
\end{subequations}
with $\s_1 $ and $\s_3$  two of the three Pauli matrices. 
The spinor indices are  raised and lowered using
the SL(2,${\mathbb R}$) invariant tensors
\bea
\ve_{\a\b}=\left(\begin{array}{cc}0~&-1\\1~&0\end{array}\right)~,\qquad
\ve^{\a\b}=\left(\begin{array}{cc}0~&1\\-1~&0\end{array}\right)
\eea
as follows:
\bea
\psi^{\a}=\ve^{\a\b}\psi_\b~, \qquad \psi_{\a}=\ve_{\a\b}\psi^\b~.
\eea
The 3D Dirac $\g$-matrices are 
\bea
\hat{\g}= (\g_a)_\a{}^\b:=\ve^{\b\g}(\g_a)_{\a\g}~,~~~~~~
\hat{\g}_a \hat{\g}_b = \eta_{ab} {\mathbbm 1} + \ve_{abc} \hat{\g}^c~.
\eea
In this representation of the $\g$-matrices, the Majorana spinors are real.

In $\cN=2$ supersymmetry, we usually deal with complex spinors. 
{\it Only} in the case of complex spinors,  
we use throughout this paper the following types of index contraction:
\bsubeq
\bea
&\psi\chi:=\psi^\a\chi_\a
~,~~~
\psi\bar{\chi}:=\psi^\a\bar{\chi}_\a
~,~~~
\bar{\psi}\chi:=\bar{\psi}^\a\chi_\a
~,~~~
\bar{\psi}\bar{\chi}:=\bar{\psi}_\a\bar{\chi}^\a
~;
\\
&(\g_a\psi)_\a:=(\g_a)_{\a\b}\psi^\b=(\psi\g_a)_\a
~,~~~
(\tilde{\g}_a\psi)^\a:=(\g_a)^{\a\b}\psi_\b=(\psi\tilde{\g}_a)^\a
~;
\\
&\psi\g_a\chi:=\psi^\a(\g_a)_{\a\b}\chi^\b
~,~~~
\psi\tilde{\g}_a\chi:=\psi_\a(\g_a)^{\a\b}\chi_\b
~.
\eea
\esubeq
In particular, contractions of 
two spinor covariant derivatives are defined as
\bea
\cD^2:=\cD^\a\cD_\a~,\qquad
\cDB^2=\cDB_\a\cDB^\a~.
\eea

Any three-vector $F_a$ can equivalently be realized as a symmetric spinor 
$F_{\a\b} =F_{\b \a}$.
The relationship between $F_a$ and $F_{\a \b}$ is as follows:
\bea
F_{\a\b}:=(\g^a)_{\a\b}F_a=F_{\b\a}~,\qquad
F_a=-\hf(\g_a)^{\a\b}F_{\a\b}~.
\label{vector-rule}
\eea
We can also describe the one-form $F_a$ in terms of its Hodge-dual two-form $F_{ab}=-F_{ba}$,
\bea
F_{ab}:=-\ve_{abc}F^c ~, \qquad
F_a=\hf\ve_{abc}F^{bc}~.
\label{hodge-1}
\eea
Then, the symmetric spinor $F_{\a\b} =F_{\b\a}$, which is associated with $F_a$, can 
equivalently be defined in terms of  $F_{ab}$: 
\bea
F_{\a\b}:=(\g^a)_{\a\b}F_a=\hf(\g^a)_{\a\b}\ve_{abc}F^{bc}
~.
\label{hodge-2}
\eea
These three algebraic objects, $F_a$, $F_{ab}$ and $F_{\a \b}$, 
are in one-to-one correspondence to each other. 
Thier inner products are related as follows:
\bea
-F^aG_a=
\hf F^{ab}G_{ab}=\hf F^{\a\b}G_{\a\b}
~.
\eea

An  equivalent form of the
commutation relations  \eqref{2.7c} and \eqref{2.7d} is 
\begin{subequations}
\bea
{[}\cD_{\a\b},\cD_\g{]}
&=&
-\ri\ve_{\g(\a}\cC_{\b)\d}\cD^{\d}
+\ri\cC_{\g(\a}\cD_{\b)}
-2\ve_{\g(\a}\cS\cD_{\b)}
-2\ri\ve_{\g(\a}\bar{R}\cDB_{\b)}
\non\\
&&
+2\ve_{\g(\a}{\bm C}_{\b)\d\r}\cM^{\d\r}
-\frac{4}{3}\big(2\cD_{(\a}\cS+\ri\cDB_{(\a}\bar{R}\big)\cM_{\b)\g}
+\frac{1}{3}\big(2\cD_{\g}\cS+\ri\cDB_{\g}\bar{R}\big)\cM_{\a\b}
\non\\
&&
+\Big(
{\bm C}_{\a\b\g}
+\frac{1}{3}\ve_{\g(\a}\big(
8\cD_{\b)}\cS
+\ri\cDB_{\b)}\bar{R}
\big)
\Big)\cJ
~,
\\
{[}\cD_{\a\b},\cDB_\g{]}
&=&
\ri\ve_{\g(\a}\cC_{\b)\d}\cDB^{\d}
-\ri\cC_{\g(\a}\cDB_{\b)}
-2\ve_{\g(\a}\cS\cDB_{\b)}
+2\ri\ve_{\g(\a}R\cD_{\b)}
\non\\
&&
+2\ve_{\g(\a}\bar{{\bm C}}_{\b)\d\r}\cM^{\d\r}
-\frac{4}{3}\big(2\cDB_{(\a}\cS-\ri\cD_{(\a}{R}\big)\cM_{\b)\g}
+\frac{1}{3}\big(2\cDB_{\g}\cS-\ri\cD_{\g}{R}\big)\cM_{\a\b}
\non\\
&&
-\Big(
\bar{{\bm C}}_{\a\b\g}
+\frac{1}{3}\ve_{\g(\a}\big(8\cDB_{\b)}\cS-\ri\cD_{\b)}{R}\big)
\Big)\cJ
~.
\eea
\end{subequations}
These relations are very useful for actual calculations. 

In three dimensions, the Weyl tensor is identically zero, 
and the Riemann tensor $\cR_{abcd}$ is related 
to the Einstein tensor by the simple rule 
\bea
\frac{1}{4}  \ve^{acd}\ve^{bef} \cR_{cd}{}_{ef} = \cG^{ab}: =\cR^{ab} - \hf \eta^{ab} \cR~,
\qquad \cR_{ab}{}_{cd}=\ve_{abe}\ve_{cdf}\cG^{ef}
~.
\label{Einstein-Tensor}
\eea
As a consequence, the Riemann tensor is expressed in term of the 
Ricci tensor $\cR_{ab} := \cR^c{}_{a cb} $ and the scalar curvature 
$\cR:= \eta^{ab} \cR_{ab}$ as follows:
\bea
\cR_{abcd} &=& \eta_{ac} \cR_{bd} -\eta_{ad} \cR_{bc} 
+\eta_{bd} \cR_{ac} -\eta_{bc} \cR_{ad} 
-\hf (\eta_{ac}\eta_{bd} - \eta_{ad} \eta_{bc})\cR~.
\eea
The Cotton tensor is defined as follows
\bea
\cW_{ab}:=\hf\ve_{acd}\cW^{cd}{}_{b} = \cW_{ba}
~,\qquad
\cW_{abc}
=2{\mathfrak D}_{[a}\cR_{b]c}
+\frac{1}{2}\eta_{c[a}{\mathfrak D}_{b]}\cR~.
\label{def-Cotton}
\eea
A spacetime is conformally flat if and only if $\cW_{ab} = 0$ 
\cite{Eisen}
(see \cite{BKNT-M1} for a modern proof).

\section{Superconformal sigma model}\label{AppendixB}

In this appendix we consider an alternative parametrization of the supergravity-matter  
system \eqref{s-m} and reduce it to components without gauge fixing 
the Weyl, local U$(1)_R$ and $S$-supersymmetry transformations. 

In the new parametrization, the matter sector of the theory is described in terms of 
several covariantly chiral superfields $\f^{i} = (\f^0 , \f^I)$
of super-Weyl weight $w=1/2$, 
\bea
\bar \cD_\a \f^i=0~, \qquad \cJ \f^{i}= -\hf \f^{i} ~,\qquad
\d_\s\f^{i}=\hf\s\f^{i}~.
\eea
The action is defined to be 
\bea\label{suco}
S&=&  \int {\rm d}^3x {\rm d}^2 \q \rd^2 \bar \q
\,E\,
{N}\big(\phi^{i}, \bar {\phi}^{\bar {j}}\big)
+  \Big{\{} \int {\rm d}^3x {\rm d}^2 \q \,\cE\,
P(\phi^{i})
\, +\,{\rm c.c.}~\Big{\}}    \non \\
&\equiv &S_{\rm kinetic} +S_{\rm potential}
  ~
\label{conf-sigma-action-0}
\eea
and may naturally be interpreted as a locally supersymmetric 
$\s$-model. 
For the action to be super-Weyl and U(1)$_R$ invariant, 
the K\"ahler potential $N $
and the superpotential $P $ should obey the 
homogeneity conditions
\begin{subequations}
\bea
&&
\sum_{i} \f^{i} N_{i}=\sum_{\bar{{i}}} \bar \f^{\bar{{i}}} N_{\bar{{i}}}=N~,
\label{homo-N}
\\
&& \sum_{i} \f^{i} P_{i}=4P~.
\eea
\end{subequations}
Eq. \eqref{homo-N} means that the $\s$-model target space is a K\"ahler cone \cite{GR}. 

Before reducing the action to components, we introduce 
several standard $\s$-model  definitions. 
As usual, multiple derivatives of the K\"ahler potential are denoted as
\bea
N_{i_1 \dots i_p}{}_{{\bar j}_1 \dots {\bar j}_q }:=
\frac{\pa^{(p+q)}  }{\pa\f^{i_1} \dots \pa\f^{i_p} \pa \fb^{ {\bar j}_1} 
\dots \pa \fb^{{\bar j}_q }}N~.
\eea
The K\"ahler metric\footnote{We do not assume the K\"ahler metric to be positive  
definite.}
$N_{i}{}_{\bar{j}} = N_{\bar j i}$ 
is assumed to be nonsingular, with 
 its inverse  being denoted $N^{\bar{i}}{}^{j} =N^{j \bar i}$, 
\bea
N_{i\bar{k}}N^{\bar{k}}{}^{j}=\d_{i}^{j}
~,~~~
N^{\bar{i}}{}^{k}N_{{k}\bar{j}}=\d^{\bar{i}}_{\bar{j}}
~.
\eea
The Christoffel symbols $\g_{ ij}^{k}$ are
\bea
\g_{ij}^{k}:=N_{ij\bar{l}}N^{\bar{l}k}
~,~~~~~~
\g_{\bar{i}\bar{j}}^{\bar{k}}:=N_{l\bar{i}\bar{j}}N^{l\bar{k}}~,
\eea
and the Riemann curvature $\frak{R}_{{i}{\bar{{k}}} {j}{\bar{{l}}}}$ is
\bea
\frak{R}_{{i}{\bar{{k}}} {j}{\bar{{l}}}}
=
\frak{R}_{{i}{\bar{{k}}} {j}}{}^{p}N_{{p}{\bar{{l}}}}=
\big(\pa_{{\bar{{k}}}}\g^{p}_{{ij}}\big)N_{{p}{\bar{{l}}}}
~.
\eea

We  define the component fields of $\f^i$ as follows:
\begin{subequations}
\bea
\r_\a^{i} &:=& \cD_\a \f^{i}|
~,\\
{\cal F}^{i} &:=&
-\frac{1}{4} \big{[} \cD^2 \f^{i} + \g^{i}_{{j}{k}} (\cD^\a \f^{j}) \cD_\a \f^{k}\big{]}|
~. \label{B.8b}
\eea
\end{subequations}
The physical scalar  $\f^i |$ will be denoted by the same symbol as 
the chiral superfield $\f^i$ itself. 

To reduce the kinetic term in \eqref{conf-sigma-action-0} to components, 
we associate with it the antichiral Lagrangian
\bea
\bar{\cL}_c
&=&
-\frac{1}{4}(\cD^2-4\bar{R})N
\eea
and make use of the action principle \eqref{comp-ac-1}.
The resulting component Lagrangian  is 
\bea
L_{\rm kinetic}
&=&
-\frac{1}{8}\Big{[}
\,\cR
+\frac{\ri}{2}\ve^{abc}\big(\psi_a{\bm{\bar{\psi}}}_{bc}+{\bar{\psi}}_a{\bm\psi}_{bc} \big)
\Big{]}
N
\non\\
&&
+N_{{i}{\bar{{j}}}}\Big{[}
{\cal F}^{i}\bar{{\cal F}}^{\bar{{j}}}
-(\bD^a\f^{i})\bD_a\fb^{\bar{{j}}}
-\frac{\ri}{4}\bar{\r}^{\bar{{j}}}\g^a \widetilde{\bD}_a \r^{i}
-\frac{\ri}{4}\r^{i}\g^a\widetilde{\bD}_a \bar{\r}^{\bar{j}}
+\frac{\ri}{2}\psi_a\r^{i}\bD^a\fb^{\bar{{j}}}
\non\\
&&~~~~~~~~
-\frac{\ri}{2}{\bar{\psi}}_a\bar{\r}^{\bar{{j}}}\bD^a\f^{i}
+\frac{\ri}{2}\ve^{abc}\big({\bar{\psi}}_a\g_b\bar{\r}^{\bar{{j}}} \bD_c\f^{i}-\psi_a\g_b\r^{i}\bD_c\fb^{\bar{{j}}}\big)
-\frac{1}{8}\psi^a{\bar{\psi}}_a\,\r^{i}\bar{\r}^{\bar{{j}}}
\non\\
&&~~~~~~~~
+\frac{1}{8}\psi^a\g_b{\bar{\psi}}_a\,\r^{i}\g^b\bar{\r}^{\bar{{j}}} 
+\frac{1}{8}\ve^{abc}\big(
\psi_a\g_b{\bar{\psi}}_c\,\r^{i}\bar{\r}^{\bar{{j}}}
+\psi_a{\bar{\psi}}_b\,\r^{i}\g_c\bar{\r}^{\bar{{j}}}
\big)
\Big{]}
\non\\
&&
+\frac{1}{8}\ve^{abc}\Big{[}
{\bm{\bar{\psi}}}_{ab}\g_c\bar{\r}^{\bar{{i}}} N_{\bar{{i}}}
-{\bm\psi}_{ab}\g_c\r^{i}N_{i}
\Big{]}
+\frac{\ri}{4}\ve^{abc}\psi_{a}{\bar{\psi}}_{b}\Big{[}N_{i}\bD_c\f^{i}-N_{\bar{{i}}}\bD_c\fb^{\bar{{i}}}\Big{]}
\non\\
&&
+\frac{1}{16} {\frak{R}}_{{i}{\bar{{k}}} {j}{\bar{{l}}}}\,\r^{i}\r^{j}\,\bar{\r}^{\bar{{k}}}\bar{\r}^{\bar{{l}}}
~,
\label{conf-comp-action}
\eea
where we have introduced the target-space covariant derivative
\bea
\widetilde{\bD}_a \r^{i}_\a&:=&
\bD_a \r^i_\a
+ \g^i_{jk}
\,\r^{j}_\a   \bD_a \f^{k}~.
\label{B.11}
\eea
A short calculation of the component Lagrangian
corresponding to  $S_{\rm potential}$ gives
\bea
L_{\rm potential}&=&
\cF^{{i}}{P}_{{i}}
-\frac{1}{4}\big(
P_{jk}
-\g^{{i}}_{{{j}}{{k}}}{P}_{{i}}
\big)\r^{j}\r^{k}
+\frac{\ri}{2}{\bar{\psi}}_a\g^a\r^{j}{P}_{j}
-\hf\ve^{abc}{\bar{\psi}}_a \g_b{\bar{\psi}}_c P
~+~{\rm c.c.}~~~~
\label{conf-act-superpotential}
\eea
Both Lagrangians \eqref{conf-comp-action} and 
\eqref{conf-act-superpotential} are quite compact. 

Now, we relate the theory under consideration to the $\sigma$-model \eqref{s-m}.
 We assume that the chiral scalar $\f^0$ from the set 
 $\f^{i} = (\f^0 , \f^I)$ is nowhere vanishing, 
 $\f^0 \neq 0$, and therefore it may be chosen to play the role of conformal compensator.  
We introduce a  new parametrization of the dynamical  chiral superfields defined by 
\bea
\f^0 = \F
~, \qquad
\f^{I}=\F\vf^I
~.
\label{cone-parametrization}
\eea
Here the chiral scalars $\vf^I$ are neutral under the super-Weyl and U$(1)_R$ 
transformations. Since $\F$ is nowhere vanishing, 
$N(\f,\bar{\f})$ and $P(\f)$ may be represented in the form:
\bea
N(\f,\bar{\f})
&=&
-4\Fb \,\re^{-\frac{1}{4} K(\vf,\vfb)} \F
~,~~~~~~
P(\f)=\F^4 W(\vf)~.
\label{cone-parametrization-potentials}
\eea
We assume that $K(\vf,\vfb)$ is the K\"ahler potential of a K\"ahler manifold
with positive definite metric 
$ g_{I\bar J} := K_{I\bJ}$.

Let us express the geometric objects in terms of the new coordinates
introduced.  
A  short calculation gives
\bsubeq
\bea
N_{{i}{\bar{{j}}}}&=&
\re^{-\frac{1}{4} K}\left(
\begin{array}{ccc}
-4
&~~~&
\Fb K_{\bJ}
\\
\F K_{I}
&~~~&
\F\Fb\hat{K}_{I\bJ}
\end{array}
\right)
~, \label{B.15a}
\eea
where we have denoted 
\bea
K_I:=\frac{\pa K}{\pa\vf^I}~,~
\qquad
\hat{K}_{I\bJ}:={g}_{I\bJ}-\frac{1}{4}K_IK_\bJ
~.~~~~~~
\eea
\esubeq
It follows from \eqref{B.15a} that the conditions 
$\det (N_{i\bar j} )\neq 0 $ and $\det (g_{I\bar J}) \neq 0$ are equivalent. 
For the inverse metric we obtain
\bsubeq
\bea
&N^{{\bar{{i}}} {j}}=\re^{\frac{1}{4} K}
\left(
\begin{array}{ccc}
-\frac{1}{4}\big(1-\frac{1}{4} K^L K_L \big)
&~~~&
\frac{1}{4\F} K^{J}
\\
\frac{1}{4\Fb} K^{\bI}
&~~~&
\frac{1}{\F\Fb} K^{\bI J}
\end{array}
\right)
~,
\eea
where we have denoted
\bea
K^I:= g^{I \bJ } K_\bJ~, \qquad
K^\bI:= g^{\bI J} K_J~.
\eea
\esubeq
For the Christoffel symbols $\g^i_{{k}{l}}$  we read off
\bea
\g^0_{{k}{l}}=
\left(
\begin{array}{ccc}
0
&&
0
\\
0
&&
\frac{1}{4}\F\big(
\G_{KL}^I K_I
- K_{KL}
-\frac{1}{4} K_{K} K_L
\big)
\end{array}
\right)
~,~~
\g^I_{{k}{l}}=
\left(
\begin{array}{ccc}
0
&~&
\frac{1}{\F}\d^I_L
\\
\frac{1}{\F}\d^I_K
&~&
\G_{KL}^I
-\frac{1}{2} K_{(K}\d_{L)}^I
\end{array}
\right)~,~~~~~~
\eea
where
$\G_{KL}^I$
is the Christoffel symbol 
for the K\"ahler metric $g_{I\bar J}$. 
Since $\pa_{\bar 0} \g^{k}_{{i}{j}}  =0$, the Riemann tensor
is characterized by the properties 
\bea
\frak{R}_{0{\bar{{k}}} {j}{\bar{{l}}}}=
\frak{R}_{{i}\bar{0} {j}{\bar{{l}}}}=
\frak{R}_{{i}{\bar{{k}}} 0{\bar{{l}}}}=
\frak{R}_{{i}{\bar{{k}}} {j}\bar{0}}=0
~.
\eea
Thus the only nonzero components of the Riemann tensor are
\bea
\frak{R}_{I\bK J\bL}
=
\F\Fb\re^{-\frac{1}{4} K}\Big(
R_{I\bK J \bL}
-\frac{1}{4}( K_{I\bK} K_{J\bL}
+ K_{J\bK} K_{I\bL})
\Big)
~,~~~
R_{I\bK J\bL}=  g_{P\bL} \pa_{\bK}\G_{IJ}^P 
~.~~~~~~
\eea

Our next step is to express the auxiliary fields
$\cF^{i}$, eq. \eqref{B.8b}, and the spinor fields
$\widetilde{\bD}_a \r^{i}_\a$, eq. \eqref{B.11}, 
in terms of the component fields of $\F$ and $\vf^I$.
We recall that
the component fields of $\vf^I$ are defined in \eqref{vf-comfields}. 
We do not introduce special names for the component fields of $\F$; 
we simply write them as $\F$, $\cD_\a \F$ and $\cD^2 \F$, with the bar-projection 
being always assumed here and in what follows. 
For  the auxiliary fields $\cF^{i}$ we get
\bsubeq
\bea
{\cal F}^0 &=&
-\frac{1}{4} \cD^2 \F 
-\frac{1}{16}\F\Big(\G^I_{KL} K_I- K_{KL}-\frac{1}{4} K_K K_L\Big) \l{}^K \l^L
~,
\\
{\cal F}^I &=&
F^I
-\frac{1}{2\F}   \l^{\a I}\cD_\a \F 
+\frac{1}{8} \l^I\l^J K_{J}
~.
\eea
\esubeq
For the spinor fields
$\widetilde{\bD}_a\r_\a^{i}$ we derive
\bsubeq
\bea
\widetilde{\bD}_a \cD_\a\F&=&\widetilde{\bD}_a \r_\a^0 =\bD_a \cD_\a\F
+\frac{1}{4}\F\Big(\G^I_{KL} K_I- K_{KL}-\frac{1}{4} K_K K_L\Big)  \l_\a^K  \bD_a X^L
~,~~~~~~~~
\\
\widetilde{\bD}_a \l_\a^I&=&
\bD_a \l^{{{I}}}_\a
+\G^{I}_{JK}\l^{J}_\a   \bD_a X^K
+\frac{2}{\F}   (\cD_\a\F)  \bD_a X^I
-\frac{1}{2} K_{J}\l_\a^{(I}  \bD_a X^{J)}
~.
\eea
\esubeq

Using the above results, 
for the kinetic term \eqref{conf-comp-action} we obtain 
{\allowdisplaybreaks
\bea
L_{\rm kinetic} &=&
\Big{[}\,
\frac{1}{2}\cR
+\frac{\ri}{4}\ve^{abc}\big(
\psi_a{\bm{\bar{\psi}}}_{bc}
+{\bar{\psi}}_a{\bm\psi}_{bc} \big)
\Big{]}
\Fb\re^{-\frac{1}{4} K}\F
\non\\
&&
-4\Big{[}\,
\cF^0\bar{\cF}^{\bar{0}}
-(\bD^a\F)\bD_a\Fb
-\frac{\ri}{4}\Big(
(\cDB\Fb)\g^a \widetilde{\bD}_a \cD\F
+(\cD\F)\g^a\widetilde{\bD}_a \cDB\F \Big)
\non\\
&&~~~~~
+\hf\psi_a(\cD\F)\bD^a\Fb
+\hf{\bar{\psi}}_a(\cDB\Fb)\bD^a\F
+\hf\ve^{abc}\Big(
{\bar{\psi}}_a\g_b(\cDB\Fb) \bD_c\F
-\psi_a\g_b(\cD\F) \bD_c\Fb
\Big)
\non\\
&&~~~~~
+\frac{1}{8}\ve^{abc}\Big(
\psi_a\g_b{\bar{\psi}}_c\,(\cD\F)\cDB\Fb
+\psi_a{\bar{\psi}}_b\,(\cD\F)\g_c\cDB\Fb
\Big)
\non\\
&&~~~~~
-\frac{1}{8}\psi^a{\bar{\psi}}_a\,(\cD\F)\cDB\Fb
+\frac{1}{8}\psi^a\g^c{\bar{\psi}}_a\,(\cD\F)\g_b\cDB\Fb
\Big{]}
\re^{-\frac{1}{4} K}
\non\\
&&
+\Big{[}\,
\cF^0\bar{\cF}^\bI
-(\bD^a\F)\bD_a\bar{X}^\bI
-\frac{\ri}{4}\Big(
\lb^\bI \g^a\widetilde{\bD}_a \cD\F
+(\cD\F)\g^a\widetilde{\bD}_a\lb^\bI
\Big)
\non\\
&&~~~
+\hf\psi_a(\cD\F)\bD^a\bar{X}^\bI
+\hf{\bar{\psi}}_a\lb^\bI\bD^a\F
+\hf\ve^{abc}\Big(
{\bar{\psi}}_a\g_b\lb^\bI \bD_c\F
-\psi_a\g_b(\cD\F) \bD_c\vfb^\bI
\Big)
\non\\
&&~~~
+\frac{1}{8}\ve^{abc}\Big(
\psi_a\g_b{\bar{\psi}}_c\,\lb^\bI\cD\F
-\psi_a{\bar{\psi}}_b\,\lb^\bI\g_c\cD\F
\Big)
\non\\
&&~~~
-\frac{1}{8}\psi^a{\bar{\psi}}_a\,\lb^\bI\cD\F
-\frac{1}{8}\psi^a\g^b{\bar{\psi}}_a\,\lb^{\bI}\g_b\cD\F
\Big{]}\re^{-\frac{1}{4} K}\Fb K_\bI
\non\\
&&
+\Big{[}\,
\cF^I\bar{\cF}^{\bar{0}}
-(\bD^aX^I)\bD_a\Fb
-\frac{\ri}{4}\Big(
\l^I\g^a\widetilde{\bD}_a\cDB\Fb
+(\cDB\Fb)\g^a\widetilde{\bD}_a\l^I
\Big)
\non\\
&&~~~
+\hf\psi_a\l^I\bD^a\Fb
+\hf{\bar{\psi}}_a(\cDB\Fb)\bD^aX^I
+\hf\ve^{abc}\Big(
{\bar{\psi}}_a\g_b(\cDB\Fb) \bD_cX^I
-\psi_a\g_b\l^I \bD_c\Fb
\Big)
\non\\
&&~~~
+\frac{1}{8}\ve^{abc}\Big(
\psi_a\g_b{\bar{\psi}}_c\,\l^{I}\cDB\Fb
+\psi_a{\bar{\psi}}_b\,\l^I\g_c\cDB\Fb
\Big)
\non\\
&&~~~
-\frac{1}{8}\psi^a{\bar{\psi}}_a\,\l^{I}\cDB\Fb
+\frac{1}{8}\psi^a\g^b{\bar{\psi}}_a\,\l^{I}\g_b\cDB\Fb
\Big{]}\re^{-\frac{1}{4} K}\F K_I
\non\\
&&
+\Big{[}\,
\cF^I\bar{\cF}^\bJ
-(\bD^aX^I)\bD_a\bar{X}^\bJ
-\frac{\ri}{4}\Big(
\lb^\bI\g^a\widetilde{\bD}_a \l^I
+\l^I\g^a\widetilde{\bD}_a\lb^\bJ
\Big)
\non\\
&&~~~~
+\hf\psi_a\l^I\bD^a\bar{X}^\bJ
+\hf{\bar{\psi}}_a\lb^\bJ\bD^aX^I
+\hf\ve^{abc}\Big(
{\bar{\psi}}_a\g_b\lb^\bJ \bD_cX^I
-\psi_a\g_b\l^I \bD_c\bar{X}^\bJ
\Big)
\non\\
&&~~~~
+\frac{1}{8}\ve^{abc}\Big(
\psi_a\g_b{\bar{\psi}}_c\,\l^{I}\lb^\bJ
+\psi_a{\bar{\psi}}_b\,\l^I\g_c\lb^\bJ
\Big)
\non\\
&&~~~~
-\frac{1}{8}\psi^a{\bar{\psi}}_a\,\l^{I}\lb^\bJ
+\frac{1}{8}\psi^a\g^b{\bar{\psi}}_a\,\l^{I}\g_b\lb^{\bJ}
\Big{]}\re^{-\frac{1}{4} K}\F\Fb\hat{K}_{I\bJ}
\non\\
&&
-\frac{1}{4}\ve^{abc}\Big{[}\,
\Fb\Big(
\ri\psi_{a}{\bar{\psi}}_{b}\big(4\bD_c\F-\F K_I\bD_cX^I\big)
-\hf{\bm\psi}_{ab}\g_c\big(4\cD\F-\F K_I\l^I\big)\Big)
+{\rm c.c.}
\Big{]}\re^{-\frac{1}{4} K}
\non\\
&&
+\frac{1}{16} \Big{[}R_{I\bK J\bL}-\hf g_{I\bK} g_{J\bL}\Big{]}\l^{I}\l^J\,\lb^\bK\lb^{\bL}
~.
\label{ac-333}
\eea
}
The potential term  \eqref{conf-act-superpotential} becomes
\bea
L_{\rm potential}&=&
\F^4\Big{[}\,
F^IW_I
-\F^{-1}W \cD^2 \F 
-3\F^{-2}W(\cD\F)\cD\F
-2\F^{-1}W_I \l^{I}\cD \F
\non\\
&&~~~~~
-\frac{1}{4}\l^{I}\l^{J}\big(W_{IJ}-\G^K_{IJ}W_K\big)
+\frac{\ri}{2}{\bar{\psi}}_a\g^a\big(
4\F^{-1}W\cD\F
+W_I\l^{I}
\big)
\non\\
&&~~~~~
+\hf W\ve^{abc}{\bar{\psi}}_a\g_b {\bar{\psi}}_c
\Big{]}
~+~{\rm c.c.}~
\label{sup-333}
\eea

The sum of the expressions \eqref{ac-333} and \eqref{sup-333}
constitutes the component Lagrangian of the theory 
\eqref{s-m} with no gauge condition 
on the chiral compensator $\F$ imposed.
Looking at the explicit form of  \eqref{ac-333}, it is  easy to understand 
why the gauge conditions  \eqref{F-Weyl} have been chosen. 
First of all, it is seen from the first line of \eqref{ac-333} 
the canonically normalized 
Hilbert-Einstein gravitational Lagrangian 
corresponds to 
the Weyl gauge condition \eqref{Weyl-gauge}.
Secondly, consider the terms in  \eqref{ac-333} which involve
 the gravitino field strength coupled to the matter fermions.
These consist of
\bea
\Fb{\bm\psi}_{ab}\g_c\Big(
 \re^{-\frac{1}{4} K} \cD_\a\F 
-\frac{1}{4}  \F\re^{-\frac{1}{4}} K_I  \l^I_\a \Big)
={\bm\psi}_{ab}\g_c
\cD_\a \Big( \Fb\re^{-\frac{1}{4} K} \F\Big)
\eea
and its complex conjugate. 
To eliminate these cross terms, we 
have to impose  the $S$-supersymmetry gauge condition \eqref{S-susy-gauge}.
Finally, the U$(1)_R$ gauge condition \eqref{U(1)-gauge}
eliminates an overall phase factor in the superpotential 
\eqref{sup-333}. 
In the gauge \eqref{F-Weyl},
the only remaining field in $\F$ 
occurs at the  $\q^2$ component.  
It  can be defined in the K\"ahler invariant way \eqref{MbarM}.

In the gauge \eqref{F-Weyl}, the following useful relations hold
\bsubeq
\bea
\cD_\a\F
&=&
\frac{1}{4}\re^{\frac{1}{8}K}\l_\a^LK_L
~,
\\
\cF^0 &=&
-\frac{1}{4}\re^{-\frac{1}{8} K}\mathbb{M}
+\frac{1}{4}\re^{\frac{1}{8} K} F^IK_I
~,
\\
\cF^I &=&
F^I
~.
\eea
\esubeq
As usual, the bar-projection is assumed here. 
Using these relations
one may obtain the component 
Lagrangians \eqref{Type-I-comp} and \eqref{Type-I-comp-potential}
from \eqref{ac-333} and \eqref{sup-333}.


\section{Vector multiplet model}

In this  appendix we present the component Lagrangian for the model of an 
Abelian vector multiplet coupled to conformal supergravity.
As in subsection 6.1,
we denote by $G$ the gauge prepotential of the vector multiplet, 
and by  $\mathbb G$ the corresponding  
gauge invariant field strength. 
The vector multiplet action  is
\bea
S_{\text{VM}}
=-4\int {\rm d}^3x \rd^2\q\rd^2\qb
\,E\,
\Big(  \mathbb G \ln \mathbb G - 4G \cS 
-\k G \mathbb G \Big)
~,
\label{SVM000000}
\eea
with $\k$ a  constant parameter.

We define the  component fields of the vector multiplet as follow:
\bea
\cY&:=&{\mathbb G}\big|
~,
\\
\upsilon_\a &:=&\cD_\a {\mathbb G}\big|
~,
\\
\cZ&:=&\ri \cD^\a \bar \cD_\a {\mathbb G}|
~,
\\
\cB^{a}
&:=&
-\hf(\g^a)^{\a\b}{[}\cD_{(\a},\cDB_{\b)}{]}\mathbb G|
=
\cH^a 
-\ve^{abc} \cY \bar  \psi_b \psi_c
-\ri \ve^{abc} \big( \psi_b\g_c \u + \bar \psi_b\g_c{\bar{\u}} \big)
~.
\eea
As in section 6,  $\cH^a$ denotes the Hodge-dual of 
the component field strength,  
\bea
\cH^a = \hf \ve^{abc}\cH_{bc}~, \qquad 
\cH_{ab} = {\mathfrak D}_a a_b -  {\mathfrak D}_b a_a - \cT_{ab}{}^c a_c~.
\eea
We also choose the WZ gauge \eqref{6.12} for the vector multiplet. 
Then, the other component fields of $G$ are:
\bea
\big[ \cD_{ (\a } , \bar \cD_{\b)} \big] G| &=&\hf a_{\a\b}
~,
\\
\cD^2\cDB_\a G|
&=&
2\ri\u_\a
~,
\\
\cDB^2\cD^2G|
&=&
-2\ri\bD_{b}a^{b}
-2\big(\psi_a\g^a \u+\bar{\psi}_a\g^a\bar{\u}\big)
-2\ri\cY\bar{\psi}_a\psi^{a}
-\bar{\psi}_a\g^b\psi^{a}a_b
-2\cZ
~.~~~~~~
\eea

The component Lagrangian corresponding to the action \eqref{SVM000000} is
\bea
L_{\rm VM}&=&
\frac{1}{4}\cY^{-1}\cB^a\cB_a
-a_a\cF^a
-\cY^{-1}(\hat{\bD}^a\cY)\hat{\bD}_a\cY
-\frac{1}{2}\cY\cR+\frac{1}{4}{\cY}^{-1}\cZ^2
\non\\
&&
-\ri\cY^{-1}\big(\bar{\u}\g^a\hat{\bD}_a\u+\u\g^a\hat{\bD}_a\bar{\u}\big)
-\hf\cY^{-2}\cB_a\bar{\u}\g^a\u
-\frac{\ri}{2}{\cY}^{-2}\cZ\bar{\u}\u
-\frac{1}{2}\cY^{-3}\u^2\,\bar{\u}^2
\non\\
&&
+\k\Big{[}\,-\cY\cZ
-\frac{1}{4}a_a\cB^b
-\ri\bar{\u}\u
+\frac{\ri}{2}\ve^{abc}\cY^2{\bar\psi}_a\g_b\psi_c
-\frac{1}{4}\ve^{abc}\cY a_c{\bar\psi}_a \psi_b
\Big{]}
\non\\
&&
+\Big\{
-\frac{1}{4}\ve^{abc}\big(2\bar{\u}\g_a{\bm{\bar\psi}}_{bc}+\ri\cY \psi_{a}{\bm{\bar\psi}}_{bc}\big)
-\hf\cY^{-1}\psi_a\g^a\tilde{\g}^b\u\Big(\hat{\bD}_b \cY-\frac{\ri}{2}\cB_b\Big)
\non\\
&&~~~~
+\frac{1}{4}\cY^{-1}\cZ\psi_a\g^a\u
-\frac{\ri}{4}\cY^{-2}\psi_a\g^a\bar{\u}\,\u^2
+\frac{1}{4}\cY^{-1}\ve^{abc}\psi_a\g_b\psi_c\,\u^2
\non\\
&&
~~~~
+\k\Big{[}\,
\cY\psi_a\g^a\u
-\frac{\ri}{4}\ve^{abc}a_a\psi_b\g_c \u
\Big{]}
+{\rm c.c.}
\Big\}
~.
\label{SLVM111111}
\eea
Here we have introduced new covariant derivatives
\bea
\hat{\bD}_a\cY
&:=&
\bD_a \cY
-\hf\psi_a\u
-\hf\bar{\psi}_a\bar{\u}
~,
\\
\hat{\bD}_a\u^\a
&:=&
\bD_a\u^\a
-\frac{\ri}{2}\bar{\psi}_a{}^\a\cZ
+\frac{\ri}{2}(\bar{\psi}_a\tilde{\g^b})^{\a}\Big(
\hat{\bD}_b \cY
-\frac{\ri}{2}\cB_b
\Big)
~.
\eea


\section{The action for conformal supergravity}\label{AppendixC}

In the family of  $\cN=2$ locally supersymmetric  theories  in three dimensions, 
conformal supergravity \cite{RvN} is one of the oldest.  Originally 
it was constructed by gauging the 3D $\cN=2$ superconformal algebra,  ${\mathfrak{osp}}(2|4)$,
in ordinary spacetime, as a direct generalization of the formulation for $\cN=1$ 
conformal supergravity \cite{vN}
(the latter theory being a natural reformulation of topologically massive $\cN=1$ supergravity \cite{DK,Deser}). 
The construction in \cite{RvN} was soon generalized
to the case of $\cN$-extended conformal supergravity \cite{LR89}.
In accordance with \cite{LR89}, 
the action for  $\cN$-extended conformal supergravity is a 
locally supersymmetric completion of
the gravitational and SO$(\cN)$ gauge Chern-Simons terms.
This action is on-shell for $\cN \geq 3$, and therefore 
its applications are rather limited.\footnote{Recently, 
off-shell conformal supergravity actions 
have been constructed for the cases $\cN=3,4,5$ \cite{BKNT-M2} and 
$\cN=6$ \cite{NT,KNT-M}. Upon elimination of the auxiliary fields, these actions reduce
to those proposed in \cite{LR89} only for $\cN=3,4,5$.  In the case $\cN=6$, however, 
the on-shell version of the off-shell action in \cite{NT,KNT-M} contains an additional U(1) 
gauge Chern-Simons term as compared with \cite{LR89}.}
As concerns the off-shell $\cN=1$ and $\cN=2$ component actions \cite{RvN,vN}, it 
appears useful to re-cast them in a superfield form, simply because all $\cN=1$ and $\cN=2$
locally supersymmetric  matter
systems are naturally formulated in superspace. 

As mentioned in the introduction, Refs.   \cite{KLT-M11,KT-M11} described 
the most general matter couplings to conformal supergravity in the cases $1\leq \cN \leq 4$, 
 including the off-shell formulations for Poincar\' e and AdS supergravity theories. 
But no conformal supergravity action was considered in these publications, 
due to the fact that 
an alternative action principle is required in order to describe pure conformal supergravity. 
Building on the earlier incomplete results in 
 \cite{GGRS,ZP88,ZP89},
the action for $\cN=1$ conformal supergravity 
has recently been constructed  in terms of the 
superfield connection as a superspace integral \cite{KT-M12}.
However, such a construction becomes impossible starting at $\cN=2$.\footnote{If 
a prepotential formulation is available, the conformal supergravity action may be written 
as a superspace integral in terms of the prepotentials. Currently, the prepotential formulations 
are known only for the cases $\cN=1$ \cite{GGRS} and $\cN=2$ \cite{Kuzenko12}.}
This is because (i)  the spinor and vector parts of the superfield connection 
have  positive dimension equal to $1/2$ and 1 respectively; 
and (ii) the dimension of the full superspace measure is  $(\cN- 3)$. 
As a result, it is not possible to construct 
contributions to the action 
that are cubic in the superfield connection for $\cN\geq 2$. 

Nevertheless, it was argued in \cite{KT-M12} that off-shell conformal supergravity actions 
(assuming their existence) may be realized in terms of the curved superspace geometry  given in
\cite{HIPT,KLT-M11} (also known as SO$(\cN)$ superspace)
provided one makes use of the superform approach for the construction of supersymmetric invariants. 
Such a realization was explicitly worked out in \cite{KT-M12} for the case $\cN=1$, 
and a general method  of constructing conformal supergravity actions for $\cN>1$ was outlined. 
However, it turns out that SO$(\cN)$ superspace \cite{HIPT,KLT-M11}
is not an optimum setting to carry out this program, see \cite{BKNT-M1} for a detailed discussion.   
${}$From a technical point of view, the derivation of the conformal supergravity actions
greatly simplifies if one makes use 
of the so-called $\cN$-extended conformal superspace of \cite{BKNT-M1},
which is a novel formulation for conformal supergravity. 
The SO$(\cN)$ superspace of \cite{HIPT,KLT-M11} is obtained from 
the $\cN$-extended conformal superspace by gauge fixing certain local symmetries, 
see \cite{BKNT-M1} for more details. In conformal superspace, the action for $\cN=2$ 
conformal supergravity is simply the Chern-Simons term associated with  ${\mathfrak{osp}}(2|4)$
\cite{BKNT-M2}. Below we re-formulate this action in SO(2) superspace.

\subsection{Conformal superspace and SO(2) superspace}

Conceptually,  the $\cN=2$  conformal superspace of \cite{BKNT-M1}  
corresponds to a certain gauging 
of the superconformal algebra  ${\mathfrak{osp}}(2|4)$ in superspace \cite{BKNT-M1}.
The corresponding covariant derivatives $\nabla_A$ 
include two types of connections: (i)  
 the Lorentz and U(1)$_{R}$ connections (as in SO(2) superspace); and 
 (ii) those associated with 
the dilatation ($\mathbb{D}$), 
special conformal ($K_a$) and $S$-supersymmetry ($S_\a,\,\bar{S}^\a$) generators 
of the $\cN=2$ superconformal algebra. 
To emphasize this grouping, the covariant derivatives $\de_A$ 
can be written in the form\footnote{The connections in \eqref{C.1}  differ in sign from those used in 
 \cite{BKNT-M1}.}
\bsubeq \label{C.1}
\bea
\de_A&=&\hat{\cD}_A
+B_A\mathbb{D}
+\frak{F}_A{}^bK_b
+\frak{F}_A{}^\b S_\b
+\widetilde{\frak{F}}_A{}_\b \bar{S}^\b~,
\label{conf-dev}
\eea
where we have denoted 
\bea
\hat{\cD}_A&=&E_A{}^M\pa_M-\hat{\O}_A{}^a\cM_a+\ri \,\hat{\F}_A\cJ
~.
\label{degauged-dev}
\eea
\esubeq
By construction, the operators $\de_A$ are subject to certain covariant constraints \cite{BKNT-M1} 
such that the entire algebra of covariant derivatives is expressed in terms of a single primary 
superfield -- the super Cotton tensor $\cW_{\a\b}$. 

As demonstrated in \cite{BKNT-M1}, the conformal superspace is intimately related to the SO(2) superspace via a 
de-gauging procedure.
The crucial observation  here is that the local special conformal and $S$-supersymmetry gauge freedom 
can be used to switch off the dilatation connection, $B_A=0$.
In this gauge, there remains no residual special conformal and $S$-supersymmetry gauge freedom, 
but the covariant derivatives \eqref{conf-dev} still include  
the  connections associated with the generators $K_b$, $S_\b$ and $\bar S^\b$. 
These connections are tensor superfields with respect to the remaining local Lorentz and U(1)$_{R}$
symmetries. From the constraints obeyed by the conformal covariant derivatives, 
one may deduce that the operators $\hat{\cD}_A$ look like
\bea
\hat{\cD}_a=\cD_a+\ri \, \cC_a\cJ~,\qquad
\hat{\cD}_\a=\cD_\a
~,\qquad
\hat{\cDB}^\a=\cDB^\a ~,
\label{degauged-cov-2-1}
\eea
where $\cD_A$ are the covariant derivatives of the SO(2) superspace, 
as defined in section \ref{geometry},
and $\cC_a$ is one of the corresponding torsion superfields.
The connections $\frak F$'s are uniquely determined as functionals of the torsion 
superfields of the SO(2) superspace. 
In terms of the one-forms ${\frak{F}}^\a := E^B {\frak{F}}_B{}^\a$ and 
$\widetilde{\frak{F}}_\a :=E^B \widetilde{\frak{F}}_{B\a}$, one obtains
\bsubeq
\bea
\frak{F}^\a
&=&
E^b
\Big{[}
-\hf(\g_b)_{\b\g} {\bm C}^{\a\b\g}
+\frac{1}{6}(\g_b)^{\a\b}\big(\ri\cDB_\b\bar{R} +  \cD_{\b}\cS\big)
\Big{]}
-E^\a\bar{R}
+\bar{E}_\b\Big{[}
 \cC^{\b\a} +\ri\ve^{\b\a}\cS
\Big{]}
~,~~~~~~
\\
\widetilde{\frak{F}}_\a
&=&
E^b
\Big{[}
-\hf(\g_b)^{\b\g}\bar{{\bm C}}_{\a\b\g}
+\frac{1}{6}(\g_b)_{\a\b}\big( \ri\cD^\b R- \cDB^{\b}\cS\Big)
\Big{]}
-E^\b\Big{[}
\cC_{\b\a} 
+\ri\ve_{\b\a} \cS
\Big{]}
-\bar{E}^\a R
~.~~~~~~~~
\eea
\esubeq

\subsection{Curvature two-forms}

In SO(2) superspace,  there exists a two-parameter freedom to define the vector covariant derivative.
Instead of dealing with $\cD_{a}$, one may  work equally well with a deformed covariant derivative
${\mathfrak D}_{a} $ defined by
\bea
{\mathfrak D}_{a} = \cD_{a} +\l \cS M_{a} + \r \ri \,\cC_{a} \cJ ~,
\label{vcd}
\eea
where $\l$ and $\r$ are real parameters.
A natural question is the following: What is special about the deformation \eqref{degauged-cov-2-1}?
Here we answer this question. 

Let us introduce the torsion and curvature tensors for the covariant derivatives 
\eqref{degauged-cov-2-1}, 
\bea
[\hat{\cD}_A,\hat{\cD}_B\}
=
\hat{T}_{AB}{}^C\hat{\cD}_C
-\hat{R}_{AB}{}^c\cM_c
+\ri\hat{R}_{AB}\cJ
~.
\label{degauged-algebra}
\eea
Associated with the Lorentz and U(1)$_R$ curvature tensors are the 
following two-forms: 
$\hat{R}^c=\hf E^B\wedge E^A \hat{R}_{AB}{}^c$ and 
$\hat{R}=\hf E^B\wedge E^A \hat{R}_{AB}$.
The explicit expressions for these two-forms are:
{\allowdisplaybreaks\bsubeq\bea
\hat{R}^c&=&\hf E^\b\wedge E^\a\Big{[}
4\bar{R}(\g^c)_{\a\b}
\Big{]}
+\bar{E}_\b\wedge E^\a\Big{[}
-4\ri\cS(\g^c)_{\a}{}^{\b}
-4\d_{\a}^{\b}\cC^c
\Big{]}
+\hf \bar{E}_\b\wedge \bar{E}_\a\Big{[}
-4{R}(\g^c)^{\a\b}
\Big{]}
\non\\
&&
+E^\b\wedge E^a\Big{[}
(\g_a)_\b{}^{\g}(\g^c)^{\d\r}{\bm C}_{\g\d\r}
+\frac{1}{3}\big(\d_\b^\g\d_a^c+2\ve_{ab}{}^{c}(\g^b)_{\b}{}^{\g}\big)
\big(
2{\cD}_{\g}\cS
+\ri{\cDB}_{\g}\bar{R}
\big)
\Big{]}
\non\\
&&
+\bar{E}_\b\wedge E^a\Big{[}
(\g_a)^{\b\g}(\g^c)^{\d\r}\bar{{\bm C}}_{\g\d\r}
+\frac{1}{3}\big(\ve^{\b\g}\d_a^c+2\ve_{ab}{}^{c}(\g^b)^{\b\g}\big)\big(
2{\cDB}_{\g}\cS
-\ri{\cD}_{\g}R
\big)
\Big{]}
\non\\
&&
+\hf E^b\wedge E^a\,\ve_{abd}\Big{[}\,
\frac{1}{4}(\g^d)^{\a\b}(\g^c)^{\t\d}
\big(\ri\cD_{(\t}\bar{{\bm C}}_{\d\a\b)}
+\ri\cDB_{(\t} {\bm C}_{\d\a\b)}\big)
-4\cC^{d}\cC^c
\non\\
&&~~~~~~~~~~~~~~~~~~~~~
+\eta^{cd}\Big(
\frac{2\ri}{3}({\cD}^\a{\cDB}_{\a}\cS)
+\frac{1}{6}({\cD}^2 R+{\cDB}^2 \bar{R})
-4\cS^2
-4\bar{R}R
\Big)
\Big{]}
\label{degauged-2-form-L}
~,\\
\hat{R}&=&
\bar{E}_\b\wedge E^\a\Big{[}
 4\ri \cC_{\a}{}^{\b}
+4\d_{\a}^{\b}\cS
\Big{]}
+E^\b\wedge E^a\Big{[}
\ri(\g_a)^{\g\d}{\bm C}_{\b\g\d}
-\frac{1}{3}(\g_a)_\b{}^{\g}\big(
{\cDB}_{\g}\bar{R}
-2\ri\cD_{\g}\cS
\big)
\Big{]}
\non\\
&&
+\bar{E}_\b\wedge E^a\Big{[}
\ri(\g_a)_{\g\d}\bar{\bm C}^{\b\g\d}
-\frac{1}{3}(\g_a)^\b{}_{\g}\big(
{\cD}^{\g}R
+2\ri{\cDB}^{\g}\cS
\big)
\Big{]}
-\hf E^b\wedge E^a\Big{[}
\ve_{abc} \cW^c 
\Big{]}
~,~~~
\label{degauged-2-form-J}
\eea
\esubeq
}with $\cW^c $ the super Cotton tensor, eq. \eqref{Cotton-2}. 
For completeness, we also reproduce
the expressions for the two-forms 
${R}^c=\hf E^B\wedge E^A R_{AB}{}^c$ and
${R}=\hf E^B\wedge E^A R_{AB}$, 
where the curvature tensors are those which appear in \eqref{algebra}. 
Direct calculations give
\bsubeq\bea
\hat{R}^c&=&R^c
~,
\label{degauged-2-form-L-2}
\\
\hat{R}&=&R
+\bar{E}_\b\wedge E^\a\Big{[}
2\ri \cC_{\a}{}^{\b}
\Big{]}
+E^\b\wedge E^a\Big{[}
\frac{\ri}{2}(\g_a)^{\g\d}{\bm C}_{\b\g\d}
-\frac{1}{6}(\g_a)_\b{}^{\g}\big(
{\cDB}_{\g}\bar{R}
+4\ri\cD_{\g}\cS
\big)
\Big{]}
\non\\
&&
+\bar{E}_\b\wedge E^a\Big{[}
\frac{\ri}{2}(\g_a)_{\g\d}\bar{{\bm C}}^{\b\g\d}
-\frac{1}{6}(\g_a)^\b{}_{\g}\big(
{\cD}^{\g}R
-4\ri{\cDB}^{\g}\cS
\big)
\Big{]}
\non\\
&&
-\hf E^b\wedge E^a\Big{[}
\ve_{abc}\ve^{cef}{\cD}_{e}\cC_{f}
\Big{]}
~.
\label{degauged-2-form-J-2}
\eea
\esubeq

The unique feature of the deformation \eqref{degauged-cov-2-1}
is that the top component of the U(1) curvature two-form 
\eqref{degauged-2-form-J} is a primary superfield equivalent to the super-Cotton 
tensor.\footnote{SMK and GT-M are grateful to Joseph Novak for this observation.} 

\subsection{Closed three-form}

In $\cN=2$ conformal superspace, the Chern-Simons three-form ${\bm\S}_{\rm CS}$ 
is characterized by the following properties \cite{BKNT-M2}: (i) it is closed,
$\rd \, \! {\bm\S}_{\rm CS}=0$; and (ii) under the gauge transformations, it is invariant modulo
  exact terms. This three-form generates the off-shell action for $\cN=2$ conformal 
 supergravity. In this subsection, we follow the de-gauging procedure of \cite{BKNT-M1} to obtain
an expression for this closed three-form in SO(2) superspace, 
$\frak{J}:={\bm\S}_{\rm CS}|_{\rm de-gauged}$.
The calculations are straightforward, and we present only the final result.

The three-form $\frak{J}$ turns out to be
\bea
\frak{J} &= &
- {\hat{R}}^a \wedge \hat{\Omega}_a 
+  \frac{1}{6} \hat{\Omega}^c \wedge \hat{\Omega}^b \wedge \hat{\Omega}^a \ve_{abc}
-2{{\hat{R}}}\wedge \hat\Phi
-8\ri E^a \wedge \frak{F}^\a \wedge \bar{\frak{F}}_\b (\g_a)_{\a}{}^{\b} 
~.
\label{conf-CS-degauged-final}
\eea
The expression for $\frak{J}$ is naturally written in terms of the deformed covariant  
derivatives $\hat{\cD}_A$.
Making use of \eqref{degauged-cov-2-1}, it is a simple exercise 
to rewrite \eqref{conf-CS-degauged-final}
in terms of the original covariant derivatives $\cD_A$. 

It is interesting to note that the closed  three-form $\frak{J}$ can be written as
\bsubeq \label{C.9}
\bea
\frak{J} &= &\hat{\S}_{\rm CS} -\S_{ T}~,
\eea
where we have introduced 
\bea
\S_{ T}&=&-8\ri E^a \wedge \frak{F}^\a \wedge \widetilde{\frak{F}}_\b (\g_a)_{\a}{}^{\b} 
~,\\
\hat{\S}_{\rm CS}&=&
 {\hat{R}}^a \wedge \hat{\Omega}_a 
-  \frac{1}{6} \hat{\Omega}^c \wedge \hat{\Omega}^b \wedge \hat{\Omega}^a \ve_{abc}
+2{{\hat{R}}}\wedge \hat\Phi
~.
\eea
\esubeq
The three-form $\hat{\S}_{\rm CS}$ is a sum of the Lorentz and U(1)$_R$ Chern-Simons 
three-forms associated with the covariant derivatives $\hat{\cD}_A$.
The components of $\S_{ T}$ are functions of the torsion tensor and its covariant derivatives only;
this is why  $\S_{ T}$ was called the torsion induced three-form in \cite{KT-M12}.
The three-forms $\hat{\S}_{\rm CS}$ and $\S_{T}$ satisfy the equations
\bea
\rd\S_{T}=\rd\hat{\S}_{\rm CS}=\hat{R}^a\wedge \hat{R}_a+2\hat{R}\wedge\hat{R}~.
\label{C.10}
\eea
By construction, the closed  three-form $\frak{J}$ is invariant under 
the super-Weyl transformations modulo exact terms. 
In fact,  the relative coefficient between  the Lorentz and U(1)$_R$ Chern-Simons  terms 
in \eqref{C.10} is  fixed by the condition that $\frak J$ be 
 super-Weyl invariant modulo exact terms.

The covariant derivatives \eqref{degauged-cov-2-1} and the closed three-form 
\eqref{C.9} constitute the unique solution to the $\cN=2$ problem posed in \cite{KT-M12}.

\subsection{Conformal supergravity action}

Using the three-form 
$\frak{J}=\frac{1}{3!}E^C\wedge E^B\wedge E^A \frak{J}_{ABC}
=\frac{1}{3!}\rd z^P\wedge \rd z^N\wedge \rd z^M \frak{J}_{MNP}$, 
we can write down the 
locally supersymmetric and super-Weyl invariant action 
($\ve^{mnp}:=\ve^{abc}e_a{}^m e_b{}^n e_c{}^p$)
\bea 
S = \int_{\cM^3} \frak{J} = \int \rd^3 x\,e  \,{}^* \frak{J} |_{\q=0}\ , \qquad 
{}^*\frak{J} = \frac{1}{3!} \ve^{mnp} \frak{J}_{mnp} 
\label{ectoS}
~.
\eea
Upon implementing the component and gauge fixing reduction described in section \ref{components},
the action becomes
\bea
S =  \frac{1}{4} \int \rd^3 x \,e \,\ve^{abc} &\Big{[}&{\cR}_{bc}{}_{fg}  \omega_{a}{}^{fg} 
+ \frac{2}{3}{\omega}_{af}{}^g {\omega}_{bg}{}^h {\omega}_{ch}{}^f 
- 4 {{\cF}}_{ab} {b}_c 
+  \ri{\bm{\bar{\psi}}}_{bc}\g_d\tilde{\g}_a\ve^{def} {\bm\psi}_{ef}
\Big{]} 
~.~~~~~~~~~
\label{C.12}
\eea
This is the component action for $\cN=2$ conformal supergravity of \cite{RvN}.

\begin{footnotesize}

\end{footnotesize}

\end{document}